\RequirePackage{fix-cm}
\documentclass[twocolumn,epjc3]{svjour3}  
\smartqed  
\journalname{Eur. Phys. J. A}

\usepackage[dvips,dvipdf]{graphicx}

\usepackage{xcolor,amssymb}
\usepackage{color}
\usepackage{epsf}
\usepackage{soul}
\usepackage{cuted}

\usepackage{morefloats}

\colorlet{darkgreen}{green!50!black}
\colorlet{brightyellow}{yellow!75!red}
\colorlet{orange}{red!50!yellow}
\colorlet{darkblue}{blue!60!black}
\colorlet{darkred}{red!80!black}

\begin{document}

\title{Comparison of \={N}N optical models}

\author{Jaume Carbonell\thanksref{e1,addr1}  
\and
Guillaume Hupin\thanksref{e2,addr1}
\and
Slawomir Wycech\thanksref{e3,addr2}  
}
\thankstext{e1}{e-mail: jaume.carbonell@ijclab.in2p3.fr}
\thankstext{e2}{e-mail: hupin@ijclab.in2p3.fr}
\thankstext{e3}{e-mail: Slawomir.Wycech@ncbj.gov.pl}
\institute{Universit\'e Paris-Saclay, CNRS/IN2P3, IJCLab, 91405 Orsay, France \label{addr1}
\and
National Centre for Nuclear Research, Warsaw, Poland\label{addr2}
}

\date{Received: date / Accepted: date}

\maketitle

\today


\bigskip
\begin{abstract}
We,   compare the  strong part of the \={N}N interaction obtained
by the Nijmegen partial wave analysis and the results of some of the
most popular \={N}N optical potentials in configuration space.
We have found severe discrepancies in most of the partial waves, especially above $p_{Lab}$=400 MeV/c
where the partial wave analysis displays  a resonant-like structure in the $^{31}$S$_0$ and $^{33}$P$_0$ waves.
Some theoretical difficulties to interpret  this behaviour  in terms of dynamical resonances are pointed pout and 
an alternative explanation is suggested.
A much better stability is observed in the low energy parameters, apart from some discrepancies due to the presence
of near-threshold quasi-bound states in particular waves.
Large deviations have also been found between the corresponding potentials, at short and medium-range ($r\gtrsim 1$ fm) distances.
\end{abstract}

\section{Introduction}

In comparison  with the Nucleon-Nucleon (NN) case, the Antinucleon-Nucleon- (\={N}N) interaction remains poorly known. 
The reason for that is, on  one hand the relatively limited number of   \={N}N low-energy  data
and on the other hand  the intrinsic difficulty of theoretically describing a system which has
hundreds of open annihilation many-body channels at rest.
See for instance 
\cite{REV_Annihilation_DGMF_PPNP29_1992,REV_LEAR_KBMR_PREP368_2002,REV_LEAR_KBR_PREP413_2005} 
and references therein.

A rigorous theoretical approach of this physical problem in  its full complexity is far beyond our  possibilities, both formal and computational,
and it will remain so probably for a long time.
There are however phenomenological  ways to model  the low energy \={N}N physics and obtain a reasonable description of the
existing experimental data, provided one renounces to describe each particular annihilation channel and by introducing a relatively large number of parameters.
A  successful example is provided by the \={N}N optical models, which date from the early days of antiproton physics \cite{Levy_NCVIII_1958}, and whose  main
properties have been recently reviewed  in \cite{REV_NNB_JMR_Frontiers_2020,REV_NNB_JMR_Handbook_2022}.

The first accurate description of the \={p}p experimental results 
was provided by the energy-dependent  partial wave analysis of Nijmegen group \cite{ZT_NNB_Nijm_PW_2012} (NPWA) 
which presents an almost perfect description ($\chi^2\approx1$) of the existing data below $p_{Lab}<$925 MeV/c,
although after applying a severe rejection criteria.
In this analysis,  the long- and medium-range \={N}N interaction is given by a  one- plus two-pion exchange potential ($V_{\pi} \equiv V_{1\pi}+V_{2\pi}$) 
at N2LO
chiral EFT detailed in  \cite{V2pi_NN_NPW_PRL82_1999}.
This potential is matched at $b$=1.2 fm to a state and energy-dependent  complex boundary conditions  
which parametrise  the short range physics, in particular the very complex annihilation dynamics.
This is realised by fixing,  for each energies E and partial wave $\alpha=\{ T,L,S,J\}$,  the logarithmic derivative 
of  the corresponding wave function at $r$=$b$: $P_{\alpha}(E)$=$b\left( \Psi'_{\alpha}/\Psi_{\alpha} \right)_{r=b}$.
The parameters of the NPWA, i.e. the low energy constants (LEC's) of $V_{\pi}$ ($c_1,c_3,c_4$) and the complex boundary conditions $P_{\alpha}$, were determined
in \cite{ZT_NNB_Nijm_PW_2012} by a fit to the p\=p scattering data.
The LEC's found in this way, were  compatible with previous determinations from $pp$  \cite{V2pi_NN_NPW_PRL82_1999}  and a combined fit of $pp$ and $pn$ 
scattering data \cite{V2pi_NN_NPW_PRC67_2003}.

The possibility of performing a PW analysis of the \={N}N data has been questioned  \cite{JMR_Comment_NIJMPW_1995,REV_NNB_JMR_Frontiers_2020}, 
as it requires the determination of, at least, twice as many parameters  as  in the NN case,
the number of \={N}N partial waves is higher than for NN,
and the available \={N}N data are orders of magnitude less abundant.
For instance, the \={N}N  S-matrix 
for a tensor uncoupled states, is no longer determined by a real parameter $\delta$ as in the unitary case ($S=e^{2i\delta}$),  
but by a complex quantity $\delta_C=\delta_R+i \delta_I$ whose (positive) imaginary part $\delta_I$ 
controls the inelastic processes through the parameter $\eta=\mid S\mid =e^{-2\delta_I}$  ($0<\eta<1$),  thus
allowing the same formal expression for the S-matrix ($S=e^{2i\delta_{C}}$).
This criticism is sound and can eventually  rise  some questions about the uniqueness of the solution,
especially when the inelasticity parameter $\eta$,  and so the S-matrix itself, are very small. 
There is, however,  no doubt that  the results presented in \cite{ZT_NNB_Nijm_PW_2012} provide an excellent description of 
 the selected data set and constitute at least one reliable solution in the domain 100 MeV/c $<p_{Lab}<$1000 MeV/c.

Once determined the parameters of the \={p}p PW analysis,  the authors of Ref.  \cite{ZT_NNB_Nijm_PW_2012} 
removed the  Coulomb potential and  the 
$n-p$ mass difference ($\Delta_0\equiv m_n-m_p$), and obtained the  strong \={N}N phase-shifts in the isospin symmetry, 
which are in fact the  non trivial and interesting part of the interaction.
These results,   which can be considered to a large extent as being model independent,
are extremely useful for a critical  comparison between the different models, without
directly relying  on the experimental observables. 
The former,  involve usually contributions of many partial waves
and can hide eventual significant disagreements among the different interaction models.
The strong \={N}N Nijmegen phase shifts constitute the basis of our further analysis.

The strong  \={N}N phase-shifts provided by the Nijmegen group were also the starting point to determine the parameters of the 
most recent \={N}N J\"{u}lich  potential \cite{Haidenbauer_JHEP_2017}
\footnote{It is worth mentioning that, under this denomination,
one can include a series of previous works on \={N}N  interactions developed since the 90's, based on the G-parity transform of the meson-exchange
NN Bonn model \cite{VNNB_Julich_PRC44_1991} and even  the N2LO version of the chiral EFT  \={N}N potential \cite{Haidenbauer_JHEP_2014}. 
For shortness of the notation, we will hereafter denote  J\"{u}lich \={N}N potential as the one described in Ref. \cite{Haidenbauer_JHEP_2017}}.
This potential is based on the G-parity transform of a previously established chiral EFT  NN potential  at  N3LO \cite{NN_N3LO_Epelbaum_2015}.
It contains contributions from one- and two-pion exchange and of contact terms with up to four derivatives. 
The annihilation part is taken into account by introducing  imaginary contact terms in each partial wave, regularized by gaussian form factors.
The  potential is inserted in a relativistic Lipmann-Schwinger equation to obtain the phase-shifts.
The low energy constants  of the pion-exchange part were taken from the pion-nucleon dynamics and
the remaining ones,
as well as the annihilation constants, were  adjusted to reproduce to the strong phase shifts and inelasticity parameters  of the Nijmegen PWA in the isospin basis.  
Supplemented with the Coulomb and $\Delta m$ term, this model provides an equally  good description  of
the \={p}p data as in the NPWA.  Furthermore it has been extended to  describe the zero energy protonium
results as well as  the existing \={n}p  data ($T=1$).

The J\"{u}lich potential constitutes nowadays  the most complete and accurate description of the \={N}N data,
would it be at the price of a considerable number of parameters ($\approx$ 90).
However, it has been derived in momentum space 
what makes difficult its implementation to study  more complex systems, in particular the very peripheral  and loosely bound hydrogenic  orbits  of the  \={p}-A systems.
These Coulomb-like states constitute  the cornerstone of  the PUMA research project  \cite{PUMA_EPJA_2022} that 
requires reliable theoretical predictions of the annihilation probabilities from some of them.
A recent application to the simple \={p}-d case has
been recently obtained \cite{LC_PLB820_2021} using a simplified (local) form of the J\"{u}lich potential. 
They led to significantly different predictions with respect to other existing models and it is not clear
which part of these differences is a genuine prediction of the  potential itself or results  from the simplifications.
On the other hand  the strong non local character of the J\"{u}lich  potential
 makes it difficult to be used in configuration space  calculations, where the few-nucleon 
scattering problem can be more easily solved.

In view of further applications, but also for the sake of a theoretical consistency, it is of the highest  interest to  
examine the predictive power of some of the most popular \={N}N optical  models formulated in configuration space, 
by comparing   them  at the level of strong phase shifts 
as well as with the recent -- phase  equivalent -- Nijmegen PW and J\"{u}lich results.
A previous comparison devoted to protonium level shifts and scattering lengths was published many years ago \cite{CIR_ZPA334_1989,CRW_ZPA343_1992}  
but to our knowledge no systematic study  has been performed at non zero energies.

Our goal in  the present paper is thus to  establish  a detailed comparison of the  Nijmegen PW analysis and J\"{u}lich results 
with some selected optical models  widely used in the literature and formulated in configuration space.
To this aim we will  consider  the  last updated version  (2009) of the Paris  potential \cite{PARIS_PRC79_2009}, 
the Dover-Richard models  \cite{DR1_PRC21_1980,DR2_RS_PLB110_1982}  published in 1980-82
and the Khono-Weise \cite{KW_NPA454_1985} potential formulated in 1986.
They represent different degrees of complexity in the theoretical description
and, apart from being formulated in configuration space, they have in common that: 
{\it (i)}  they make full use of conventional (not EFT) meson-exchange theory,
{\it (ii)} they were constructed before the NPWA \cite{ZT_NNB_Nijm_PW_2012}, 
and 
{\it (iii)} they were adjusted to a restricted data set.
At the end, they obtain a less accurate description of the experimental results than NPWA and J\"{u}lich model,
but they use a much smaller number of parameters.

As we will see in what follows, there exist huge disagreements
among the partial wave predictions of the considered optical models, often hidden when considering only integrated cross sections.
They claim for an urgent clarification of the \={N}N interaction
at the  two-body level, both from the theoretical as well as  from the experimental point of view,
before intending  a  minimal  model independent description of more complex
systems, furthermore involving off-shell properties of the interaction.

We will sketch in section  \ref{O_Models} the  main ingredients of the theoretical \={N}N  formalism
used in Refs.  \cite{ZT_NNB_Nijm_PW_2012,Haidenbauer_JHEP_2017} as well as in our own calculations with Paris, DR and KW optical models. 
Section  \ref{Results} is devoted to compare the strong phase shifts for the S and P  partial waves, low energy parameters
 and S- and P-wave protonium level shifts of these different models.
Section \ref{Conclusion} contains some concluding remarks.

\begin{figure}[htbp]
\begin{center}
\includegraphics[width=9.cm]{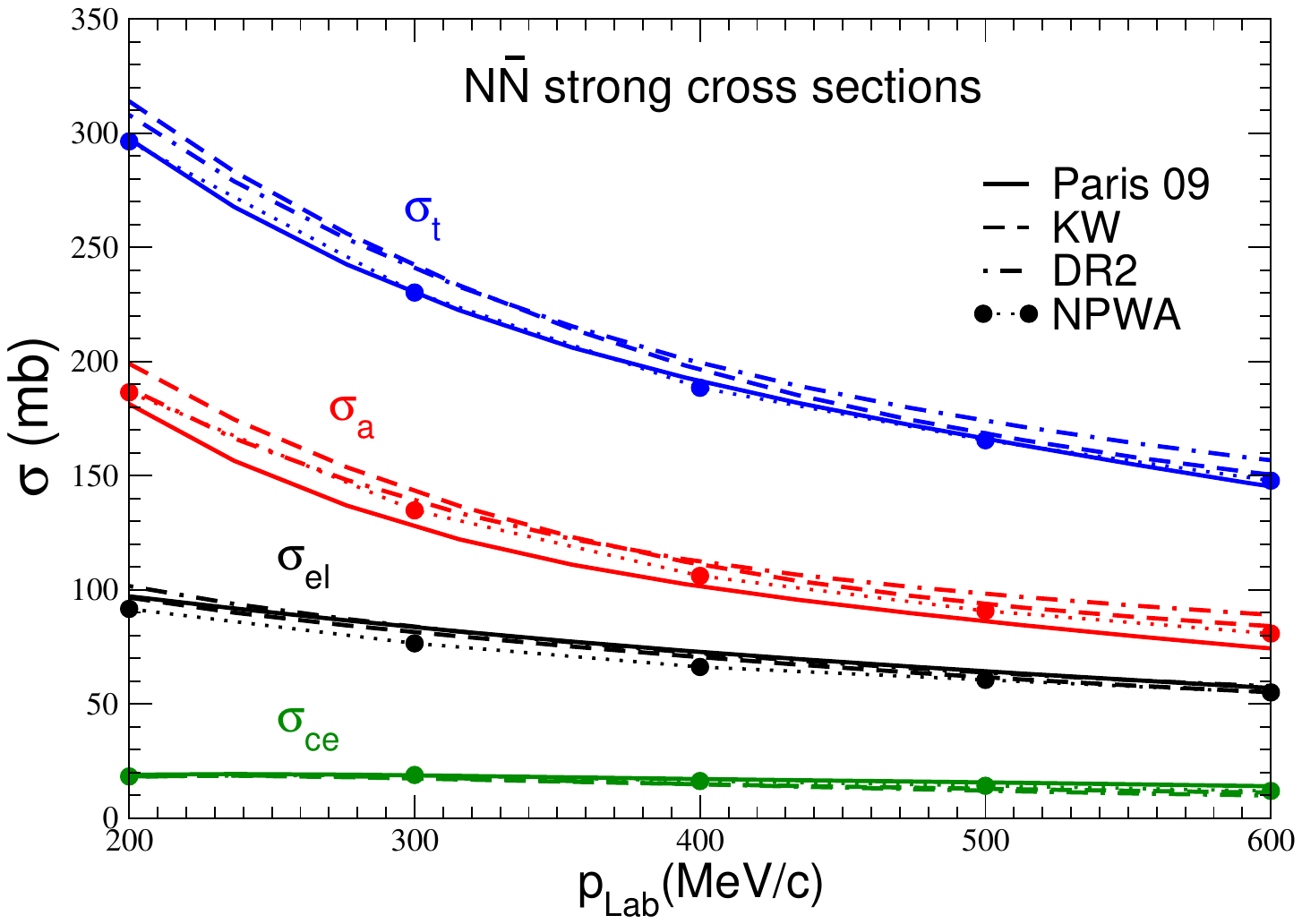} 
\caption{Integrated strong \={N}N cross sections -- elastic $\sigma_e$ (black), annihilation $\sigma_a$ (red ), charge-exchange $\sigma_{ce}$ (green) and their sum $\sigma_t$ (blue) 
--  as functions of the \={N} laboratory momenta
for DR2 (dashed dotted line), KW (dashed line)  and Paris 2009 (solid line) optical models. The results of the Nijmegen Partial Wave analysis \cite{ZT_NNB_Nijm_PW_2012}
are indicated by filled circles.}\label{sigmas}
\end{center}
\end{figure}

\section{The formalism for \={N}N optical models}\label{O_Models}
 
The strong part of the \={N}N force is  derived in the isospin  basis ($V_T$). 
However, this basis is not adapted to computing low-energy \={p}p scattering processes  due to the relevant
role of Coulomb interaction and, in a less extent, to the $n$-$p$ mass difference ($\Delta_0\equiv m_n-m_p$)
which couples, even asymptotically, the isospin states.
One uses, instead, the so called particle-basis where the  
$\mid p\bar{p}\rangle$ and $\mid n\bar{n}\rangle$ states are coupled only by the short range "charge-exchange" potential.
By adopting  the  isospin conventions \cite{Gasiorowicz_1966,CIR_ZPA334_1989} 
\begin{equation}\label{Nbar_doublet}
N = {p\choose n}  \quad   \bar{N} = {- \bar n\choose + \bar p}   \equiv  
\begin{array}{lcl}
|1/2,+1/2> &=& -|\bar n> \cr
|1/2,-1/2>  &=& +|\bar p>
\end{array}
\end{equation}
the particle basis is expressed in terms  of \={N}N  isospin  states  $\mid T,T_3>$  as
\begin{equation}\label{particle_basis}
\begin{array}{lcl c lcl}
|p\bar{p}>&=&+{1\over\sqrt2}\left\{|00> + |10>\right\}  \\
|n\bar{n}>&=&+{1\over\sqrt2}\left\{|00> - |10>\right\}  \\
|p\bar{n}>&=&-|1,+1>                                 \\
|\bar{p}n>&=&+|1,-1>           
\end{array}
\end{equation}

The $\mid p\bar{p}>$ and  $\mid n\bar{n}>$ states  can be cast into  a single state vector
\[ \mid \Psi \rangle=  \pmatrix{   \Psi_{ p\bar{p}}   \cr \Psi_{n\bar{n}}  }   \]
 which, in  the  \={N}N models that we will consider,  obeys the non-relativistic  Schrodinger equation 
\begin{equation}\label{SCHEQ_OM}
 (E- H_0 ) \mid \Psi \rangle= \hat{V}\;  \mid \Psi \rangle   
 \end{equation}
 where $E$ is the (non-relativistic)  \={p}p energy in the center of mass.
 The potential matrix
\begin{equation}\label{V_P}
 \hat{V}= \pmatrix{
V_{p\bar{p}}     & V_{ce}      \cr
V_{ce}             &       V_{n\bar{n}}                 }
\end{equation}
is expressed in terms of the isospin components $V_T$  
and the \={p}p Coulomb potential
\[ V_c(r)= - {\alpha \over r}\]
as
\begin{eqnarray}
V_{p\bar{p}}  &=& V_{N\bar N} +  V_c   \\
V_{n\bar{n}}  &=& V_{N\bar N}   + 2\Delta_0  \label{Vnnb}\\ 
2V_{N\bar N} &=&V_0+V_1 \\
2V_{ce}          &=& V_0-V_1  
\end{eqnarray}
The kinetic  energy is assumed to be channel-diagonal  with the $p$-$n$ averaged mass $m$
\begin{equation} 
H_0 = - {\hbar^2\over m} \Delta   \qquad  m={m_p+m_n\over2} 
\end{equation}

After performing the PW expansion, the reduced radial wave functions $u_i$ obey a set of
$n_c$ coupled differential equations
 \begin{equation}\label{ERR}
 u''_i + q_i^2 u_j - \sum_{j=1}^{n_c} v_{ij} u_j =0 
\end{equation} 
where $i,j$ encodes the channel indexes $\{\bar{p}p,\bar{n}n\}$ as well as the quantum number $\alpha=\{L,S,J\}$, $q_i$ the channel momenta in the center of mass
and 
\[ v_{ij}= {m}  V_{ij} \]
We  use natural units ($\hbar=c=1$) along the paper.
For the tensor uncoupled states ($^1$S$_0$,$^1$P$_1$,$^3$P$_0$,...) $n_c$=2 and for the tensor coupled ($^3$SD$_1$,$^3$PF$_2$,...) $n_c$=4.

\begin{figure*}[htbp]
\begin{center}
\vspace{-0.cm}
\includegraphics[width=8.5cm]{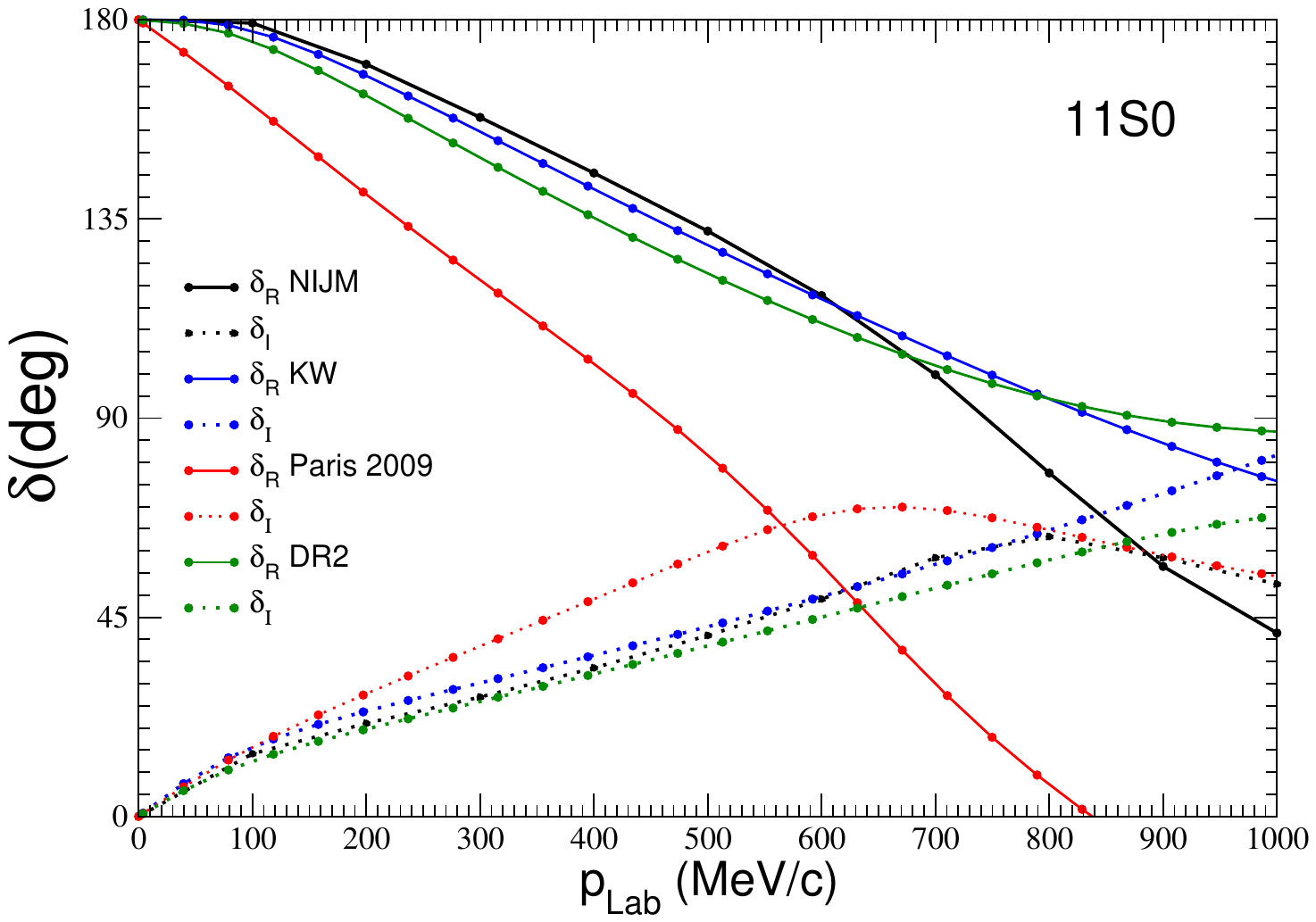} 
\vspace{-0.cm}
\includegraphics[width=8.5cm]{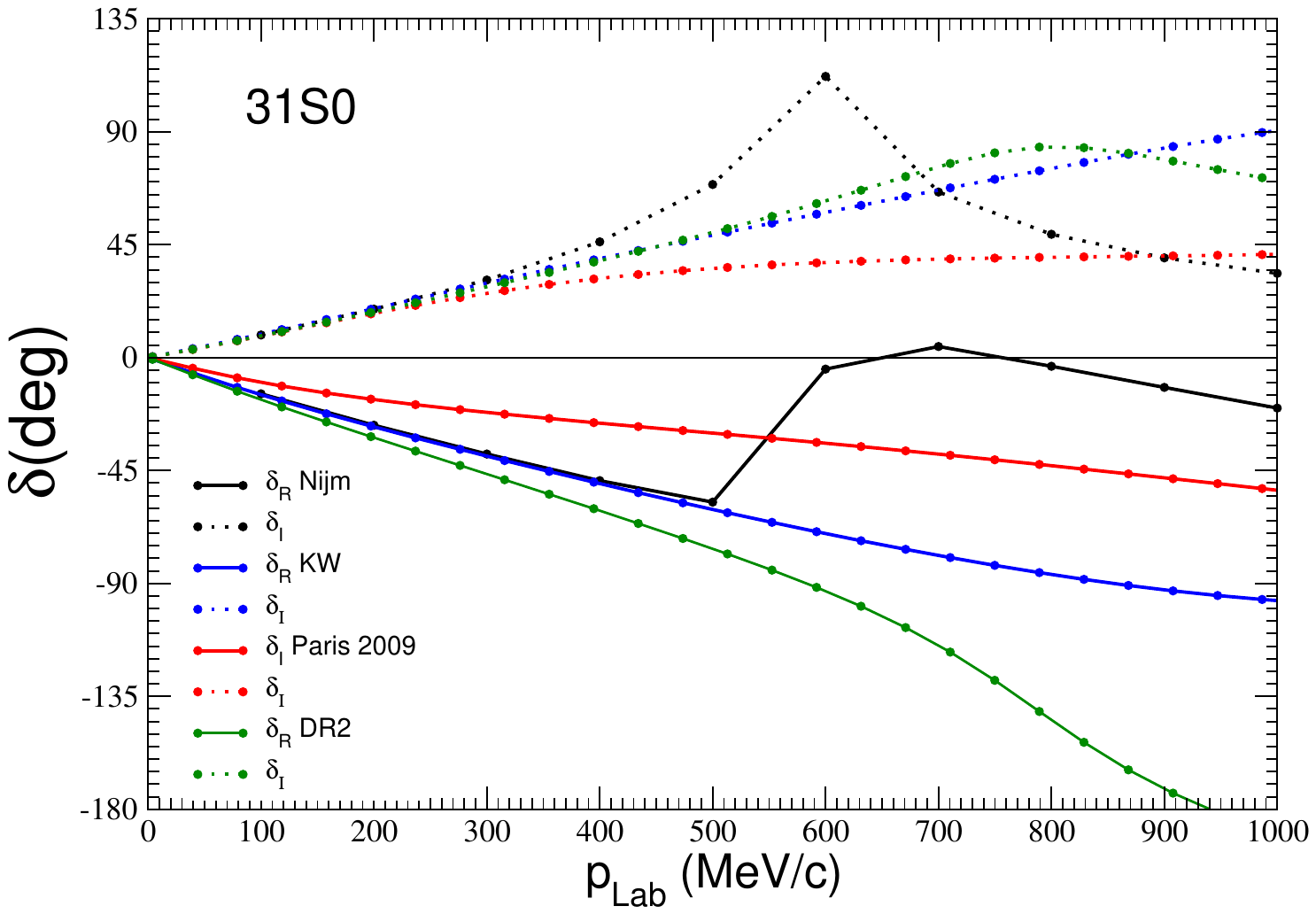}
\vspace{-0.cm}
\caption{\={N}N $^1$S$_0$  scattering  phase shifts (degres) as functions of the \={N} laboratory momenta and for different optical models.
Left panel for  T=0 state ($^{11}$S$_0$) and right one for T=1 ($^{31}$S$_0$).
Solid lines correspond to the real part $\delta_R$ and dashed lines to the (positive)  imaginary part $\delta_I$.}\label{delta_1S0}
\end{center}
\end{figure*}

The relations between the channel momenta are obtained by assuming  the same c.m. total energy ($\sqrt{s}$), which leads to:
\[ {s\over 4}= m_{\alpha}^2+ q_{\alpha}^2 \]
This gives
\begin{eqnarray}
q^2_{\bar pp} &=& m E \\
q^2_{\bar nn} &=& q^2_{\bar pp}  -  2\Delta_0 \; {m} 
\end{eqnarray}
We will denote hereafter  by $q\equiv q_{\bar{p}p}$  the c.o.m. momenta of the \={p}p driving channel.
Notice than when using the differential form (\ref{ERR}), the $n-p$ mass difference $\Delta_0$ is already
included in the channel momenta and must be removed from the potential $(\ref{Vnnb})$.

In the numerical calculations we used $m$= 938.28  MeV and $\Delta_0=m_n-m_p$=1.2933 MeV.
The \={n}n channel is  open at  the \={p}p center of mass energy $E\ge$2.5866 MeV,  i.e. $q$=0.2497 fm$^{-1}$.
The relation with the laboratory momenta is given by
\[ p_{Lab}= 2 q \sqrt{ 1 + \left({  q\over m}\right)^2 } \approx 2q \]
that is $p_{Lab}$=98.54 MeV/c.  

The strong \={N}N  potentials that we have  considered in this work take the form
\begin{equation}\label{VOM}
 V(r)=  U(r) +  W(r) 
 \end{equation}
where  real  $U$   is a  G-parity transform of a NN potential regularised below some cut-off radius $r_c$,
and $W$ is the complex potential (eventually containing also a real part) accounting for the annihilation. 
 
The Paris \={N}N model  \cite{PARIS_PRC79_2009,CLLMV_PRL_82,PLLV_PRC50_1994,ELLV_PRC59_1999}  
is based on the G-parity transform of the Paris NN potential \cite{Paris_NN_Lacombe_PRD12_1975,Paris_NN_PRC21_1980}.
It contains one- and two-pion exchange (the latter via dispersion relations), plus $\omega$ and $A_1$ potentials as part of three-pion exchange.
The real part $U$ has a central, spin-spin, spin-orbit, tensor and quadratic spin-orbit terms.
The first two are energy-dependent, what results into seven scalar amplitudes for a given isospin $T$.
They are regularized below some distance $r_c$ ($r_c$=0.84 fm or $r_c$=1.0 fm) by a cubic polynomial, whose coefficients introduce adjustable parameters. 
All together this gives nine parameters for each isospin.
The annihilation potential $W$, derived in \cite{Bachir_PhD_Paris_1980}, is purely imaginary and
has a similar spin structure as  the real part. It depends on six parameters for each $T$.
A particularity of the Paris potential is the short range character of W ($r_a=1/2m_N\approx$ 0.1 fm).
The total number of parameters of the model is $\approx30$ and  that ensures
a fairly good description ($\chi^2$/datum $\approx$ 5) of most of the existing data, without any selection criteria, including  differential cross sections and polarization observables.

\begin{figure*}[htpb]
\begin{center}
\vspace{-0.cm}
\includegraphics[width=8.5cm]{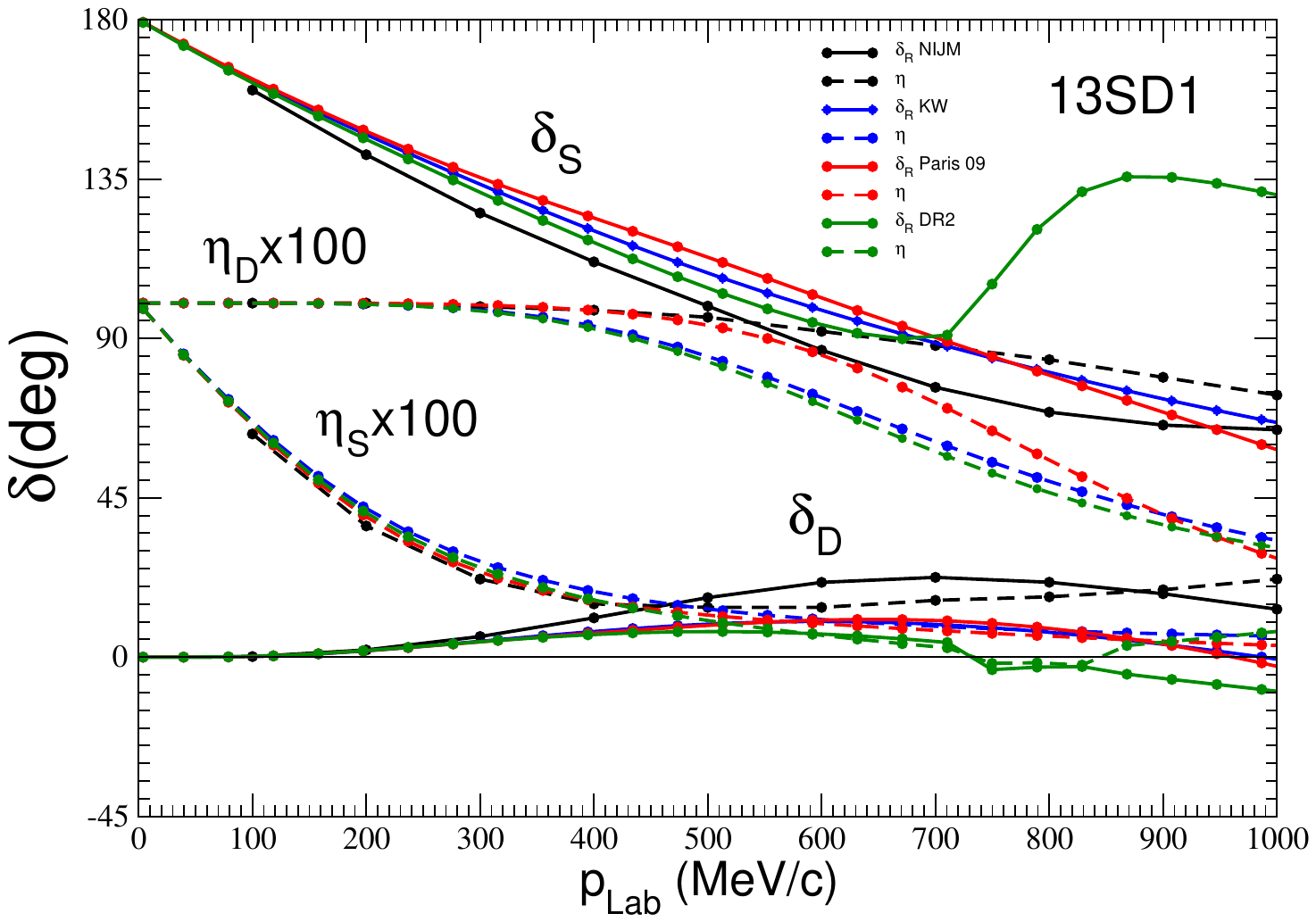} 
\includegraphics[width=8.5cm]{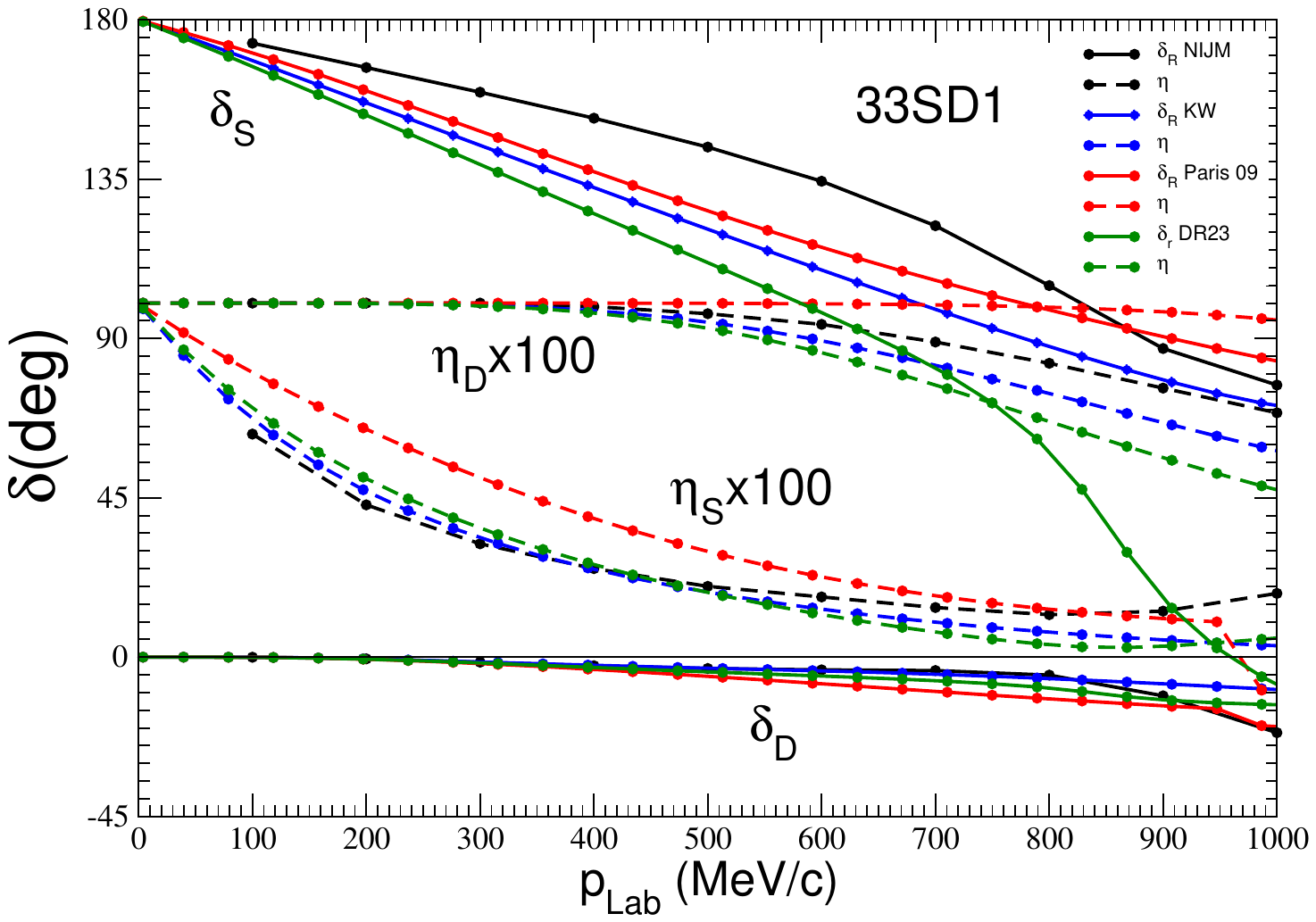}
\vspace{-.cm}
\includegraphics[width=8.5cm]{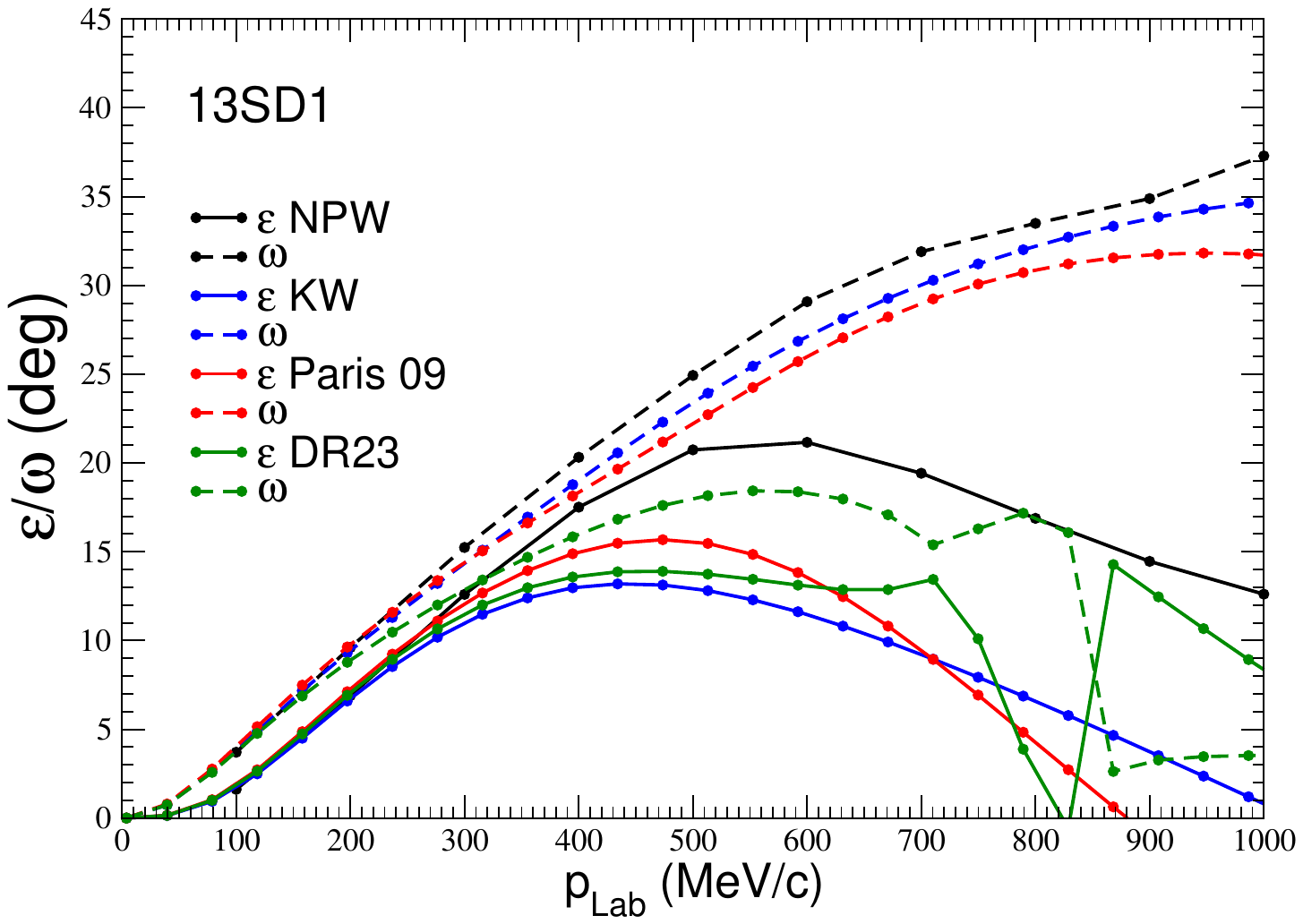} 
\vspace{-0.cm}
\includegraphics[width=8.5cm]{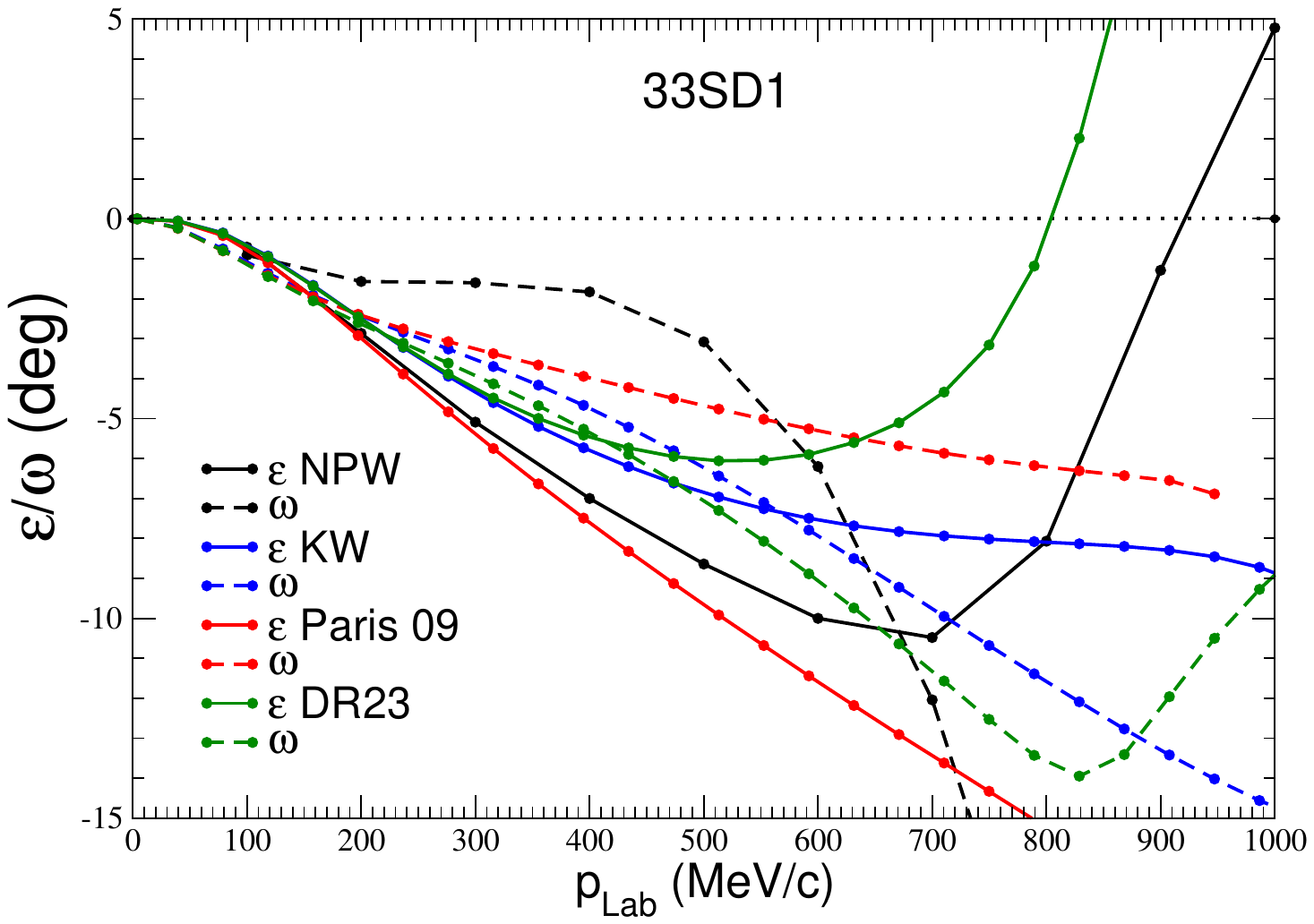}
\caption{$^3$SD$_1$ \={N}N bare phase shifts and inelasticities (upper panel) and mixing parameters  (lower panel),  
as functions of the \={N} laboratory momenta.}\label{delta_3SD1}
\end{center}
\end{figure*}

In Dover and Richard models --  DR1 version \cite{DR1_PRC21_1980}  and DR2 version  \cite{DR2_RS_PLB110_1982} -- 
$U$ is taken from a simplified version of the  NN Paris potential \cite{Paris_NN_Lacombe_PRD12_1975} 
containing $\pi,2\pi$ and $\omega$ regularized below $r_c=0.8$ fm. 
DR models were adjusted to  reproduce some  analytic parametrisations of  
the total, elastic, charge-exchange and annihilation experimental integrated cross sections in the range $0.4 < p_{Lab}<0.9$  GeV/c with  a $\chi^2$/data$\approx$0.5 for DR1.
In addition to $r_c$, there are  only four parameters which control the annihilation potential $W$.

In Kohno-Weise  model  \cite{KW_NPA454_1985}, U is taken from the NN Ueda potential \cite{U_PTP_79} with $\pi,\rho,\omega,\sigma$ meson contributions, 
regularized below $r_c$=1 fm by a $C^1$ matching to a Woods-Saxon potential.
As for DR, the parameters come only from W
and are adjusted to reproduce the \={p}p total $(\sigma_t)$, elastic $(\sigma_e)$ and charge exchange ($\sigma_{ce}$) cross sections
in the region $200<p_{Lab}<700$ MeV/c.
In this way  this model provides a good description of  the forward  \={p}p elastic differential cross sections at $p_{Lab}$=400, 500, 600 MeV/c, 
and \={p}p elastic differential cross sections at $p_{Lab}$=390,490,590 MeV/c
as well as  of differential ce at 490 and 590 MeV/c. No  $\chi^2$ is given in this analysis.

In DR and KW models,  the annihilation potential $W$ is local, energy- and state-independent. It  has the common form
\begin{equation}\label{W}
 W(r)=- {W_0\over 1+ e^{ r-R\over a} }  
\end{equation} 
with the parameters given in Table \ref{Tab_Par_OM}.
\begin{table}[h]
\begin{center}
\begin{tabular}{l l l  l}
            &     DR1       &   DR2    &  KW  \\\hline
W$_0$ (GeV) &    21+20i   &  0.5+ 0.5i             &  1.2i       \cr
R  (fm)        &      0          &    0.8           &   0.55     \cr
a   (fm)         &     0.2        &    0.2           &   0.2    \cr   
\end{tabular}
\caption{Parameters of the Dover-Richard (DR1 and DR2 versions)  and Khono-Weise (KW)  \={N}N optical models.}\label{Tab_Par_OM}
\end{center}
\end{table}

These three optical models  differ by their meson contents, the value of the cut-off radius $r_c$, the regularization procedure  
as well as by their annihilation potentials.
They generate the very different potentials presented in \ref{App_Pots}.
As  an illustrative example, let us consider Figure \ref{U_1S0} from this Appendix, corresponding 
to the real part of the $^{11}$S$_0$ potentials.
They have in common a strong attraction  (200-400 MeV at $r=0.8$ fm)  in T=0 channel, which seems not required by the NPWA.
On the other hand, the Paris potential displays a strong repulsion below $r\approx0.6$ fm 
as well as a repulsive barrier at $r\approx1$ fm that are absent in the other models.
Despite of that, they provide quite similar results for the 
integrated  elastic ($\sigma_e$), annihilation ($\sigma_a$) and charge exchange ($\sigma_{ce}$) cross sections. 
This can be seen in Figure \ref{sigmas}, where the integrated strong cross sections
of these three models (together with their sum $\sigma_t=\sigma_e+\sigma_a+\sigma_{ce}$)  are compared to each other as well as to the NPW results \cite{ZT_NNB_Nijm_PW_2012}.
The same agreement was observed  in the protonium S- P- and D- level shifts and widths as well as for the strong and \={p}p scattering lengths (see Refs. \cite{DR2_RS_PLB110_1982,Bachir_ZPA325_1986,CIR_ZPA334_1989,CRW_ZPA343_1992}).
However, no  comparison has been done at the level of phase-shifts.

For the three considered models, Paris 2009, DR2 and KW, we have computed the S-matrix in the energy range $0<p_{Lab}<1000$ MeV/c 
and for each PW state.
We have extracted  the S-matrix real parameters  and compared them with the results of the Nijmegen PW analysis (Tabs VII-IX-X from Ref. \cite{ZT_NNB_Nijm_PW_2012}).
The comparison with J\"{u}lich potential, adjusted to reproduce the  former, would be redundant
except for the low energy parameters that were not given in the  NPWA  \cite{ZT_NNB_Nijm_PW_2012} and that have been included in our discussion.

For the  uncoupled states,  the \={N}N  S-matrix 
 is  determined by  a complex phase shift $\delta_C=\delta_R+ i \delta_I$ whose (positive) imaginary part  $\delta_I$ 
is unambiguously defined by the modulus of the S-matrix, the inelasticity parameter $0<\eta<1$, according to
\begin{equation}\label{S_uncoupled}
 S=e^{2i\delta_C} = e^{2i\delta_R}  \; e^{-2\delta_I} \qquad  \delta_I =-{1\over 2}  \ln \eta
\end{equation} 
Notice that the annihilation cross section in a given PW, is entirely determined by $\eta$ as 
\begin{equation}\label{siga_1ch}
 \sigma_a= (2J+1) {\pi\over 4q^2} \left[ 1- \eta^2 \right]
\end{equation} 

For the  tensor-coupled states (e.g. $^3$SD$_1$), the 2$\times$2 S-matrix can be parametrised
by 6 real parameters --  two "bare phase shifts" $\delta_n$, 
two mixing parameters $\epsilon, \omega$
and two inelasticities $\eta_n$ --  according to the Bryan and Klarsfeld factorisation
\cite{Bryan_PRC24_1981,Klarsfeld_PLB126_1983,Bryan_PRC30_1984}
\begin{equation}\label{S_BK}
  \pmatrix{S_{11}&S_{12}\cr S_{21}&S_{22}}  =   \pmatrix{   e^{i\bar{\delta}_1} & 0 \cr 0  & e^{i\bar{\delta}_2} }  \; M \;    \pmatrix{   e^{i\bar{\delta}_1} & 0 \cr 0  & e^{i\bar{\delta}_2} }  
 \end{equation}
where
\begin{equation}\label{M_BK}
 M=  \pmatrix{ \cos\epsilon & i \sin\epsilon \cr i \sin\epsilon & \cos\epsilon}   H    \pmatrix{ \cos\epsilon & i \sin\epsilon \cr i \sin\epsilon & \cos\epsilon}  
\end{equation}
and  the matrix $H$, real and symmetric, contains the inelastic parameters $\eta_i$ as eigenvalues
\begin{equation}\label{H_BK}
 H=\pmatrix{ \cos\omega &  -\sin\omega \cr \sin\omega & \cos\omega}\pmatrix{\eta_1&0 \cr 0 & \eta_2} \pmatrix{ \cos\omega &  \sin\omega \cr -\sin\omega& \cos \omega}   
\end{equation}

Unitary models are defined by the condition 
\[ SS^{\dagger}=\bf 1\]
to be fulfilled  by (\ref{S_uncoupled}) and (\ref{S_BK}).
For   uncoupled states, this implies $\eta\equiv1$ (or equivalently $\delta_I\equiv0$).
For   tensor-coupled states, this implies $\omega=0$,  $\eta_1=\eta_2=1$ and so $H=\bf 1$. In this case, (\ref{S_BK})
takes the usual Stapp-Ypsilantis-Metropolis (SYM) form defining the standard bare phase shifts $\bar\delta_n$ and mixing parameter $\bar\epsilon$ 
of the unitary case \cite{SYM_PR105_1957}.
   
It is worth mentioning that, in the non unitary case, the definition of complex phase shifts  for the coupled-channel states ($^3$SD$_1$,$^3$PF$_2$,...) 
is not free from ambiguities.
In fact, the inelasticity parameters can be negative: they are only limited by the so-called "unitarity condition" \cite{Klarsfeld_PLB126_1983}
 \[ {\rm Tr}(1-SS^{\dagger})= 2-{\rm Tr}(M^2)= 2-\eta_1^2-\eta_2^2 >0  \]
which presumes nothing about their sign.
We have found that in some of the considered models  one of the inelasticity parameters is indeed negative.
This happen at relatively high energy, when the mixing angles $\epsilon$ and $\omega$ are large. 
The natural extension of the uncoupled case  (\ref{S_uncoupled}) to each  inelasticity parameter $\delta_{I,n}=-{1\over 2}\ln \eta_n$ poses a  problem.
There are alternative ways to define the complex  phase shifts, e.g. the straightforward extension of the SYM parametrisation with complex parameters. 
However, though being totally consistent,  the relation with respect to the previously defined parameters is not clear, even for the real part of the phase shifts.
Again, the differences appears at large values of the mixing parameters, i.e. beyond the zero energy region. Because of that,  the   definition
 of the low-energy parameters (scattering length and effective range) remains unambiguous.

To get rid of this ambiguities, and to keep as closer as possible to the results of NPWA, we have only displayed in the tensor-coupled case
the real bare shifts, together with the inelasticities and mixing parameters. 

The practical determination of these parameters is more involved than in the tensor decoupled case 
and we have followed the  procedure described in Sect VII of Ref. \cite{ZT_NNB_Nijm_PW_2012}. 

Together with the phase shifts,
the corresponding effective range functions have been also computed and  the corresponding Low Energy Parameters (LEP) have been extracted.

\section{Results}\label{Results}

\begin{figure*}[htbp]
\begin{center}
\vspace{-0.cm}
\includegraphics[width=8.5cm]{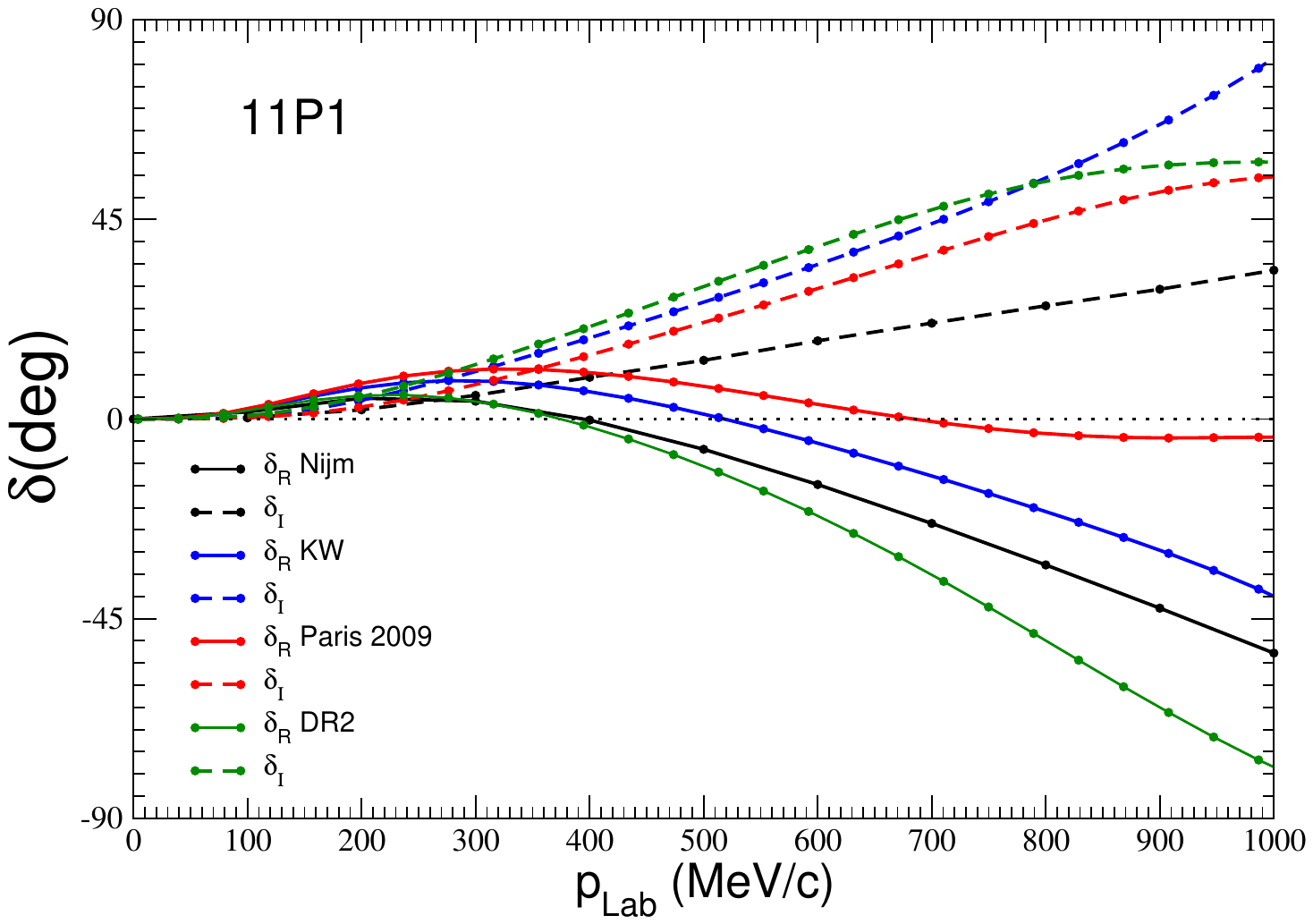} 
\vspace{-0.cm}
\includegraphics[width=8.5cm]{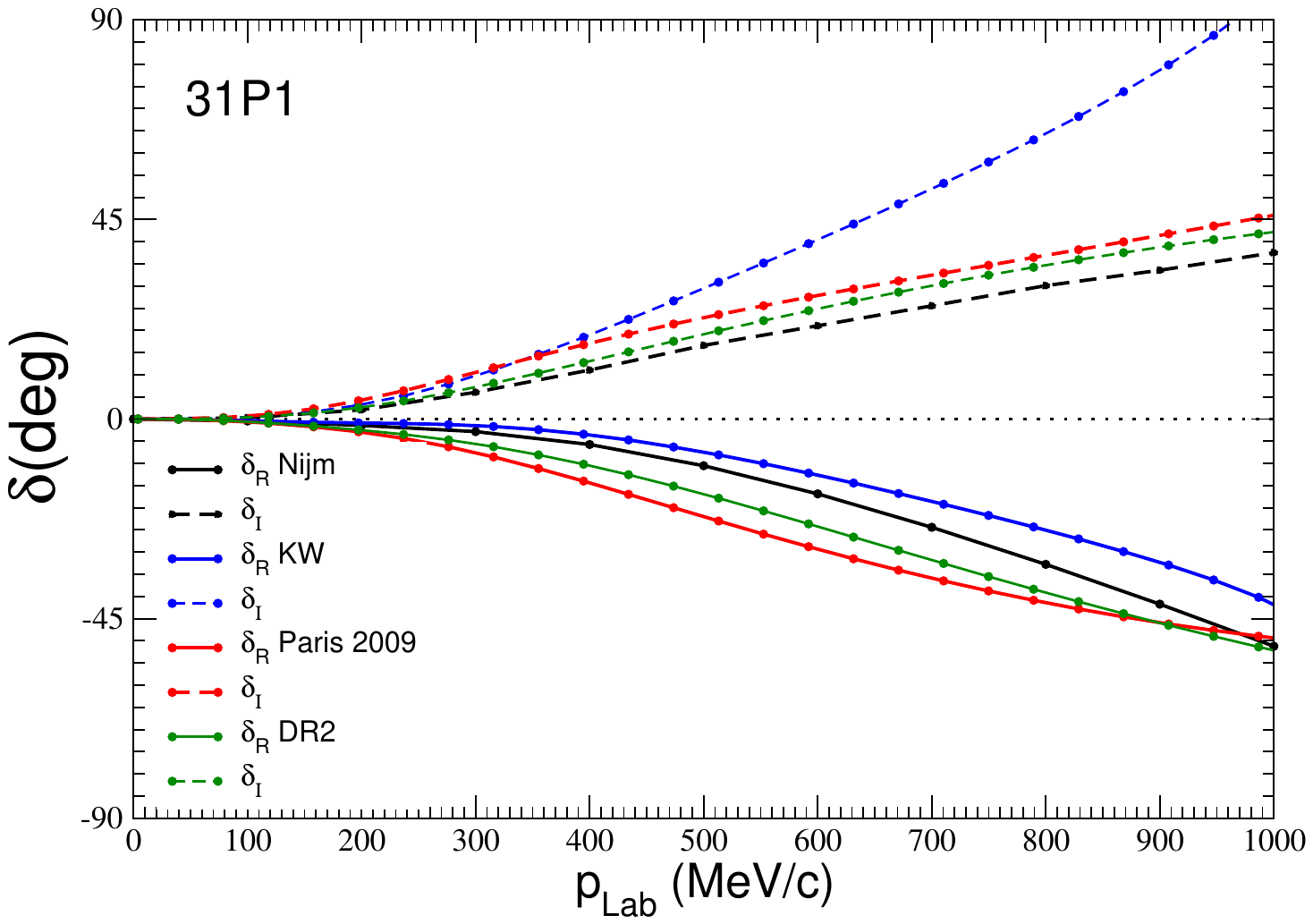}
\caption{$^1$P$_1$ \={N}N scattering  phase shifts as functions of the \={N} laboratory momenta. We use the same conventions as in Fig. \ref{delta_1S0}.}\label{delta_1P1}
\end{center}
\end{figure*}

\subsection{Phase shifts}\label{Sect_delta}

We first present the strong \={N}N complex phase shifts $\delta_C=\delta_R+ i \delta_I$ 
for the lowest partial waves as functions of the \={N} laboratory momentum $p_{Lab}$.
The different states $\alpha\equiv\{T,S,L,J\}$
 are alternatively denoted in the spectroscopic notation by $\alpha\equiv ^{2T+1,2S+1}L_J$.

The values corresponding to Nijmegen PW analysis are taken from Tab. VIII of Ref. \cite{ZT_NNB_Nijm_PW_2012}.
The other models KW, DR2 and Paris (2009) have been computed by the authors directly from the potentials, with the original model parameters. 
We emphasize that the J\"{u}lich model \cite{Haidenbauer_JHEP_2017} results
are, by construction, adjusted to  the Nijmegen PW and there is no need to included them.

There is an $\pm n\pi$ ambiguity in the definition of the phase shift $\delta$ which is formally solved by imposing
$\delta(E\to +\infty)=0$ and by keeping the same determination 
when the energy is decreased. 
Due to the sizable strengths  of the  \={N}N potentials, this recipe is however of little practical interest since 
it imposes to start with the solution at very high energy and go inwards in energy. 
In a unitary model (hermitian hamiltonian),
 another way to fix the determination is  by imposing the value at the origin  to be $\delta_{\alpha}$(E=0)=$n\pi$, 
where $n$  is the number of bound  states in channel $\alpha$. 
This impose the full knowledge of the spectrum for each partial wave.
On the other hand, the validity of this result, known as Levinson theorem, is 
not well established in the non unitary systems like the optical models we are considering in this work.
Thus, and for the sake of comparison, we have conventionally adjusted all the computed phase shifts to the determination given  in the Nijmegen PW
analysis \cite{ZT_NNB_Nijm_PW_2012}.

\begin{figure*}[htbp]
\begin{center}
\vspace{-0.cm}
\includegraphics[width=8.5cm]{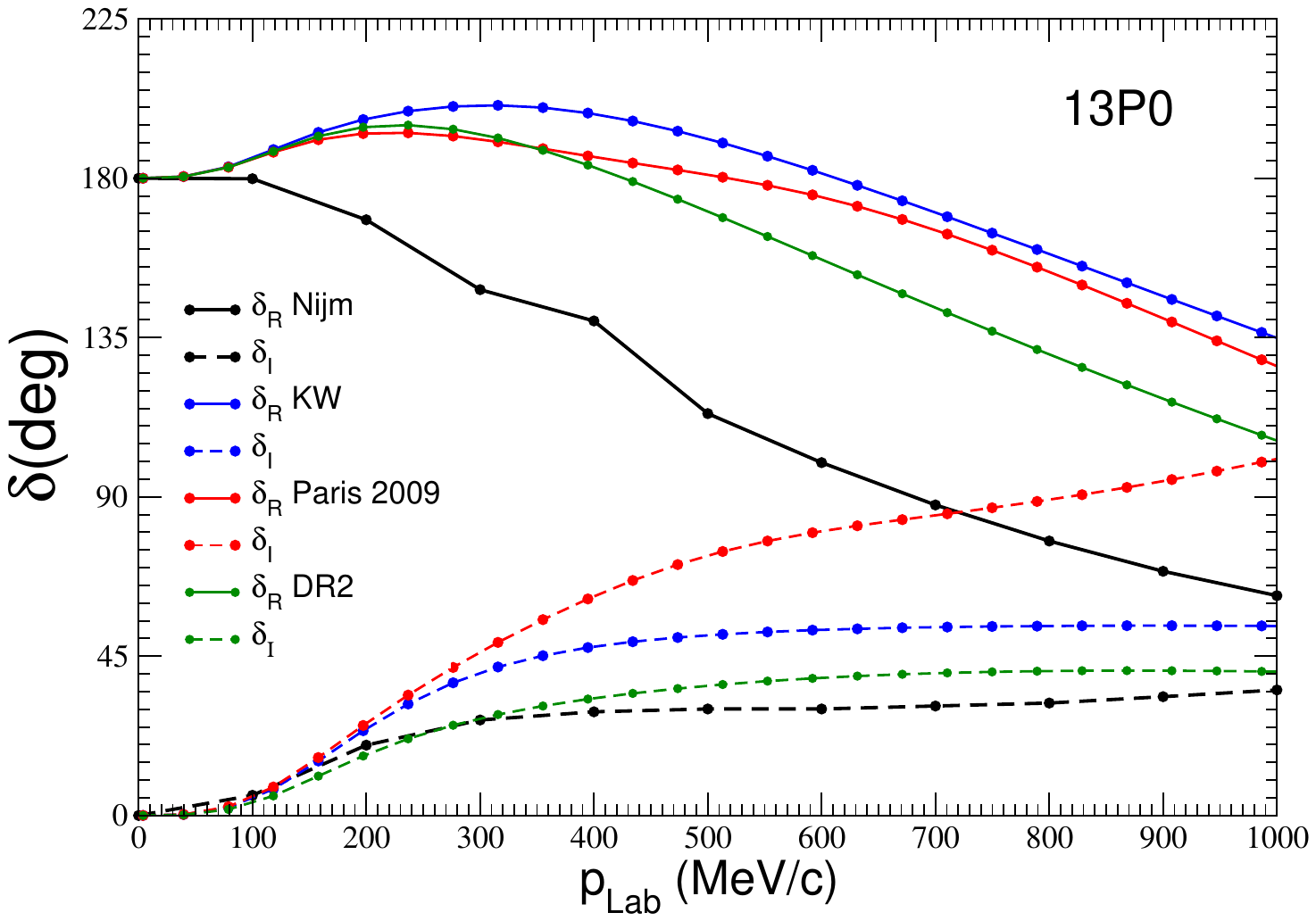} 
\vspace{-0.cm}
\includegraphics[width=8.5cm]{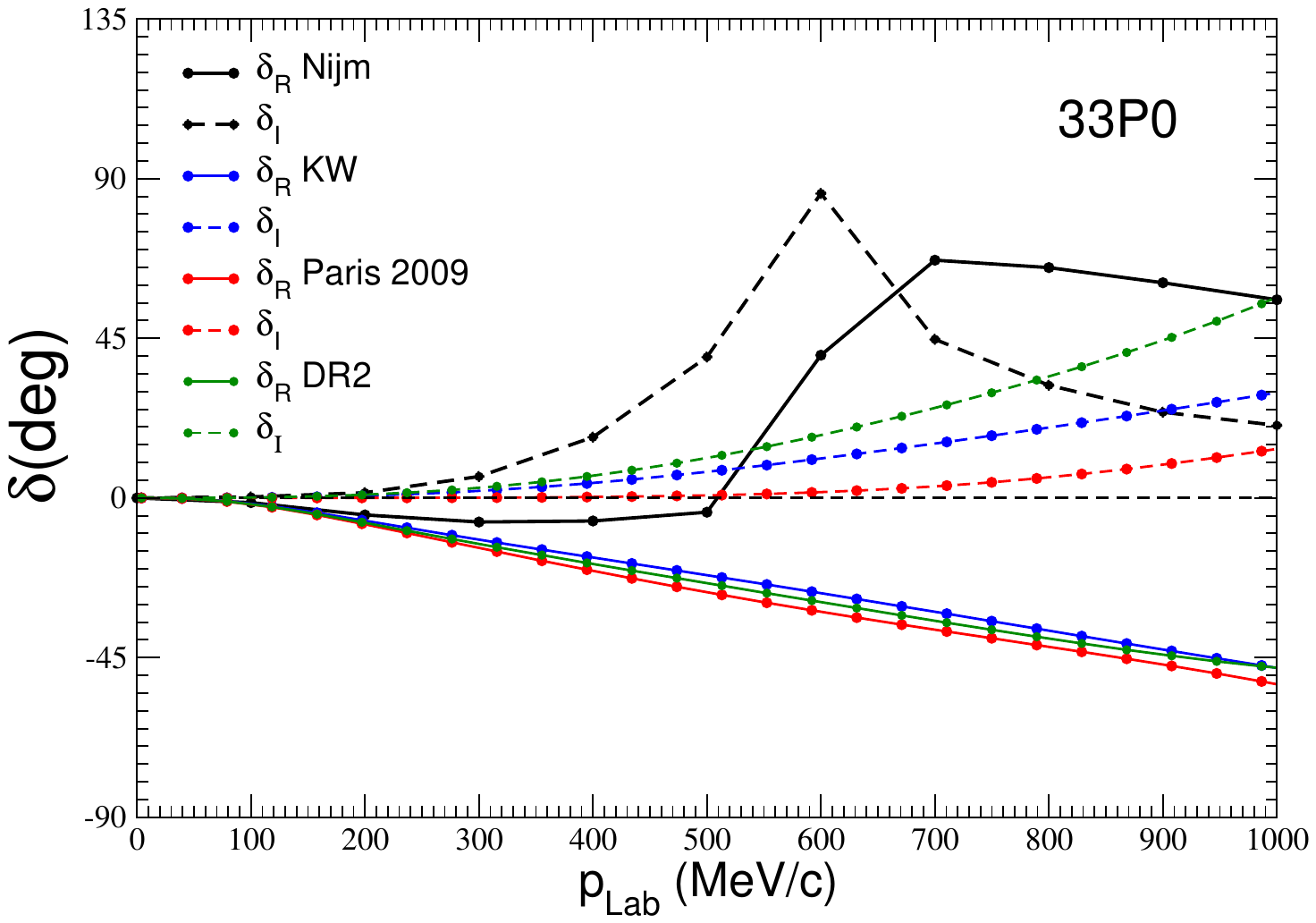}
\vspace{-0.cm}
\caption{$^3$P$_0$ \={N}N scattering  phase shifts (degres) as  functions of the \={N} laboratory momenta. We use the same conventions as in Fig. \ref{delta_1S0}.}\label{delta_3P0}
\end{center}
\end{figure*}

\bigskip
{\bf Figure \ref{delta_1S0} contains the results for the $^1$S$_0$ state}, left panel for isospin T=0 ($^{11}$S$_0$) and right panel  for T=1 ($^{31}$S$_0$).  
The real part $\delta_R$ is in solid line and the (positive) imaginary part  $\delta_I$ in dashed line, both in degres.
Different colours have been used to disentangle the different models: black for the NPW, 
blue for KW, red for Paris-2009.
As one can see, there are major differences between them, specially in $\delta_R$, which deserve some comments. 

For T=0 state (left panel), the real phase shifts of KW and DR models are close to the NPWA ones up to $p_{Lab}\approx700$ MeV/c
and they both  depart  dramatically from the Paris-2009 starting at very low energy. 
We recall the reader that the slope of the phase shift at the origin is the scattering length since $\delta_{\alpha}(q)\approx -a_{\alpha} q$
where $q$ is the center of mass momentum, related to $p_{Lab}$ by $p_{Lab}=2q$. 
This difference could be due to the fact that Paris potential
 has a near-threshold quasi-bound state in this channel, absent in KW and J\"{u}lich interactions. 
Its binding energy was $E=-4.8-26i$ MeV in Ref. \cite{PARIS_PRC79_2009}.
By using two independent methods, we confirm this state with a slightly different energy $E=-10.2- 23.2i$ MeV.
As we will see in next subsection, the existence of this  quasi-bound state is 
supported by a different sign in the corresponding scattering lengths (see Table \ref{Table_LEP_S}).
In this respect, a similar quasi-bound state is also present in DR2 model with $E=-138-320i$ MeV,  
much deeper in energy and leaving no trace in the scattering region.

It is worth emphasizing, however, that there is no  univocal relationship between the sign of the scattering length (real part) and the existence  of bound states. 
For real potentials, the positive sign can be either a consequence of a repulsive interaction or of an attractive interaction
having one (or several) bound states. The negative sign indicates always the existence of an attraction but tells us nothing about
 the existence or non-existence  of a bound state, which will actually depend on the strength of the attraction.
The situation is even more delicate when using complex  potentials and additional informations are required to draw consistent conclusions.

In particular, we would like to notice that the very existence of a quasi-nuclear state in the $^1$S$_0$ \={N}N state appears as
a consequence of the sign of the measured \={p}p scattering length \cite{YC_PAD_EPJA57_2021}. 
It was shown (See Figure 7 of this reference) that for weak (and attractive)  \={p}p interactions --  i.e.  using large values of the cutoff radius, $r_c>1.7$ fm --
 the sign of Re[$a_{p\bar p}$] scattering length is negative. It
becomes positive -- and in agreement with experiment -- only when, by decreasing $r_c$,  the interaction is strong enough to create the first \={p}p bound state.
The $a_{p\bar p}$ involves however both isospin components and,  from this single quantity,  it is not possible to conclude  in which of the component it appears.
Notice also that the properties of such states, in particular their width, strongly depend on the annihilation
dynamics. When using  short range annihilation potential, as in Paris potential and in some  coupled-channel unitary models (UCCM),  
the widths are much smaller than when using annihilation potentials type eq. (\ref{W}), as in DR2, KW and J\"{u}lich potentials.
One can find a discussion in Refs. \cite{CDPS_NPA535_91} for KW and \cite{YC_PAD_EPJA57_2021} for  UCCM.

The particular E-dependence of the Paris potential in the $^{11}$S$_0$ state is also observed 
in the imaginary phase shift $\delta_I$, which remains in a reasonable agreement with other models only up 
to $p_{Lab}\approx200$ MeV/c and displays a maximum at $p_{Lab}\approx600$ MeV/c.

When studying the $J/\Psi\to\gamma p\bar{p}$  decay of BES collaboration, the authors of Ref.  \cite{DLEW_PRC80_2009,DLW_PRC97_2018} interpreted the
first peak in the  \={p}p invariant mass in terms the  above mentioned $^{11}$S$_0$ quasi-bond state, 
and the second one in terms of a resonant state $^{11}$S$_0$  at $\approx$ 250 MeV above the threshold,
i.e. $p_{Lab}\approx$ 950 MeV/c.
The peculiar form of the Paris $^{11}$S$_0$ potential depicted in Fig. \ref{U_1S0},
displaying a deep attractive well and repulsive pocket at $r\approx$1 fm, can indeed accommodate an S-wave resonance,
but the height of the repulsive barrier is two times smaller than the supposed resonance energy.
On the other hand,  no clear evidence of this state is seen in the corresponding phase shifts.
Only a vague structure is noticeable in $\delta_R$ (red solid line) in the vicinity of $p_{Lab}\approx$ 600 MeV/c, 
with a maxima of $\delta_I$ at almost the same energy, that could be related.

For T=1 state (right panel), the same discrepancy in $\delta_R$ between the Paris result and the other models is observed.
However there is a major difference between the NPWA and the other optical models:  the sharp resonant-like structure 
that the former manifests at  $p_{Lab}\approx600$  MeV/c, both in the real and in the imaginary phase shifts.
The  existence of an S-wave  shape resonance  would require a $^{31}$S$_0$  potential  with a repulsive bump at finite distance,
like  for instance the one exhibited in Figure \ref{U_1S0} for the Paris 2009 model,  in the T=0 channel. 
But none of the considered models exhibit such a behaviour (See Appendix).

\begin{figure}[htbp]
\begin{center}
\includegraphics[width=8.5cm]{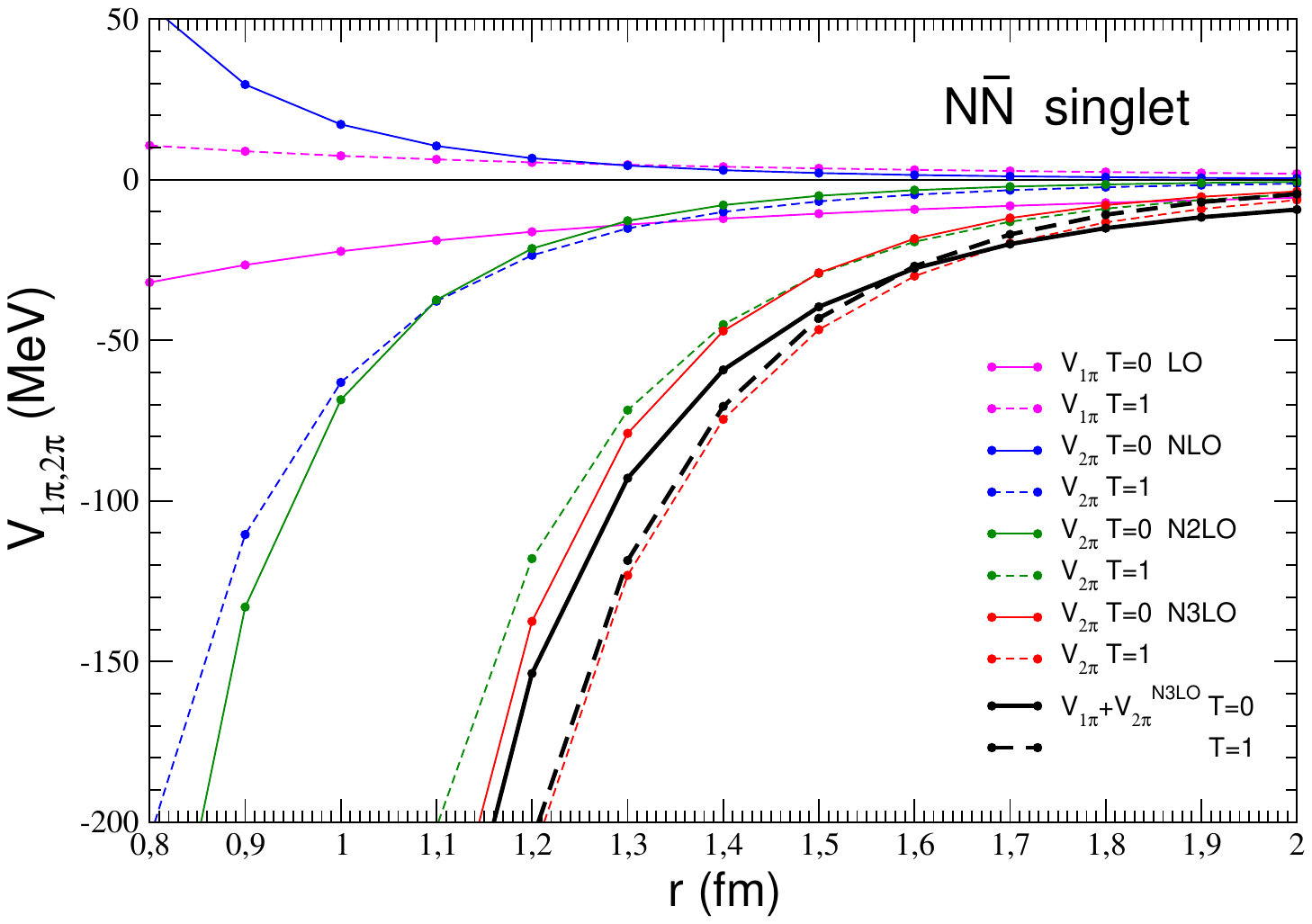}
\vspace{-0.cm}
\caption{One- ($V_{\pi}$) and two-pion ($V_{2\pi}$) exchange potentials in singlet \={N}N states. 
Results are taken from G-parity transform of the EFT inspired NN Idaho potential \cite{Machleidt_Idaho_N3LO}. 
The different orders up to N3LO, are plotted separately for both isospin (T) channels.
The sum $V_{\pi}\equiv V_{1\pi}+V_{2\pi}$ is strongly attractive. Only the T=0 states at NLO shows a short range repulsion.}
\label{V2pi}
\end{center}
\end{figure}

The possibility for the NPWA to generate a resonance in this partial wave 
with a purely  attractive   (single-channel)  potential is difficult to understand. 
Indeed, in their analysis the long- and medium-range part of the \={N}N interaction was parametrized 
by the one-pion ($V_{1\pi}$)  plus the two-pion ($V_{2\pi}$)  exchange potentials.
According to the recent work from the Idaho group on the EFT-NN interaction at N3LO \cite{Machleidt_Idaho_N3LO}, $V_{2\pi}$
for S=0 states is strongly attractive in both isospin channels, in agreement with
previous works \cite{V2pi_Kaiser_BW_NPA625_1997}. 
Let us remind that when going from NN to \={N}N system,  $V_{1\pi}$ contribution change the sign while  $V_{2\pi}$ remains unchanged.

We have displayed in Figure \ref{V2pi} the $V_{1\pi}$ and $V_{2\pi}$ contributions to \={N}N potential in spin singlet states
for both T components.
As one can see, the only repulsion appears for T=0 at NLO,  as in Paris potential, but becomes increasingly attractive at higher orders.
In NPWA, the attractive pion tail ($1\pi+2\pi$) is prolonged at $b=$1.2 fm with a boundary
condition corresponding to an also attractive square well (see Tab 1 from  \cite{ZT_NNB_Nijm_PW_2012}).
Thus, the overall $^{31}$S$_0$ potential is attractive, as it is the case
of the other (KW, DR2, Paris) examined potentials. See Figure \ref{U_1S0} in the Appendix.
Furthermore the resonant-like bump  takes place at center of mass energy of $E\approx$ 100 MeV, which is
quite a high energy for a single-channel S-wave  to produce a visible bump in the phase shifts.

It must be pointed out that other mechanism to mimic resonant like structures in S-wave exist. 
For instance when a bound state, generated by the real part of the interaction, moves into the positive energy region due to the annihilation potential. 
The trajectory from the bound state to the continuum region in the complex energy plane was
examined in our previous work with the KW potential \cite{CDPS_NPA535_91}. 
This can be illustrated by the complex energy trajectory of the $^{11}$S$_0$ state 
as function of the annihilation strength $W_0$. When $W_0=0$ there is a bound state
at E=-54.7 MeV. When the annihilation is switched on, its  imaginary part of the energy  
increases  linearly with $W_0$ and the state  is pushed out into the continuum.
It reaches Re(E)$>$0 at $W_0\approx0.47$ GeV  and has a width  $\Gamma\approx$ 230 MeV.
The energy imaginary part continues to increase until the model value $W_0=$1.2 GeV.
Thus, the widths of the positive energy states ($E\sim $100 MeV) thus obtained 
within this mechanism are very large (few hundreds of MeV) and cannot generate structures like
the one displayed in the right panel of Fig. \ref{delta_1S0}.

\begin{figure}[htbp]
\begin{center}
\includegraphics[width=8.5cm]{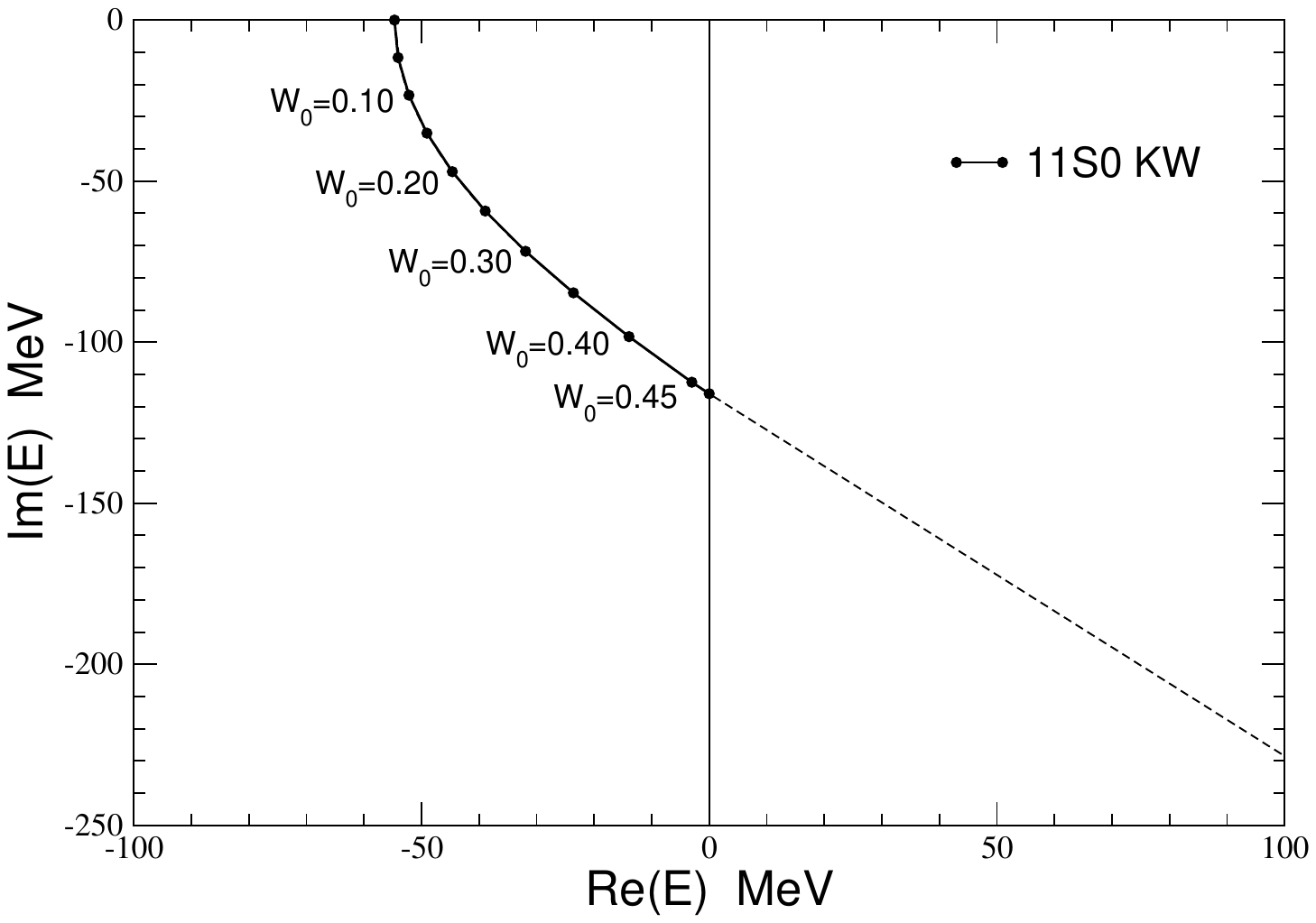}
\vspace{-0.cm}
\caption{Pole trajectory of a $^{11}$S$_0$ bound state in KW model as  function of the strength
of the imaginary part $W_0$. In absence of annihilation potential $(W_0=0)$, there is a bound state with E=-54.7 MeV.
The effect of the annihilation potential is to generate a width and to pull out the state into the continuum.
When $W_0\approx$ 0.47 GeV, one has $Re(E)=0$ and $Im(E)\approx$ 115 MeV. With the model parameter $W_0$=1.2 GeV, the width of the state is $\sim 400$ MeV.}
\label{Pole_Trajectory}
\end{center}
\end{figure}

All these reasons above, make it difficult to understand  the structure displayed in  Figure  \ref{delta_1S0} (right panel)
in terms of a resonance in the $^{31}$S$_0$ PW, especially at 100 MeV above threshold.

On the other hand, it is worth noticing that  the  J\"{u}lich potential  nicely reproduces the  $^{13}$S$_0$ phase shifts of the NPWA, 
that was attributed to an S-wave resonance, although no further explanation in terms of the underlying potential  was given in their manuscript. 
This is a non trivial achievement, that  worked also reasonably well in their N2LO version \cite{Haidenbauer_JHEP_2014}, and shows the extreme flexibility of the EFT potential.

Interestingly, the same group, came to the conclusion \cite{JPsi_Julich_PRD91_2015,JPsi_Julich_PRD98_2018}  that in order to reproduce
the BES results on the $J/\Psi\to \gamma p\bar p$ decay they were forced to slightly modify their original $^{31}$S$_1$ potential. 
Once readjusted, the corresponding phase shifts  do not reproduce anymore the NPWA structure of Figure  \ref{delta_1S0}  (right panel) but 
are very close to the -- smoothly varying -- KW results (blue curve).
This happens in the N2LO \cite{Haidenbauer_JHEP_2014} as well as in the N3LO versions  of J\"{u}lich  potential \cite{Haidenbauer_JHEP_2017}.
The authors conclude that the origin of the near-threshold peaks observed in the BES experiment,
may be explained by assuming  the existence of a \={N}N  $^1$S$_0$ quasi-bound state, but  in the T=1
rather than in the T=0 channel, as claimed in \cite{DLEW_PRC80_2009,DLW_PRC97_2018}.
Its energy was estimated to be $E=-36.9 - 47.2i$ MeV.
It could be relevant to notice that all  the examined models have indeed  a bound state in this PW when Im(V)=0. 
The imaginary part of the potential pushes  KW one into continuum while the DR2 and Paris ones remain still bound, although sizeably 
deeper than in the J\"{u}lich  potential : E=-430-346i MeV for DR2 and E=-184-171i MeV for Paris.
 
 The possible origin of the resonant-like structure manifested in the NPWA, which is also manifested in other partial waves, will be discussed later.

Our last comment on this $^1$S$_0$ state, is to remark that the imaginary phase shifts $\delta_I$ agree reasonably well with each other up to $p_{Lab}\approx400$ MeV/c,
where the resonant-like behaviour of the NPWA starts showing up.

\bigskip
{\bf Figure \ref{delta_3SD1}} displays the S-matrix bare phase shifts and inelasticities  (upper panel)  and the mixing parameters (lower panel) 
of  the triplet tensor coupled $^3$SD$_1$ state.
The $^3$S$_1$  bare phase shifts $\delta_S$ seem to be more stable than the $^1$S$_0$ ones, for both isospin states,
although in this case the best agreement is among KW, DR2 and Paris. 

For T=0,  the bare shifts $\delta_S$ are  very close to each other  up to $p_{Lab} $=700 MeV/c, where DR2 displays some structure,
crossing 90 degres in increasing, what suggest a standard  resonance driven by D-waves due to its centrifugal barrier. 
The same structure is manifested in $\delta_D$  and the  mixing parameters. 
The inelasticities are less stable: $\eta_S$ (left panel) departs sharply from NPW at $p_{Lab} \approx$500 MeV/c while
$\eta_D$ the dispersion among the models starts already at $p_{Lab} \approx$300 MeV/c.
The mixing parameter $\epsilon$ shows also strong deviations above 300 MeV/c while $\omega$'s are relatively close to each other.

For T=1 (right panel) the S-wave  inelasticity  of Paris potential departs  sensibly from the other models from the zero energy region. 
The dispersion in the  mixing parameters is huge, with  NPW and DR23 displaying  a peculiar energy dependence. 

\begin{figure*}[htbp]
\begin{center}
\includegraphics[width=8.5cm]{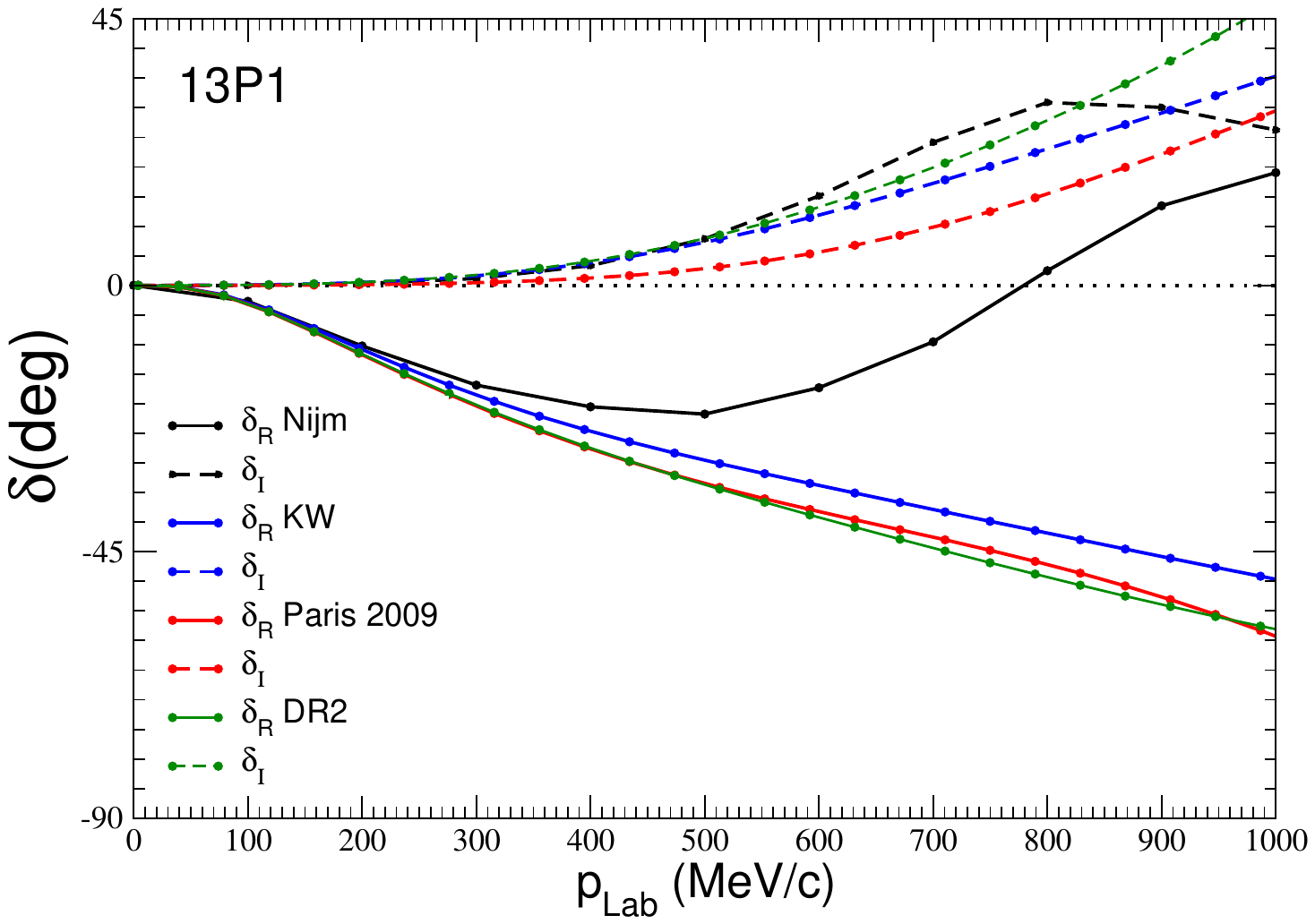} 
\vspace{-0.cm}
\includegraphics[width=8.5cm]{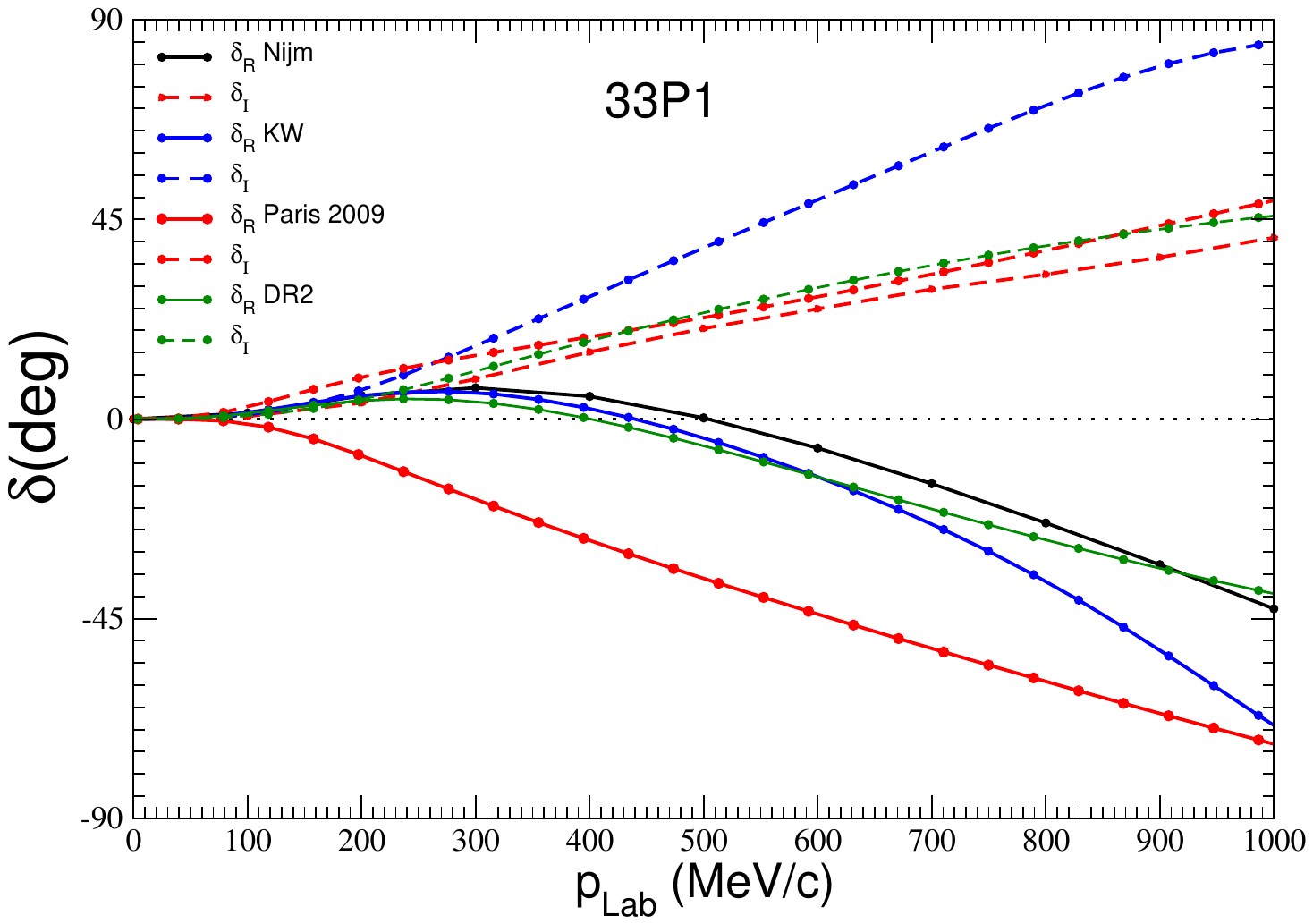}
\vspace{-0.cm}
\caption{$^3$P$_1$ \={N}N scattering  phase shifts (degres) as  functions of the \={N} laboratory momenta,
using same conventions than in Fig. \ref{delta_1S0}}\label{delta_3P1}
\end{center}
\end{figure*}
  
{\bf The $^1$P$_1$ phase shifts are displayed in Fig. \ref{delta_1P1}}, using the same line and colour conventions as in Fig. \ref{delta_1S0}.
At first glance, it seems that in this case the global behaviour of the models is quite similar, at least at low energy.
However, we will see in the next section that this is not really the case: major deviations exist from the origin but these differences
are hidden in this representation due to $q^{3}$ behaviour at small $q$.

{\bf  The $^3$P$_0$ results are displayed in Fig \ref{delta_3P0}}, with the same colour convention as in Fig. \ref{delta_1S0}
and \ref{delta_1P1}.
In this partial wave, the differences between the model predictions (KW, DR2 and Paris-2009 )  -- relatively close to each other -- and the results of  the NPWA 
are dramatic.

For T=0, the deviations in $\delta_R$ start already at $q\approx0$, displaying a different concavity. 
This corresponds to a negative effective range for the NPW (see Table \ref{Table_LEP_P}).
The imaginary phase shifts  $\delta_I$,  start differing at $p_{Lab}=200$ MeV/c and the differences  increase with the energy.

For T=1, the low energy phase shifts of all models are quite in agreement, including  NPW results, what is manifested in the LEP's displayed in Table \ref{Table_LEP_P}.
Deviations  start above $p_{Lab}=200$ MeV/c,  where  the NPW results display  the same resonant-like structure that the one observed in the $^{31}$S$_0$ state,
and practically at the same value of $p_{Lab}$.

Contrary to the $^{31}$S$_0$ case, the $^{33}$P$_0$ potential has a centrifugal barrier which can indeed acommodate a resonance, provided that the interaction is 
attractive enough. However, it is not the case in any of the examined models -- KW, DR2 and Paris -- which are globally repulsive in this channel (See Fig. \ref{U_3P0} from the Appendix).
This is in agreement with the corresponding scattering volumes in Table \ref{Table_LEP_P},  including the values of the J\"{u}lich potential :
they are very stable and have very small imaginary parts, as it corresponds to a repulsive interaction. 
Nevertheless,  the inner part of the NPW potential, used to define the boundary conditions at b=1.2 fm, is an attractive
square well with 160 MeV depth. Even if beyond $r=b$ the long range part $V_{\pi}$ is slightly repulsive, the effective potential  (V+ centrifugal) 
in the vicinity of $r=1$ fm remains globally attractive ($-80$ MeV) and we cannot exclude that a resonance could indeed be produced.

This will be in strong tension with all the potentials, but this is not the only reason to be cautious with such a possibility.
On one hand  the maximum of the centrifugal barrier  (60 MeV at $r$=1.2 fm) is sensibly smaller than the 
resonance energy  ($p_{Lab}=600$ MeV/c, $E_{cm}\approx95$ MeV). 
On the other hand, the $\delta_R$ and $\delta_I$ curves of both T=1 states, $^{31}$S$_0$ and $^{33}$P$_0$, can be overimposed in the vicinity of the peak. 
Since it is difficult to attribute such a coincidence to a dynamical effect, occurring in two independent PW at the same energy, this behaviour suggest 
to look for an alternative explanation. We will come back to this important point at the end of this section.

\bigskip
{\bf  The $^3$P$_1$ results are displayed in Fig \ref{delta_3P1}} with the colour convention of the previous Figures.
For the T=0 state (left panel), all the phase shifts display  quite a similar behaviour up to 300 MeV/c 
but  the considered optical models  start to depart  dramatically from NPWA at $p_{Lab}\approx$400 MeV/c.
Below that, it is   one of the most stable channels.

For T=1 (right panel), the deviations are more sizeable already from the zero energy region,  
specially in $\delta_R$ with Paris 2009.
This difference is due to the existence of a near-threshold quasi-bound state with $E=-3.6-12.42i$ MeV,
(slightly different form the value  E=-4.5-9.0i MeV given in \cite{PARIS_PRC79_2009}) that is absent in the other models and is responsible for a different sign of the scattering volume (see Table \ref{Table_LEP_P}).

\begin{figure*}[htpb]
\begin{center}
\vspace{-0.cm}
\includegraphics[width=8.cm]{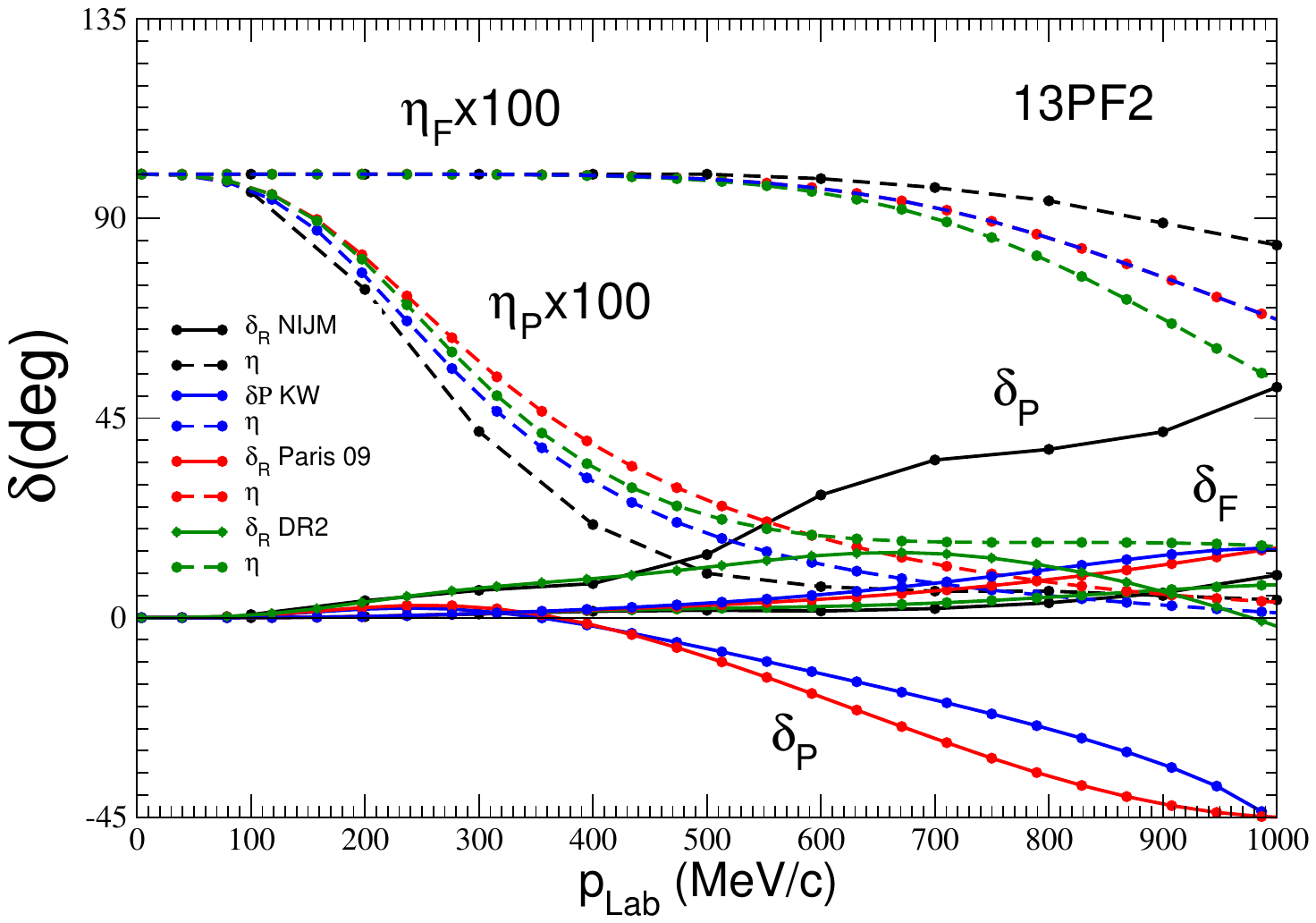} 
\includegraphics[width=8.cm]{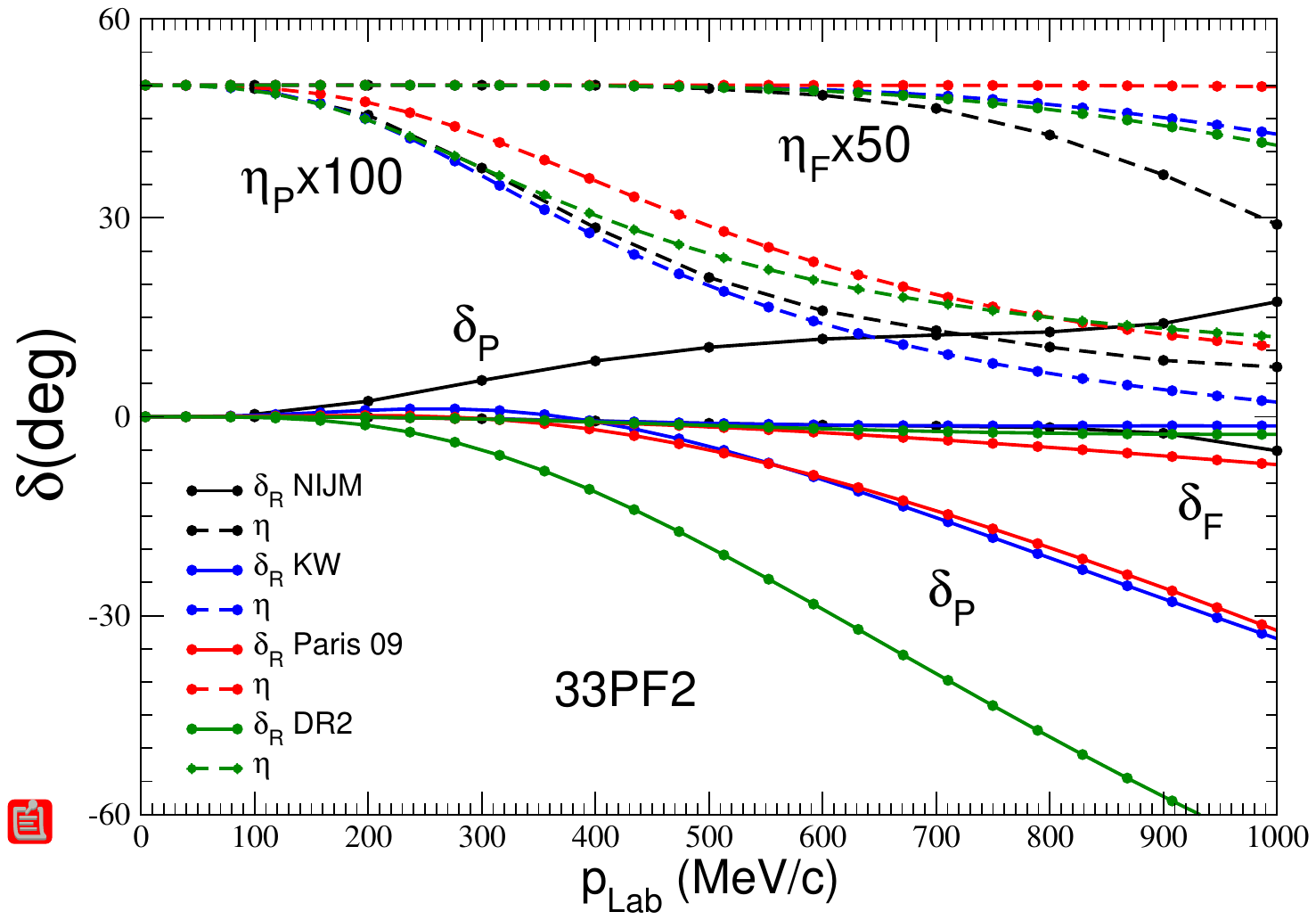}
\vspace{-.3cm}

\includegraphics[width=8.cm]{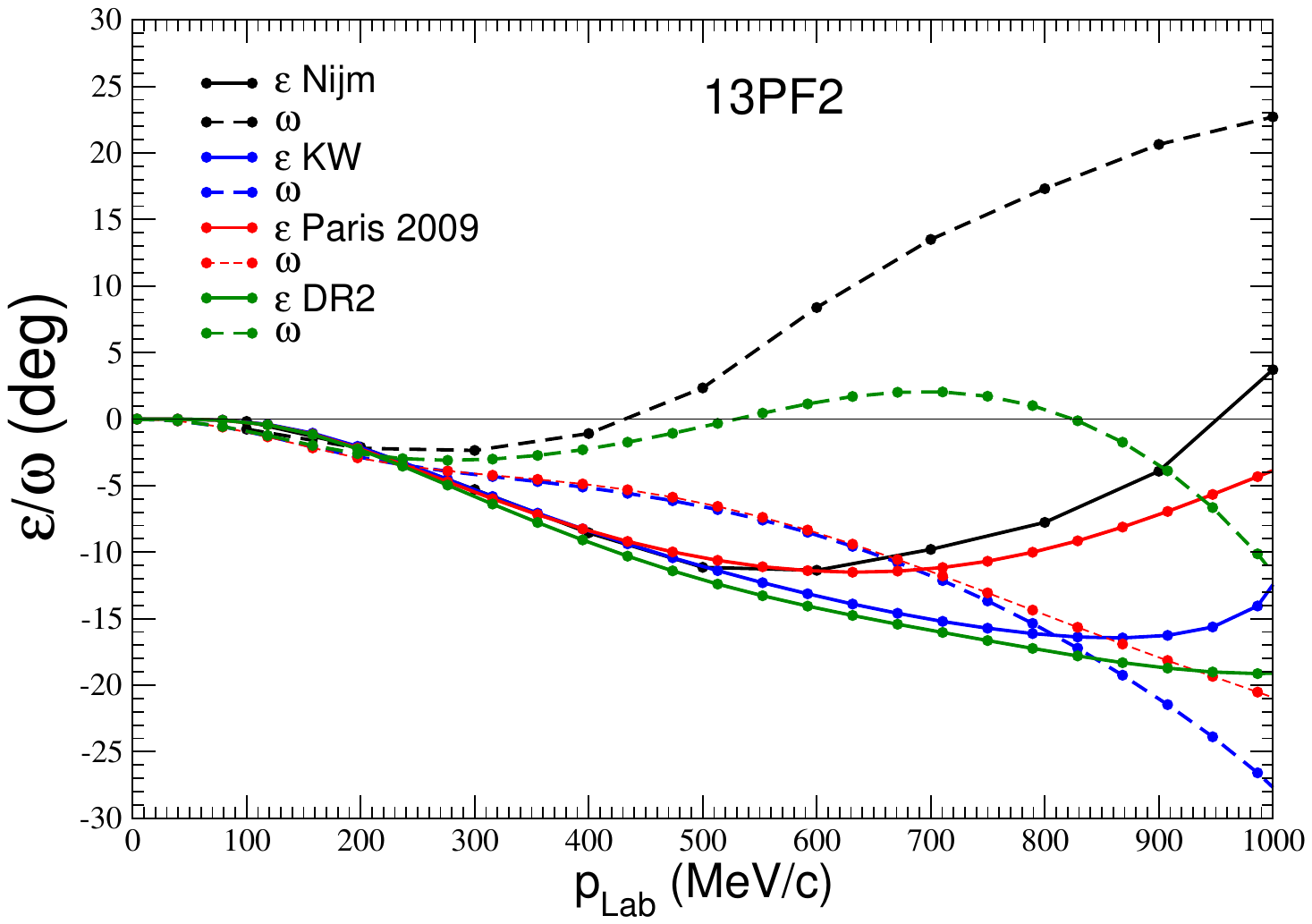} 
\vspace{-0.cm}
\includegraphics[width=8.cm]{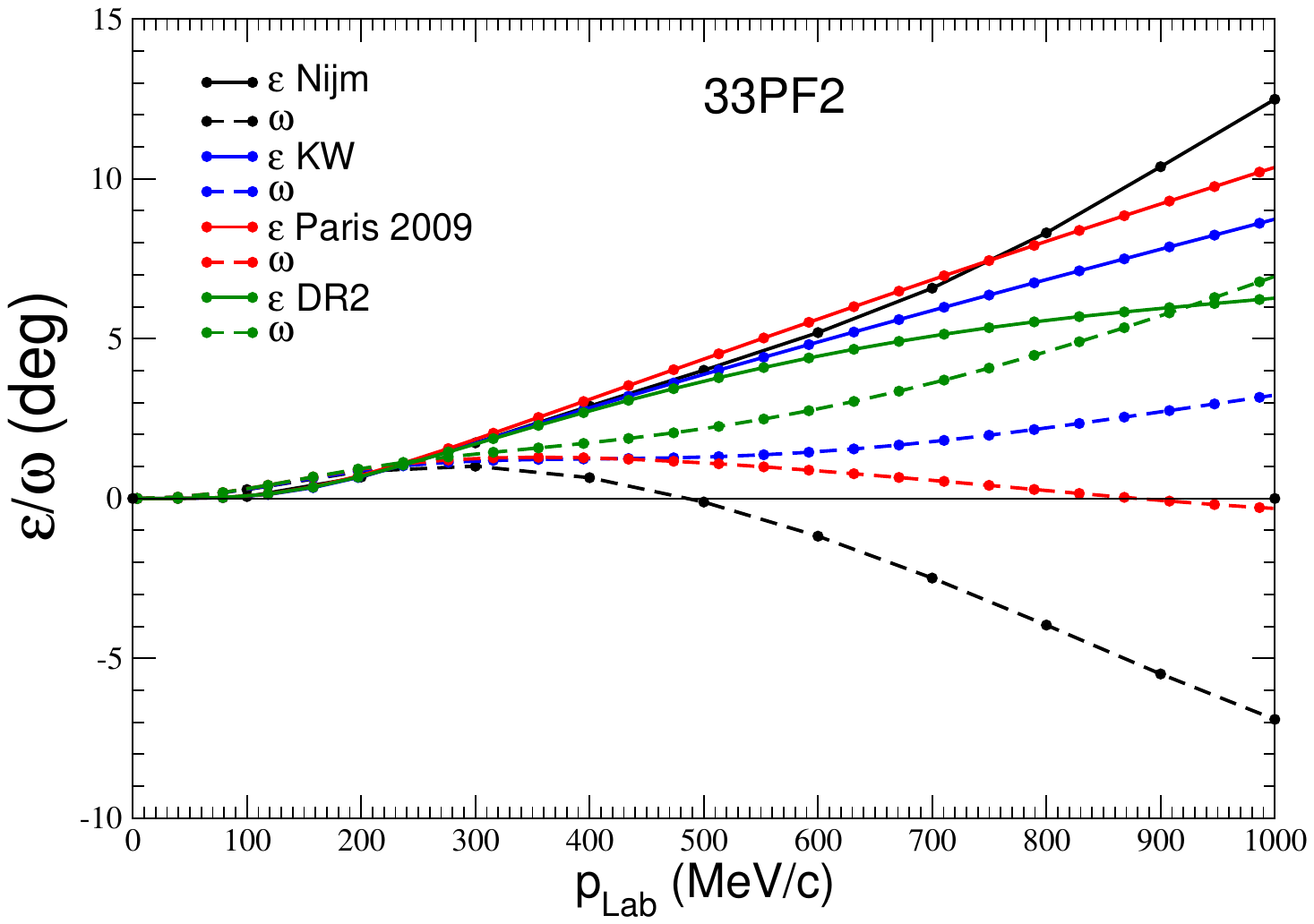}
\caption{$^3$PF$_2$ \={N}N bare phase shifts and inelasticities (upper panel) and mixing parameters  (lower panel),  
as functions of the \={N} laboratory momenta.}\label{delta_3PF2}
\end{center}
\end{figure*}

\bigskip
Finally, we show in Figure  \ref{delta_3PF2} the bare phase shifts and mixing parameters for the $^3$PF$_2$ partial wave.

For T=0, and in spite of some stability in the scattering lengths, the results of $\delta_P$ falls
in two different families:  on one side the NPW and DR2  which are attractive and on the other KW  and Paris that are repulsive. 
This qualitative difference remains such in all the considered energy domain.
The same  splitting is observed in the mixing parameters above $p_{Lab}\approx$200 MeV/c.

For T=1, a similar situation happens, with the $\delta_P$ values of NPW  evolving in the opposite direction than the rest of the models.
Remarkably, the mixing parameter $\epsilon$  remains stable up to $p_{Lab}\approx$600 MeV/c while $\omega$ 
values start diverging at 300 MeV/c.

\bigskip
As we already mentioned, the NPW results for the  $^{31}$S$_0$ and $^{33}$P$_0$ states,  display 
the same kind of non trivial structure at  $p_{Lab}\approx600$  MeV/c both for the real and the imaginary phases,
while they are absent in the examined optical models, with the exception of J\"{u}lich potential which reproduces it well.
It seems however unlikely, although not impossible, that a dynamical effect could generate two resonances
at the same energy, in different partial waves having the same parameters, one in S-wave and other in P-waves.

\begin{figure}[htbp]
\begin{center}
\includegraphics[width=9.cm]{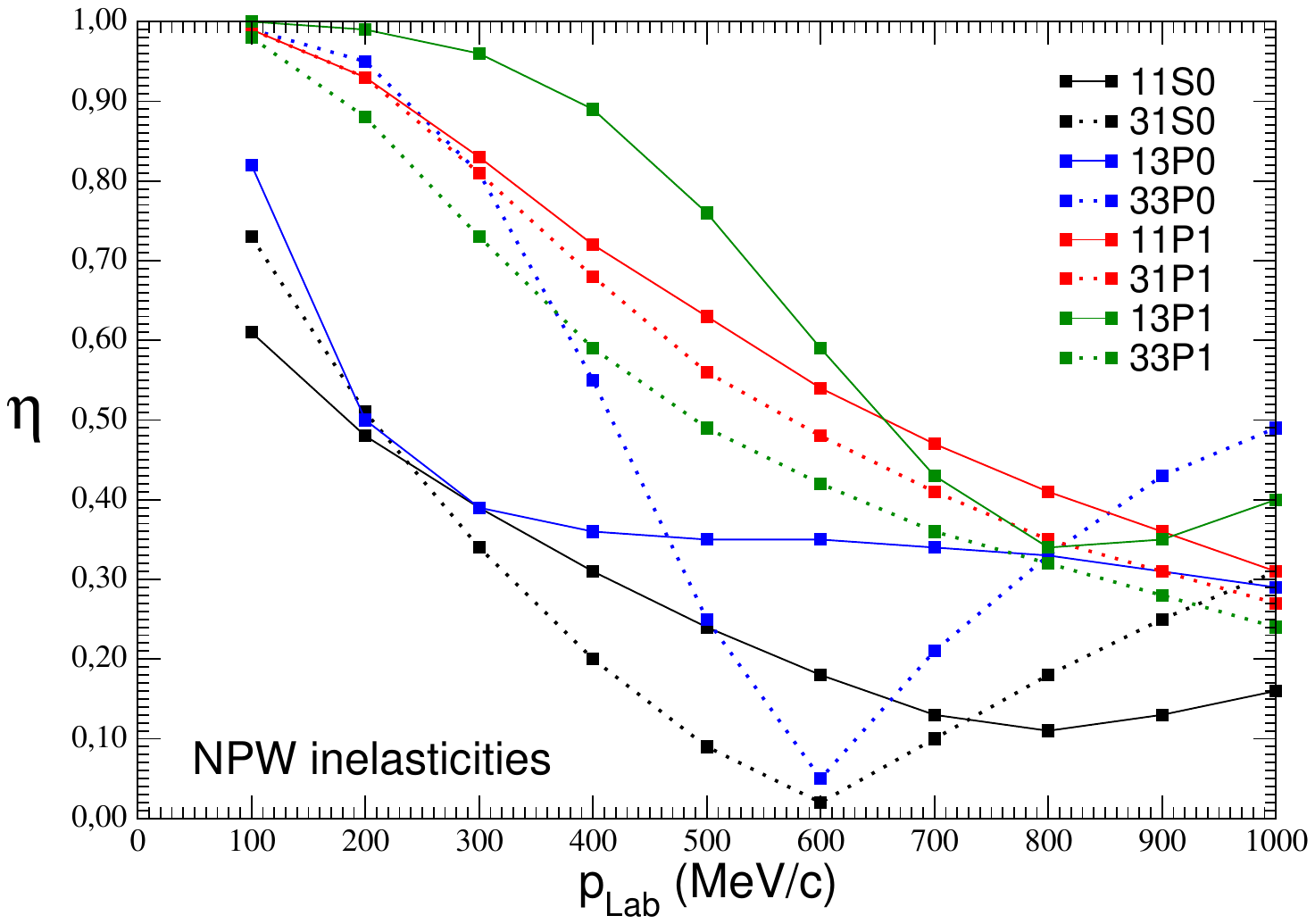} 
\caption{Inelasticity parameter $\eta$ in the NPW as  function of the \={N} laboratory momenta.}\label{eta_NPW}
\end{center}
\end{figure}

Looking for a possible explanation of these structures,
we noticed that this energy region corresponds to a sharp maximum of $\delta_I$, that is  to a minimum of the
inelasticity parameter $\eta$, which turns to be -- in this particular waves -- very small.
For the $^{31}$S$_0$, for instance, one has $\eta\approx 0.01$ at the minima, that is
one order of magnitude smaller than for the $^{11}$S$_0$. 
The same is true for the $^{33}$P$_0$ state, when compared to other P-waves.
This can be seen in Fig. \ref{eta_NPW}, where the inelasticity parameter $\eta$ is plotted 
as  function of $p_{Lab}$ for several states and where the peculiarity of the $^{11}$S$_0$ and $^{33}$P$_0$ states is manifested.

Since
\[ \mid S_{\alpha}(E)\mid=e^{-2\delta^{\alpha}_I(E)} =\eta_{\alpha} (E)\]
the scattering matrix of $^{11}$S$_0$ is in modulus $\sim 10^{-2}$. And similar values for the  $^{33}$P$_0$ one.

Given the  (not estimated) errors of the PW results in the region of the resonant-like structure,  
the  S-matrix of these particular waves is in fact compatible with zero, 
quite a different situation than for a real resonance where the S-matrix would have rather a pole.
It corresponds actually to the ''black sphere model'' scattering, 
which is quite different from a resonant scattering, although it produces indeed some structure in the phase shifts.

It remains to be seen wether this structures are an artefact of the analysis or if they
remain unavoidable conclusions of it.

In this respect, it could be pertinent to notice  that both states where they occur are $J=0$,  they have very little statistical
weight and could be affected by large errors in their determination.

Furthermore the - very small - inelasticity parameter $\eta$ enters quadratically as in the annihilation cross sections (\ref{siga_1ch})
from where it should, in principle, be extracted. However, the possibility to extract from the data analysis a significant signal of the order of 10$^{-4}$
seems unrealistic. This is specially true in an observable largely dominated by $^3$SD$_1$ and $^3$PF$_2$ partial waves, at the considered values of $p_{Lab}\approx 600$ Mev/c.

On another hand to generate a model exhibiting zeros of the S-matrix, practically at the same point of the real axis, in different partial waves
appears to be extremely artificial.

For all these reasons, we believe that the possibility for a PWA, to move from one solution to another one in the vicinity of the minima
of the inelasticity parameter, should not be disregarded.

One can thus conjecture that the phase shifts in the vicinity of the  $\delta_I$ peak, 
could, in fact, be continued without exhibiting any resonant-like behaviour, as it happens in all the discussed models, and as it was considered in Refs. 
\cite{JPsi_Julich_PRD91_2015,JPsi_Julich_PRD98_2018}  for the $^1$S$_0$ state.
This would not eliminate all the before mentioned inconsistencies among the \={N}N 
optical models but will greatly simplify the analysis in these two partial waves.

\bigskip
While  the above presentation of the phase shifts has  some interest for a global understanding of the  interaction, 
it is not very useful for a detailed comparison of the models at low energy.
Apart from the poor determination  of the phase shifts themselves, 
the  $L>0$  states have a low-energy behaviour like $\delta (q)\approx -a_Lq^{2L}$
which hides their contribution in this energy region.

One can remove the "centrifugal term" by redefining the reduced phase shifts $\bar{\delta}=\delta /q^{2L}$, as it was done in \cite{CDPS_NPA535_91}, but we believe it is more instructive to compare the effective range functions  $Z_{\alpha}$ and their dependence on the center of mass momenta $q$ for the different partial waves.
This will be done in the following section.

\subsection{The Effective range functions and low energy parameters}\label{Sect_Z}

\begin{figure*}[htbp]
\begin{center}
\vspace{-0.cm}
\includegraphics[width=8.5cm]{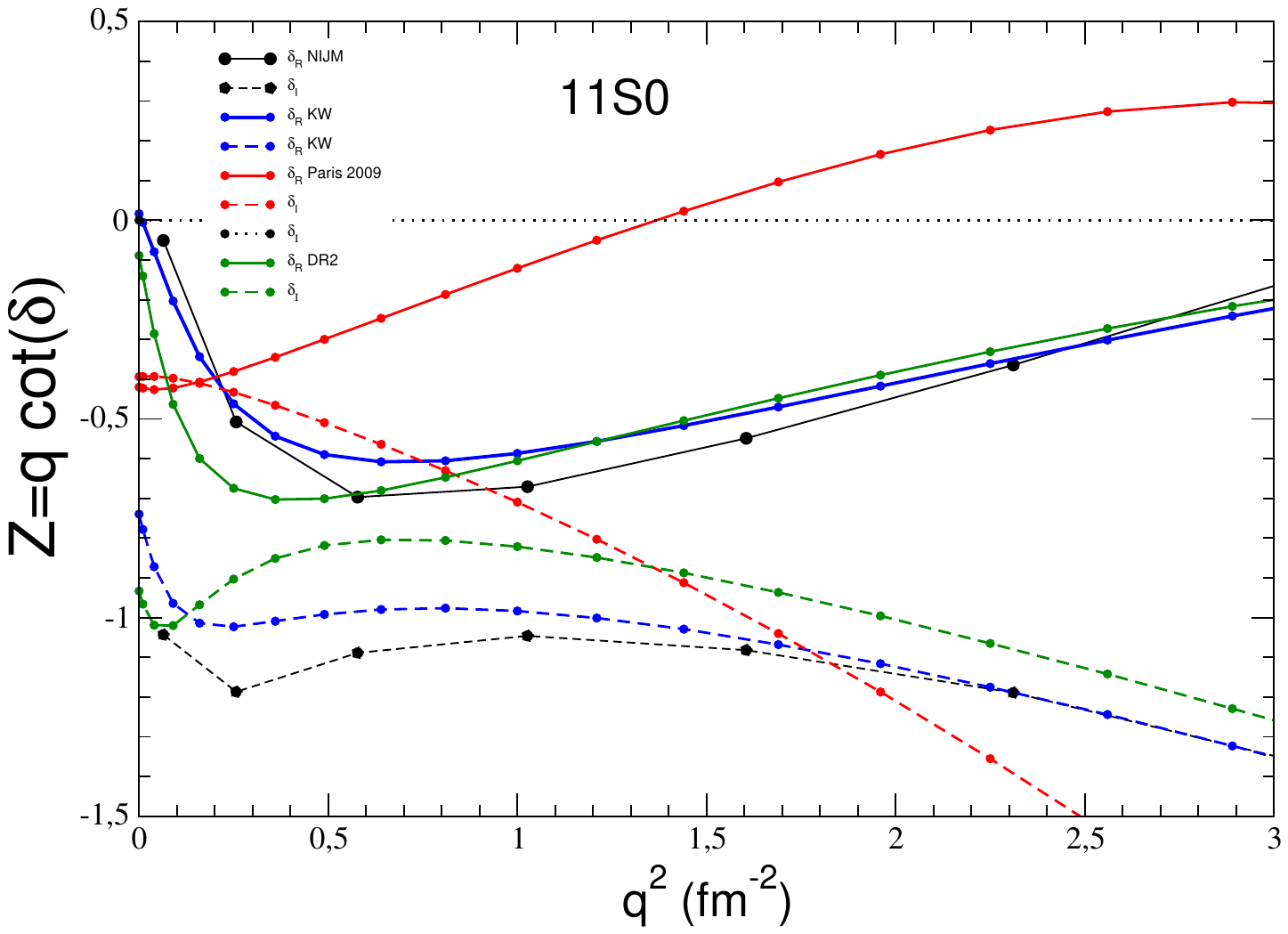} 
\vspace{-0.cm}
\includegraphics[width=8.5cm]{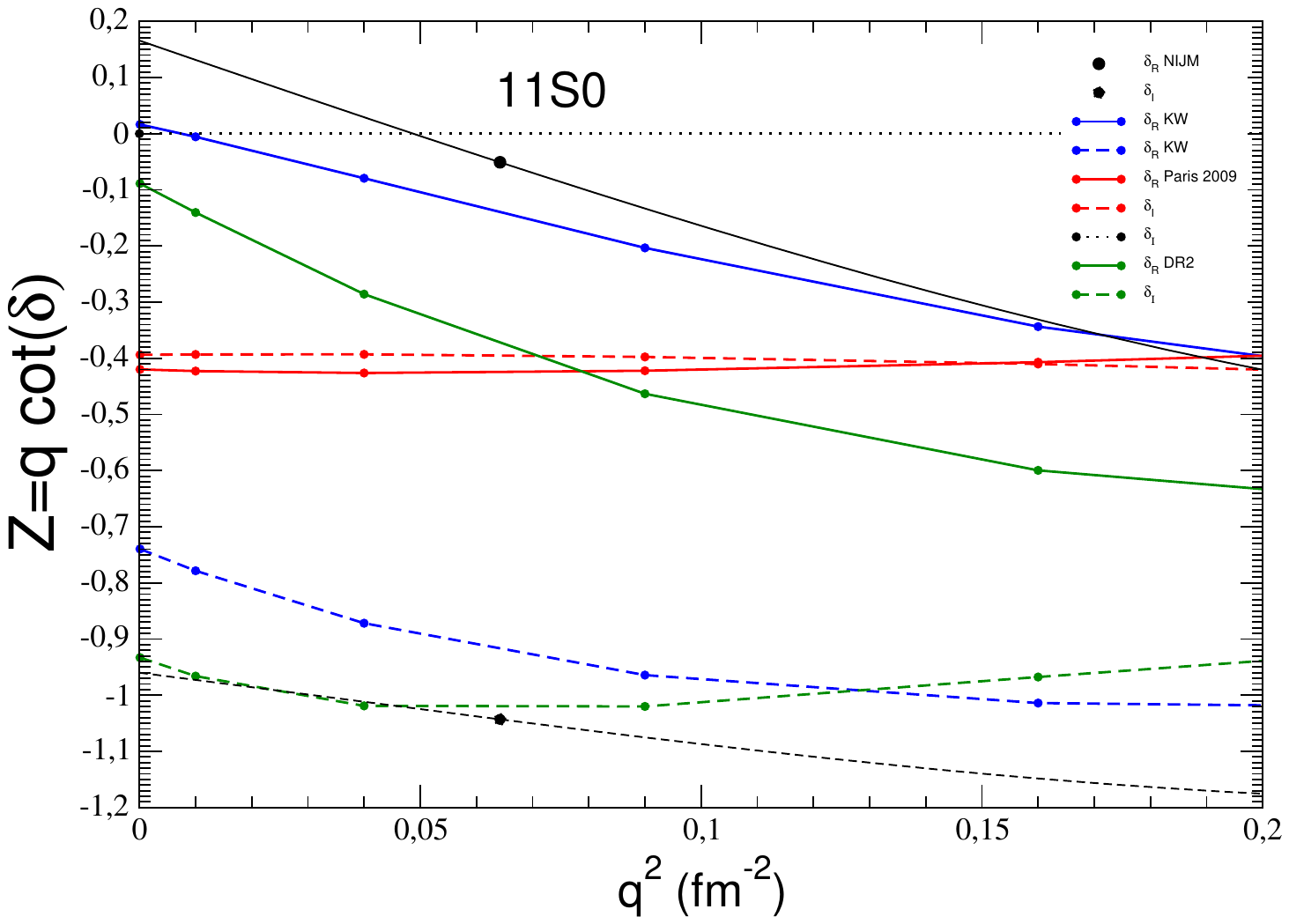}
\vspace{-0.cm}
\vspace{-.97cm}
\includegraphics[width=8.5cm]{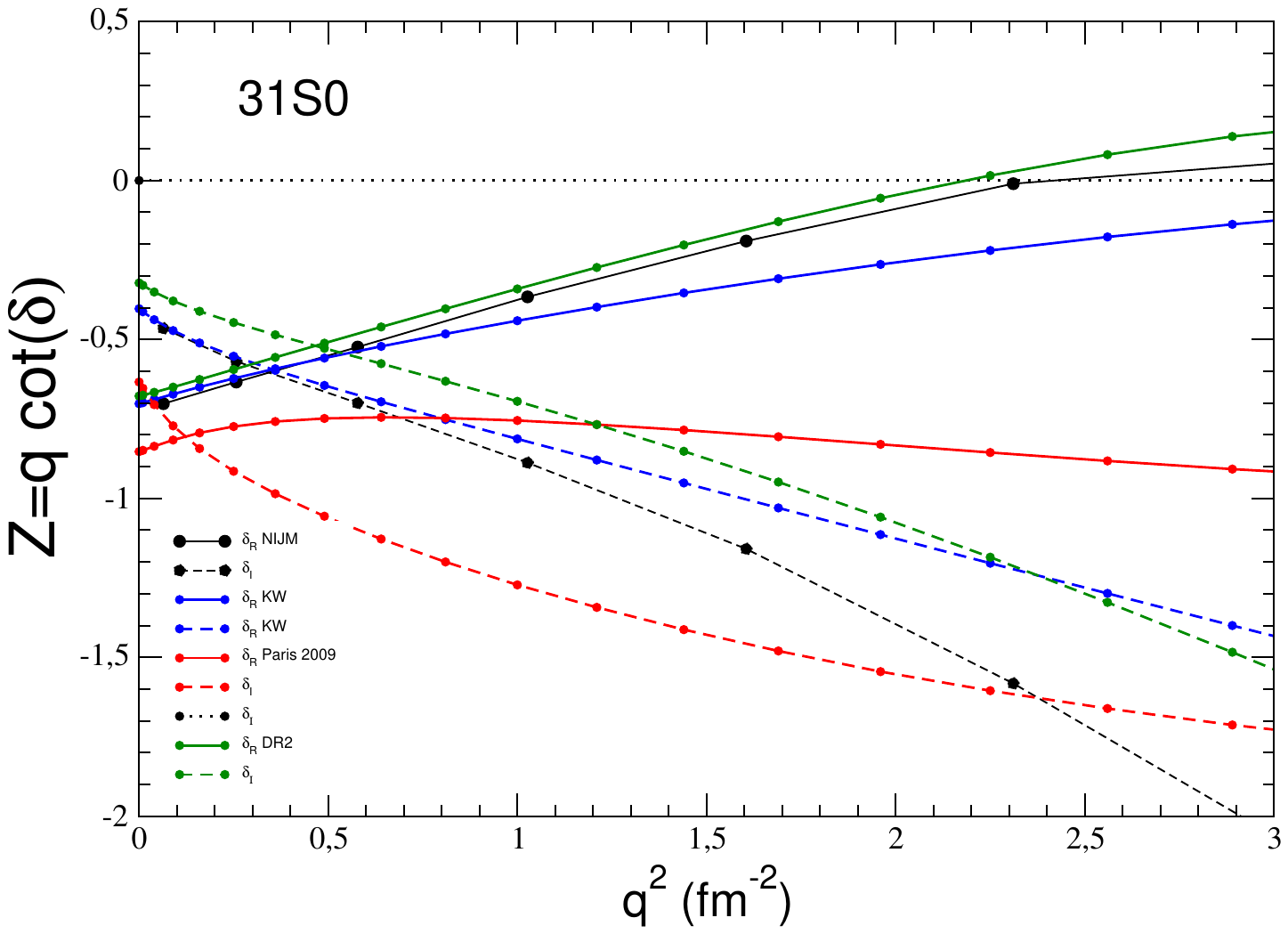} 
\vspace{-0.cm}
\includegraphics[width=8.5cm]{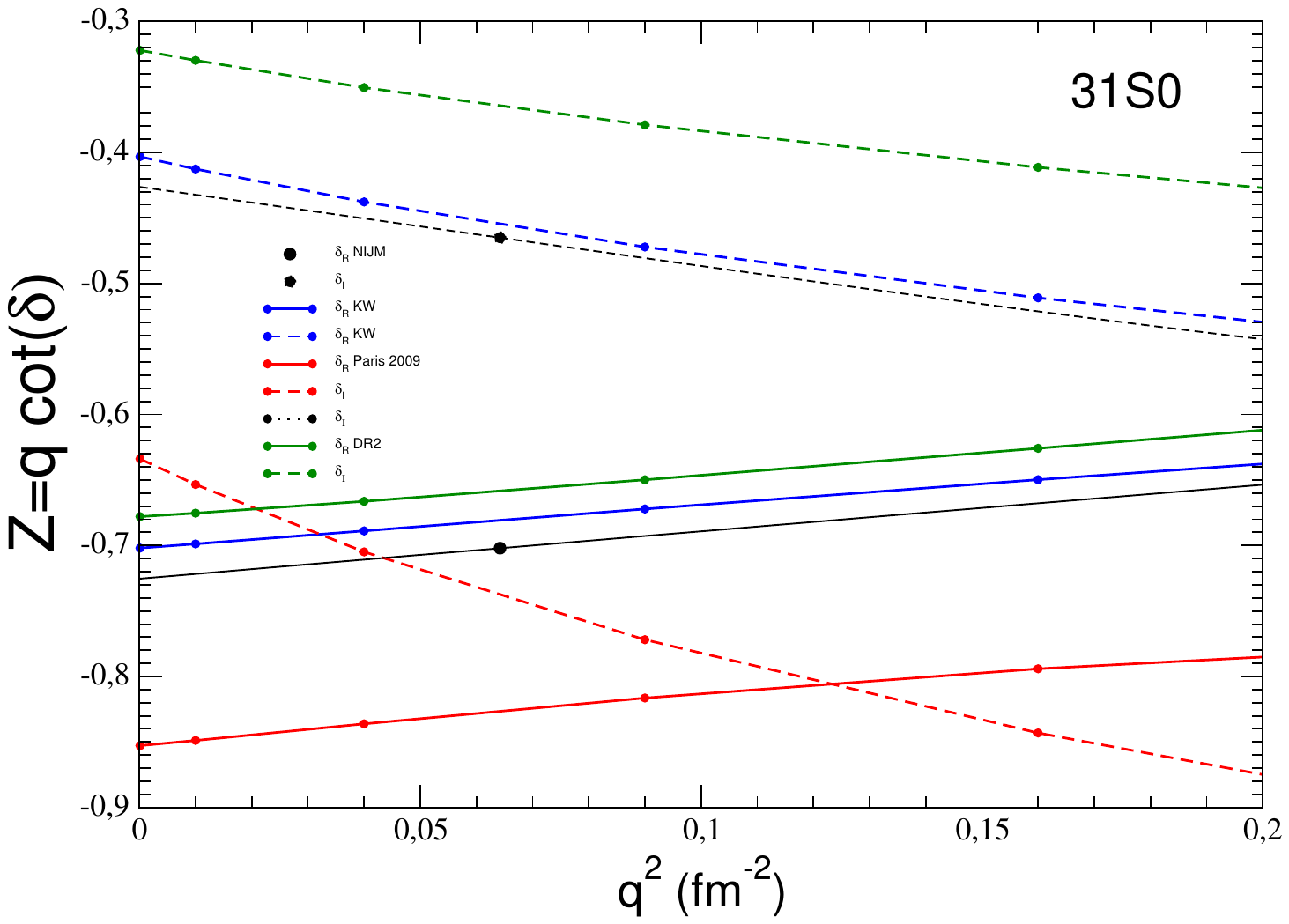}
\vspace{-0.cm}
\caption{Effective range function (\ref{Z}) for the \={N}N $^1$S$_0$ states as function of the center of mass momentum squared (in fm$^{-2}$).
Upper figures correspond to T=0 and the lower ones to T=1. In both cases, the right figure is a zoom of the left one, restricted to the low energy
domain  $q^2\in[0,0.2]$ fm$^{-2}$ ($p_{Lab}\le$ 180 MeV/c), where the ERE of (\ref{Z}) is manifested by the linear behaviour near the origin.}\label{Zq2_1S0}
\end{center}
\end{figure*}

For the tensor decoupled states  the effective range functions  $Z_{\alpha}$  take the form
\begin{equation}\label{Z}
 Z_{\alpha}(q^2)= q^{2L+1} \cot\delta_{\alpha}= -{1\over a_{\alpha}} + {1\over 2} r_{\alpha} q^2 + o(q^4)
 \end{equation}
It is interesting to plot this quantity as  function of $q^2$, for it easily allows to determine the validity region 
of the effective range expansion (ERE), explicited in eq. (\ref{Z}), and given by the linearity domain near the origin.

Notice that the value at the origin $Z(0)$ 
\begin{equation}\label{Z0}
 Z(0)= Z(0)_R + i\; Z(0)_I   =  {-a_R +  i a_I  \over \mid a\mid^2  }   
\end{equation}
is related to the scattering length as
\begin{eqnarray}\label{a_Z0}
a_R&=&  -\mid a\mid^2  Z(0)_R =    - { Z(0)_R\over \mid Z(0)\mid^2 }  \cr
a_I &=&  \mid a\mid^2  Z(0)_I    =  { Z(0)_I\over \mid Z(0)\mid^2 }  
\end{eqnarray}
In particular one has imperatively $Z(0)_I<0$.

In what follows we will display $Z_{\alpha}(q^2)$ for all the considered PW. 
First (left panel), in the full  energy domain $q^2\in[0,3]$ fm$^{-2}$, which correspond to $p_{Lab}\leq 700$ MeV/c.
Next (right panel of the same figure), presents a zoom in the low energy region $q^2\in[0,0.2]$ fm$^{-2}$ ($p_{Lab}\le$ 180 MeV/c) 
to better exhibit the linearity domain and determine the low energy parameters ($a_{\alpha}, r_{\alpha}$), abusively denoted scattering "length" and effective "range".
The scattering length is given by the $Z(0)$ value, following eq. (\ref{Z0}), and the effective range
from (twice) the slope at the origin.

The colour and model conventions are the same as those used for the phase shifts: 
real part of $Z$ is plotted by solid lines and the imaginary part by dashed lines.

The extracted LEP values are collected in Tables \ref{Table_LEP_S} and \ref{Table_LEP_P} for the  considered models and PWs.
The NPW results given in \cite{ZT_NNB_Nijm_PW_2012} were limited to $p_{Lab} \leq$ 100 MeV/c. We have quadratically
extrapolated their values at the origin by using the 3 lowest points.
They are denoted by Nijm* in Tables \ref{Table_LEP_S} and \ref{Table_LEP_P} 
and may have only an indicative value, in particular by comparing them to Julich results.
Despite of this naive extrapolation, they all fulfill $Z(0)_I<0$.

\begin{figure*}[htbp]
\begin{center}
\includegraphics[width=8.5cm]{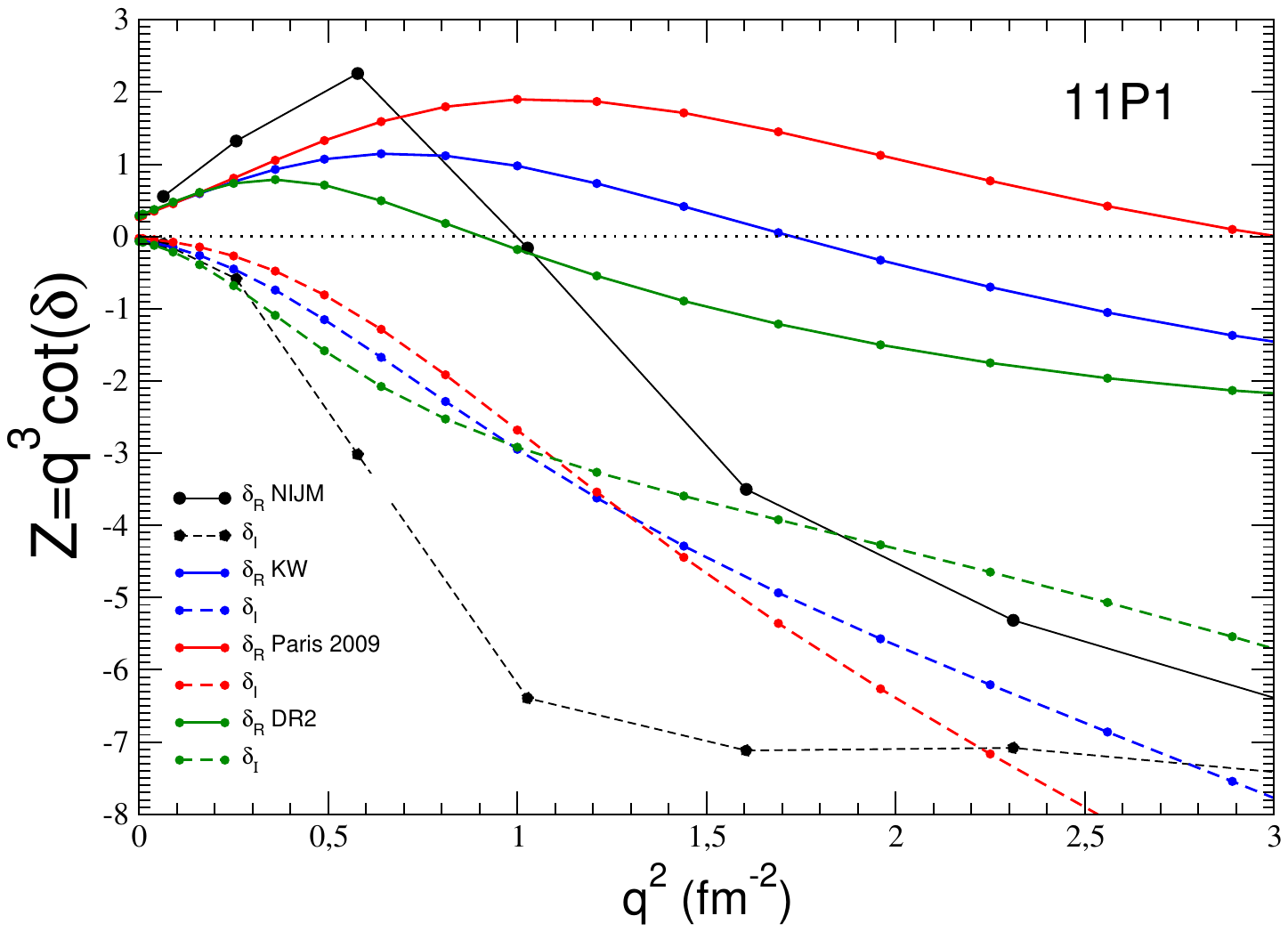} 
\includegraphics[width=8.5cm]{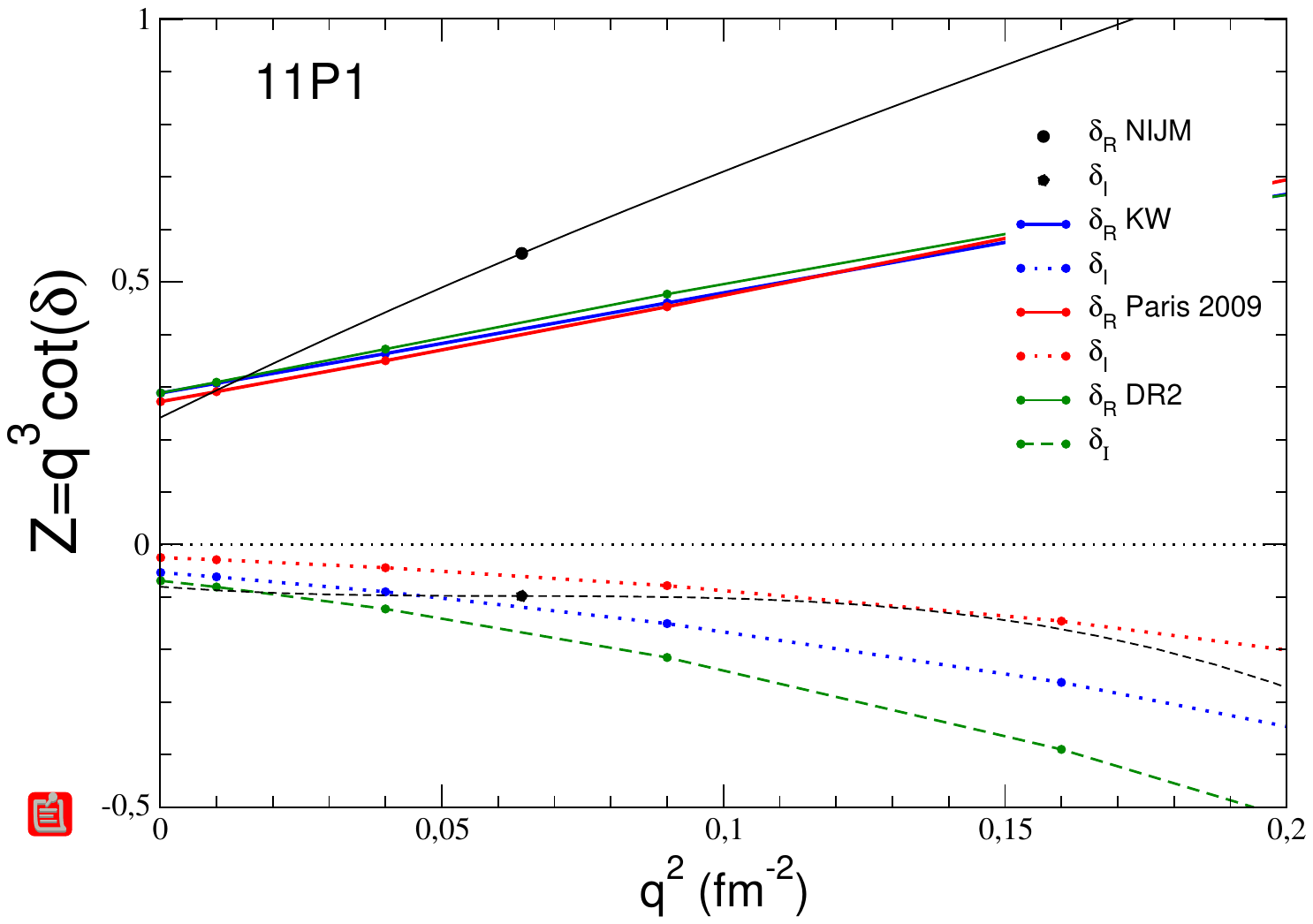}
\vspace{-.97cm}

\includegraphics[width=8.5cm]{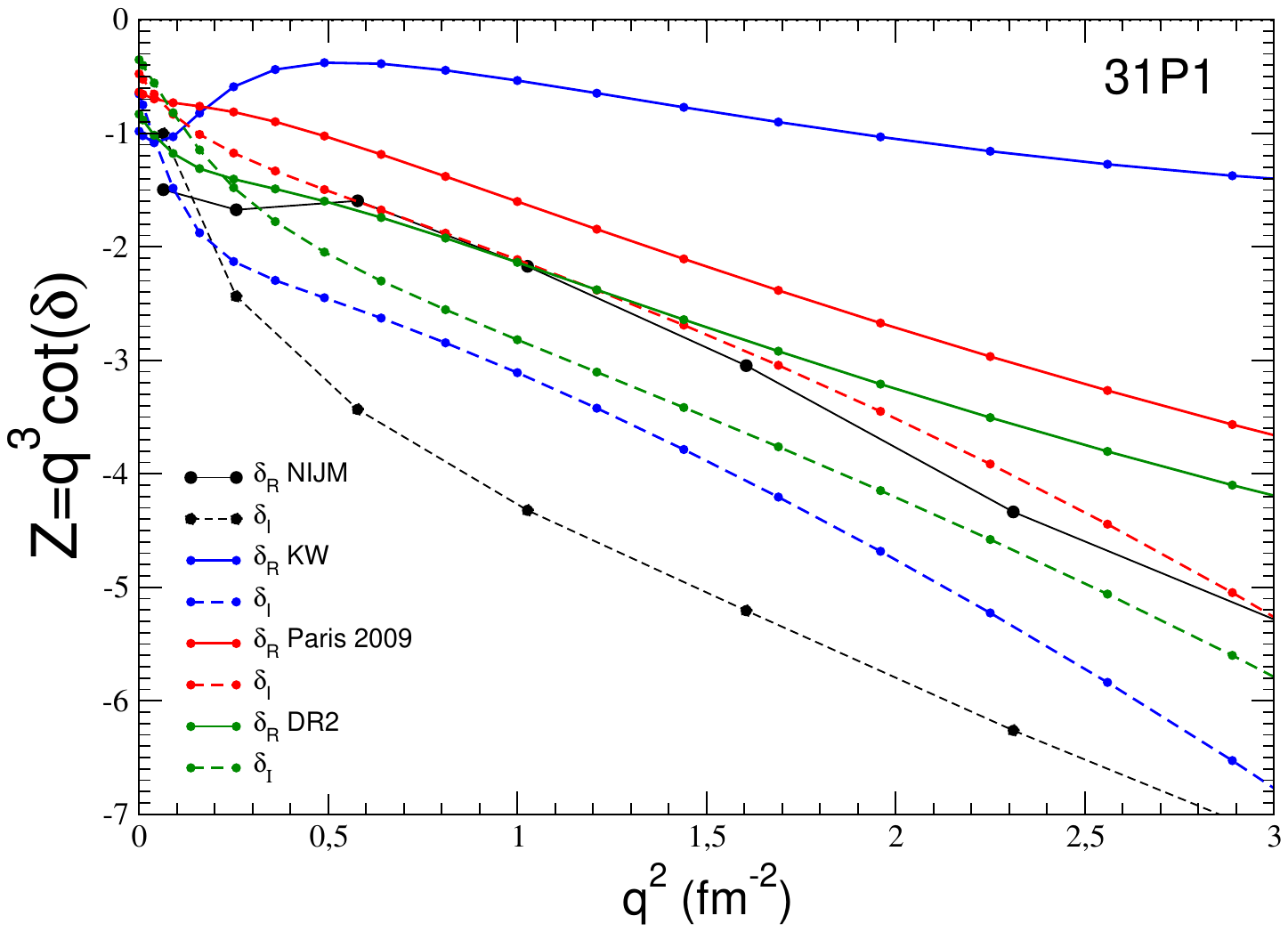} 
\includegraphics[width=8.5cm]{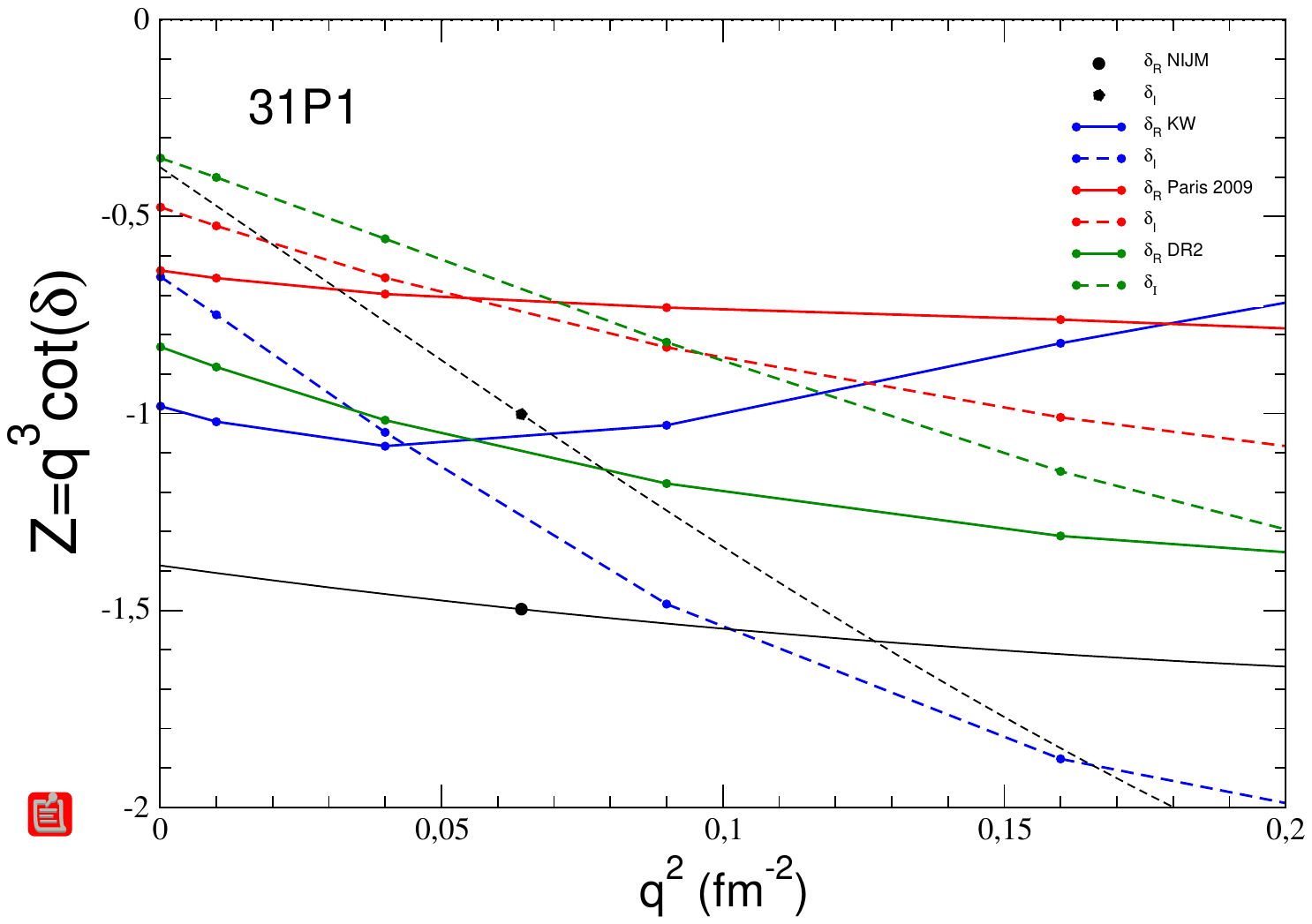}
\caption{Effective range function (\ref{Z}) for the  \={N}N   $^1$P$_1$ state as function of the center of mass momentum squared (in fm$^{-2}$),
with the same convention as in figure \ref{Zq2_1S0}.}\label{Zq2_1P1}
\end{center}
\end{figure*}

\bigskip
{\bf The results for the $^1$S$_0$} state are displayed in Fig. \ref{Zq2_1S0}. 
The upper panel correspond to T=0 and the lower one to T=1. 

 The particular behaviour of Paris results is manifested, both in the real as well as the imaginary phase shifts in all the energy domain. One finds however a qualitative agreement among the other models including NPW.

As one can see from the right upper figures,  
the $Z(q^2)$ dependence of the T=0 state for the Paris potential is totally flat in the full domain. 
For the other models, the ERE is valid only at relatively low energy $q^2\leq 0.05$ fm$^{-2}$, i.e is $p_{Lab}\approx 120$ MeV/c. 
Beyond this energy, the $q^4$  terms neglected in (\ref{Z}) become relevant,
and any linear extrapolation in $q^2$ would lead to wrong results.

The $Z(q^2)$ dependence for the T=1 state (lower panels) 
is fairly  smooth, with  no visible trace of the NPW resonant-like behaviour
manifested in the corresponding phase shift (right panel of Fig \ref{delta_1S0}) at $p_{Lab}\approx$ 600 MeV/c  ($q^2\approx$ 2.25 fm$^{-2}$).
The differences  among the models are much smaller than for T=0 (except for Paris potential) and  
lead to scattering length values which are positive and consistent to each other within 15 \% (see Table \ref{Table_LEP_S}).

\begin{figure*}[htbp]
\begin{center}
\vspace{-0.cm}
\includegraphics[width=8.5cm]{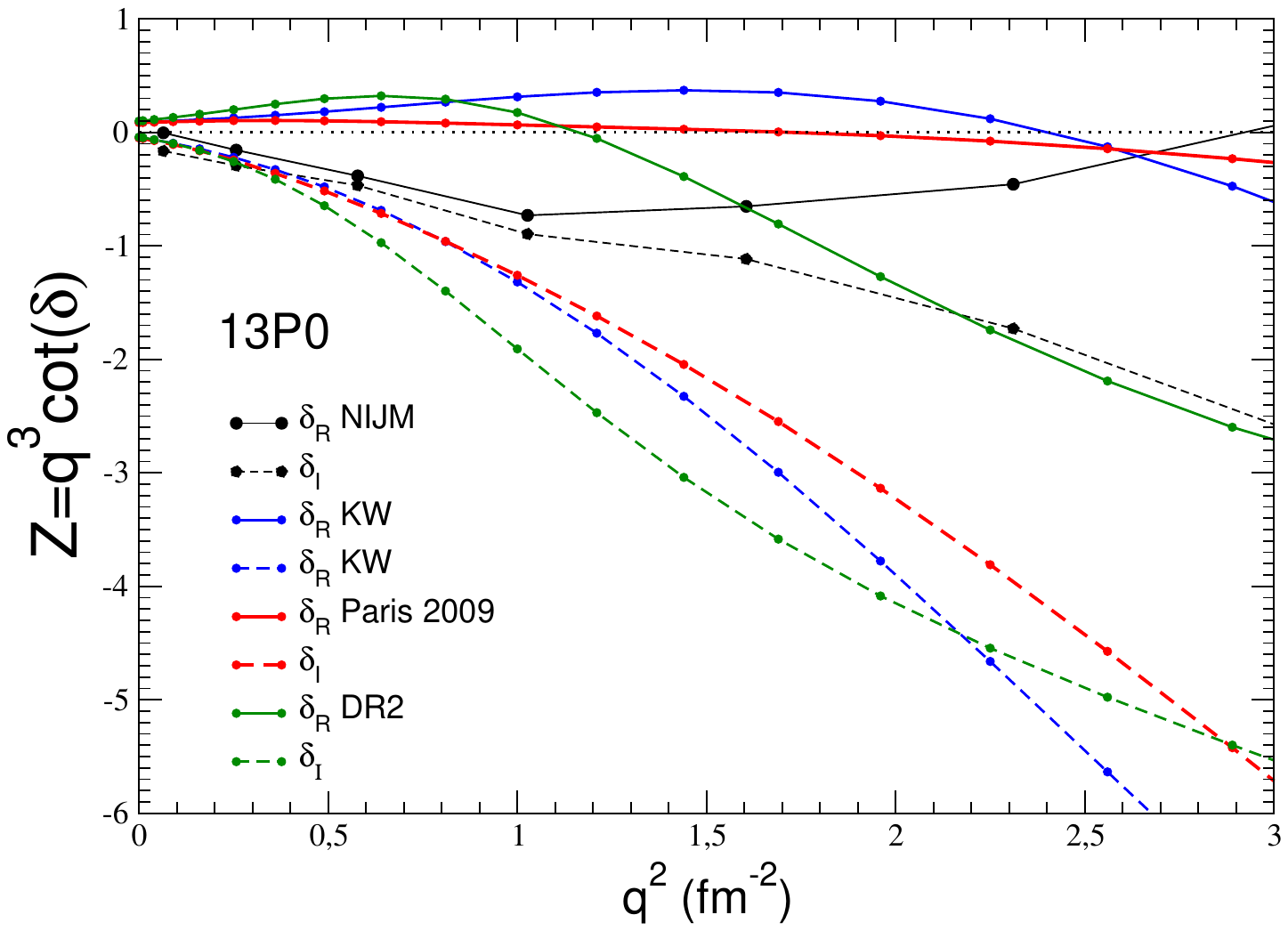} 
\includegraphics[width=8.5cm]{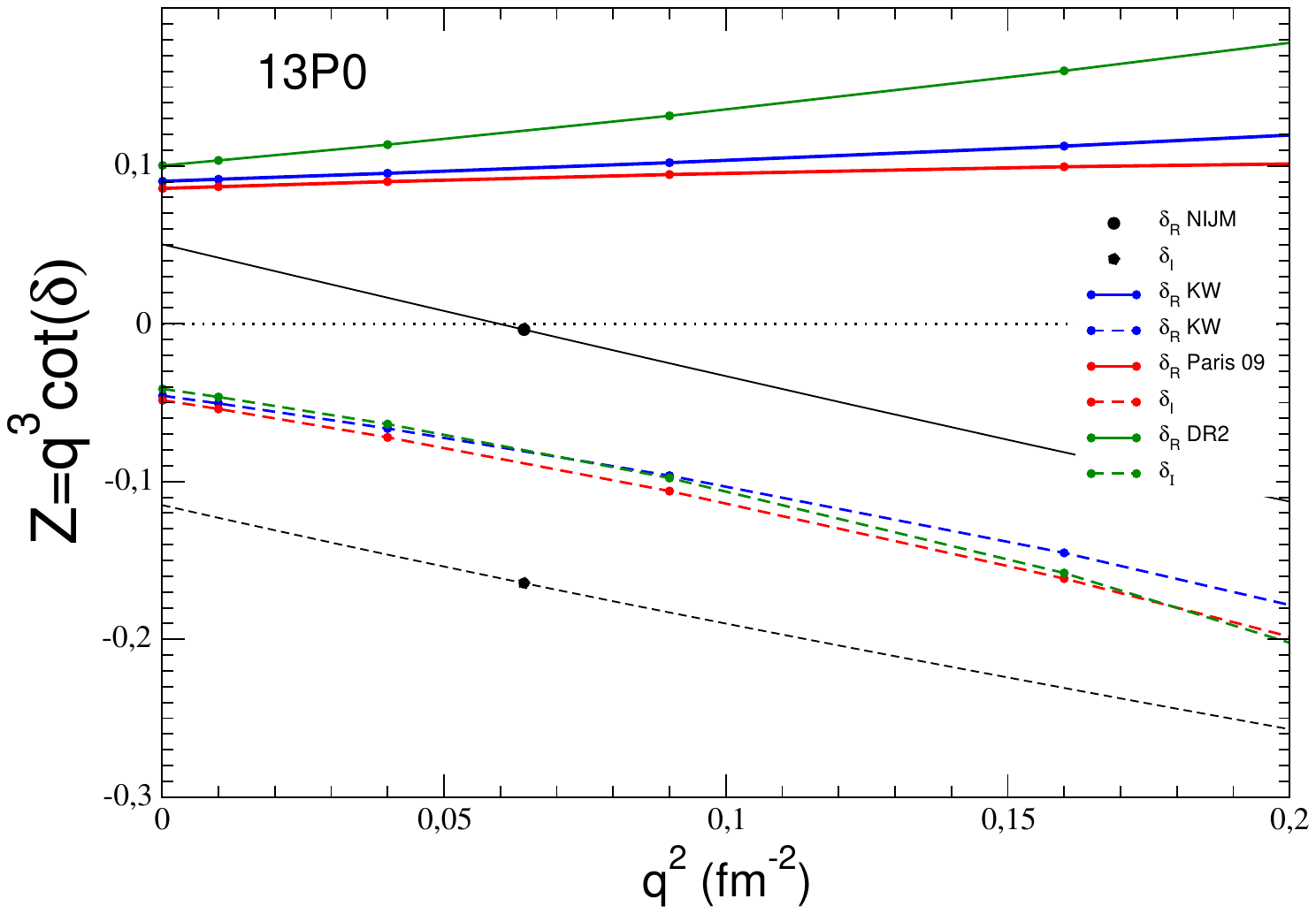}
\vspace{-.97cm}

\includegraphics[width=8.5cm]{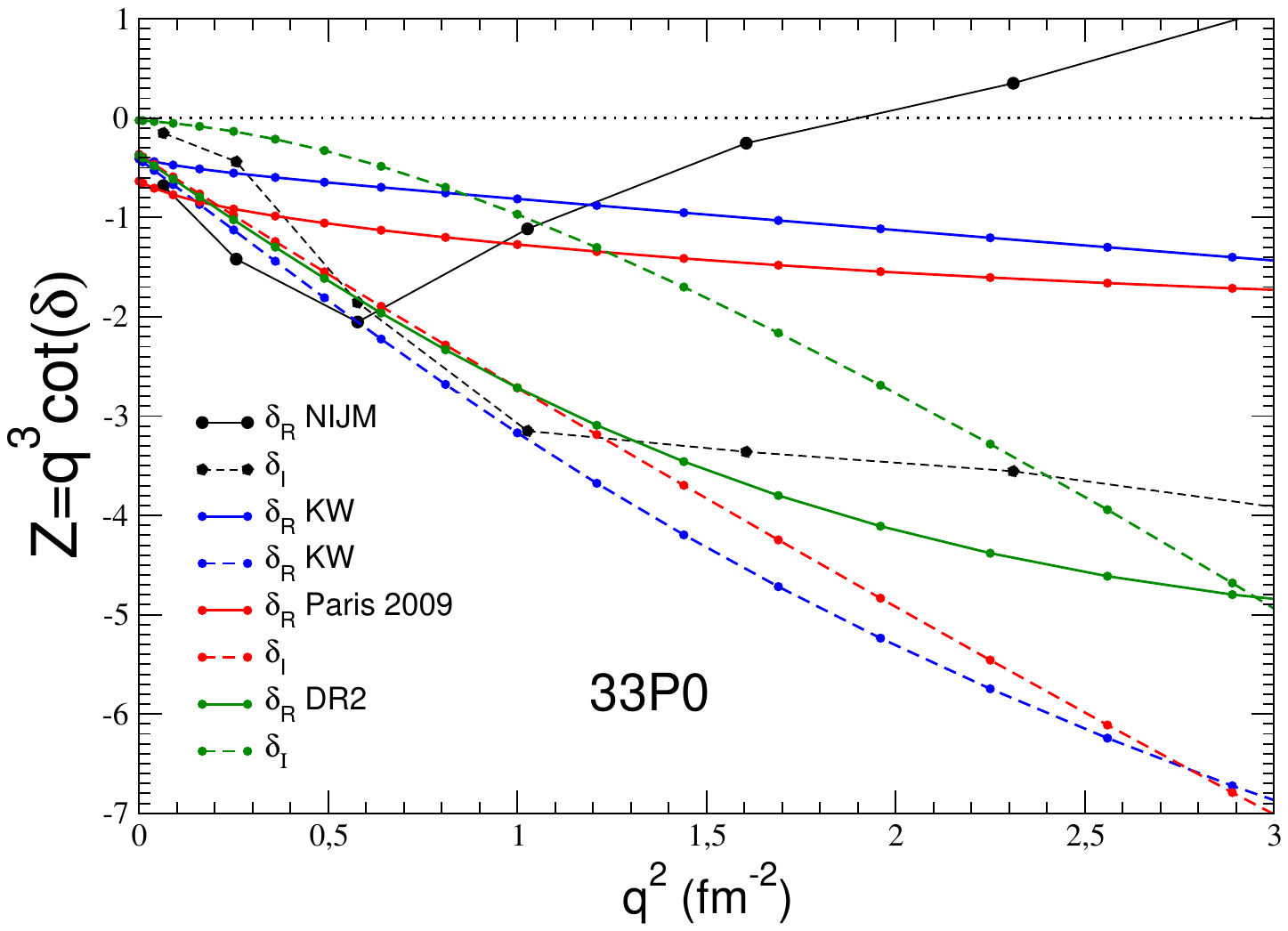} 
\includegraphics[width=8.5cm]{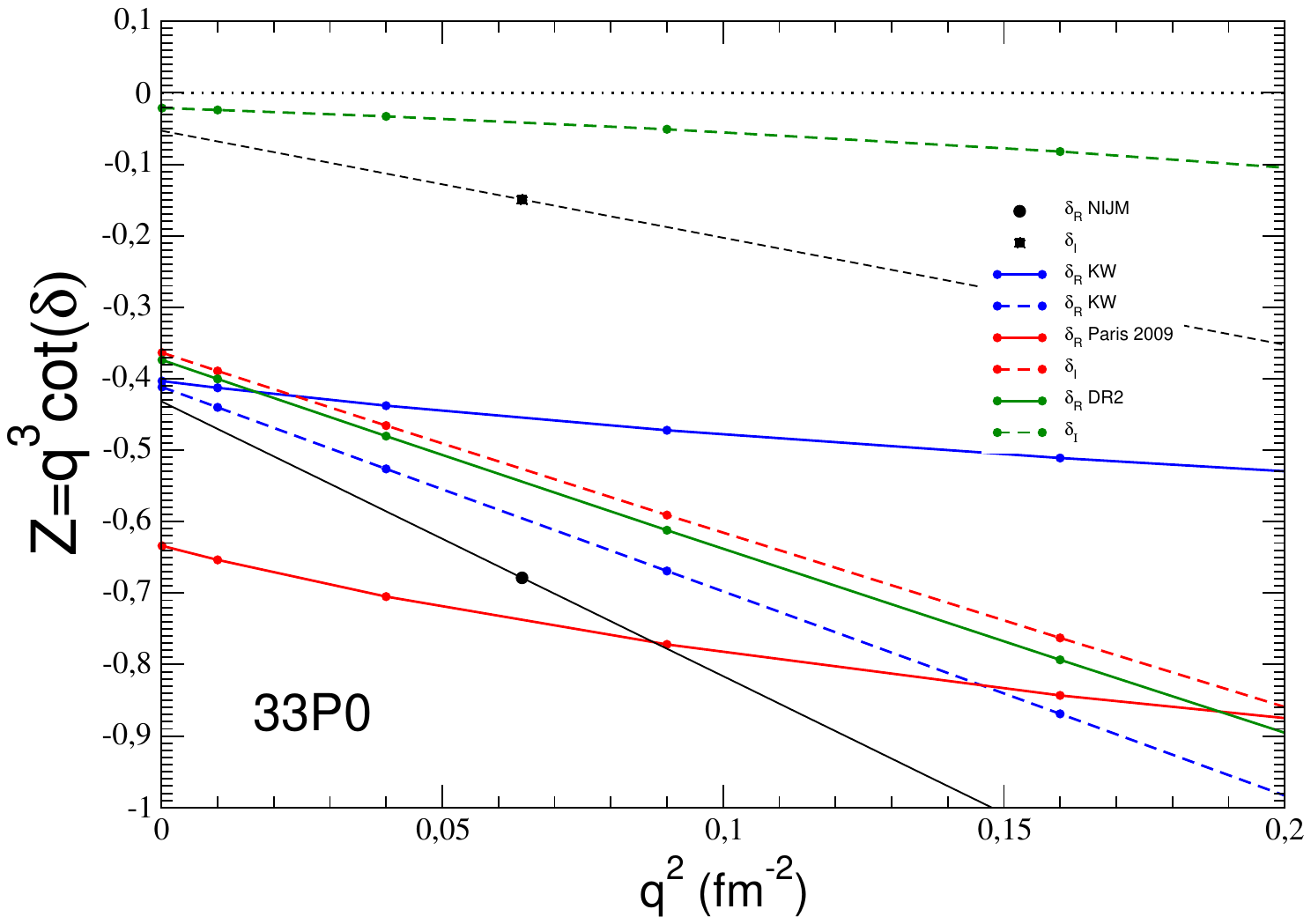}
\caption{Effective range function (\ref{Z}) for the  \={N}N   $^3$P$_0$ state as function of the center of mass momentum squared (in fm$^{-2}$) 
and the same conventions as in  figure \ref{Zq2_1S0}.}\label{Zq2_3P0}
\end{center}
\end{figure*}

\begin{figure*}[htbp]
\begin{center}
\vspace{-0.cm}
\includegraphics[width=8.5cm]{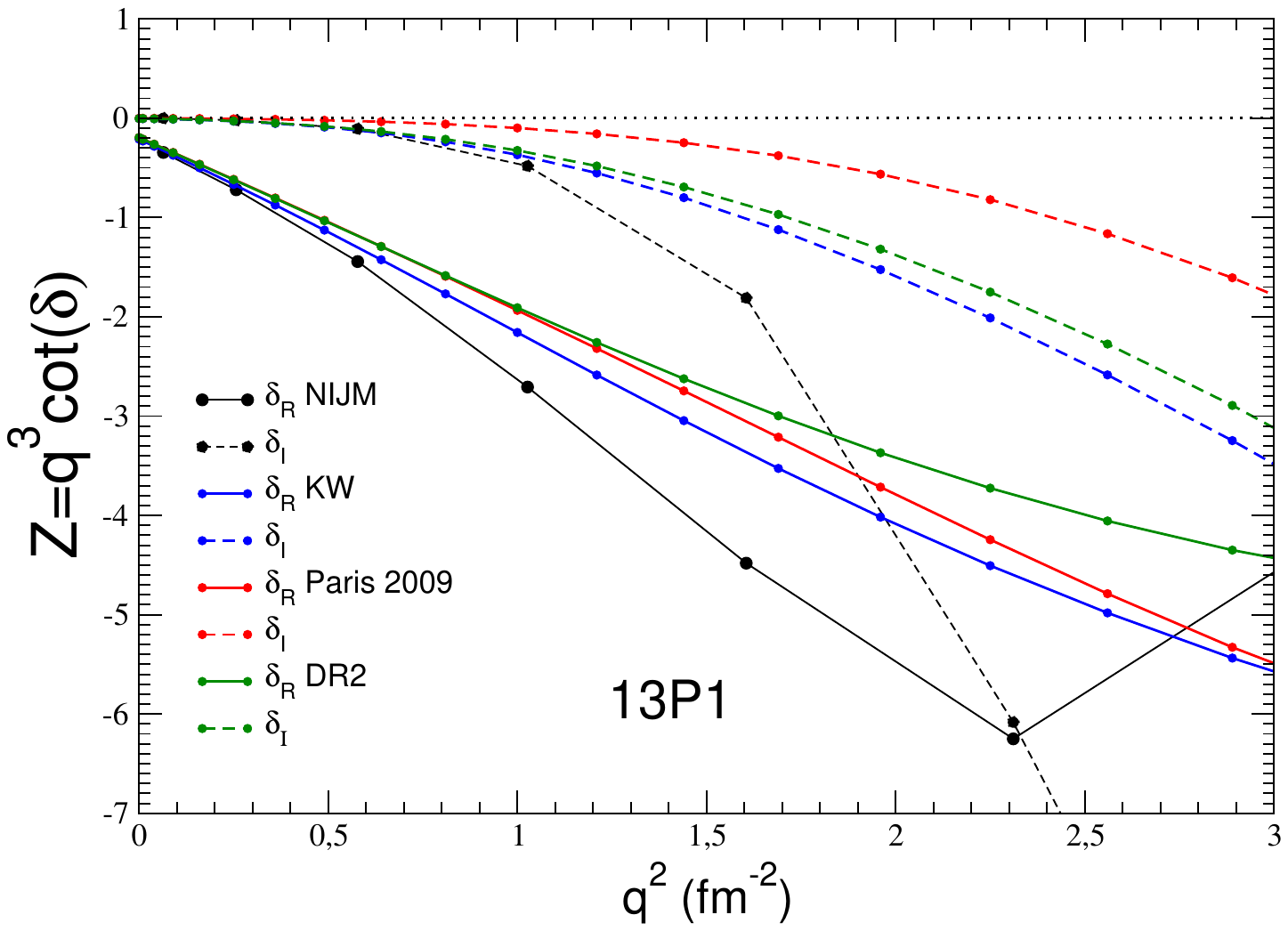} 
\includegraphics[width=8.5cm]{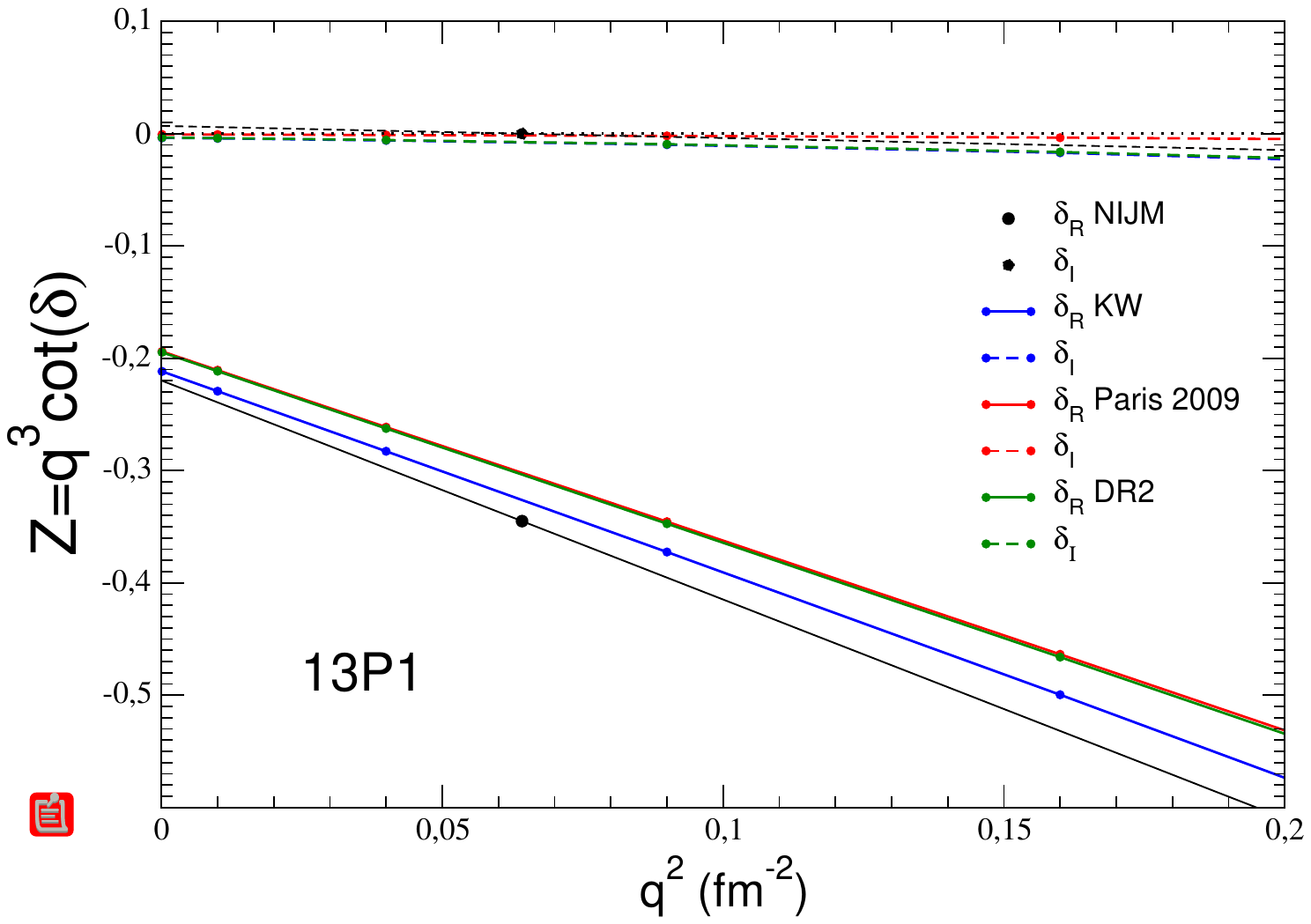}
\vspace{-0.97cm}

\includegraphics[width=8.5cm]{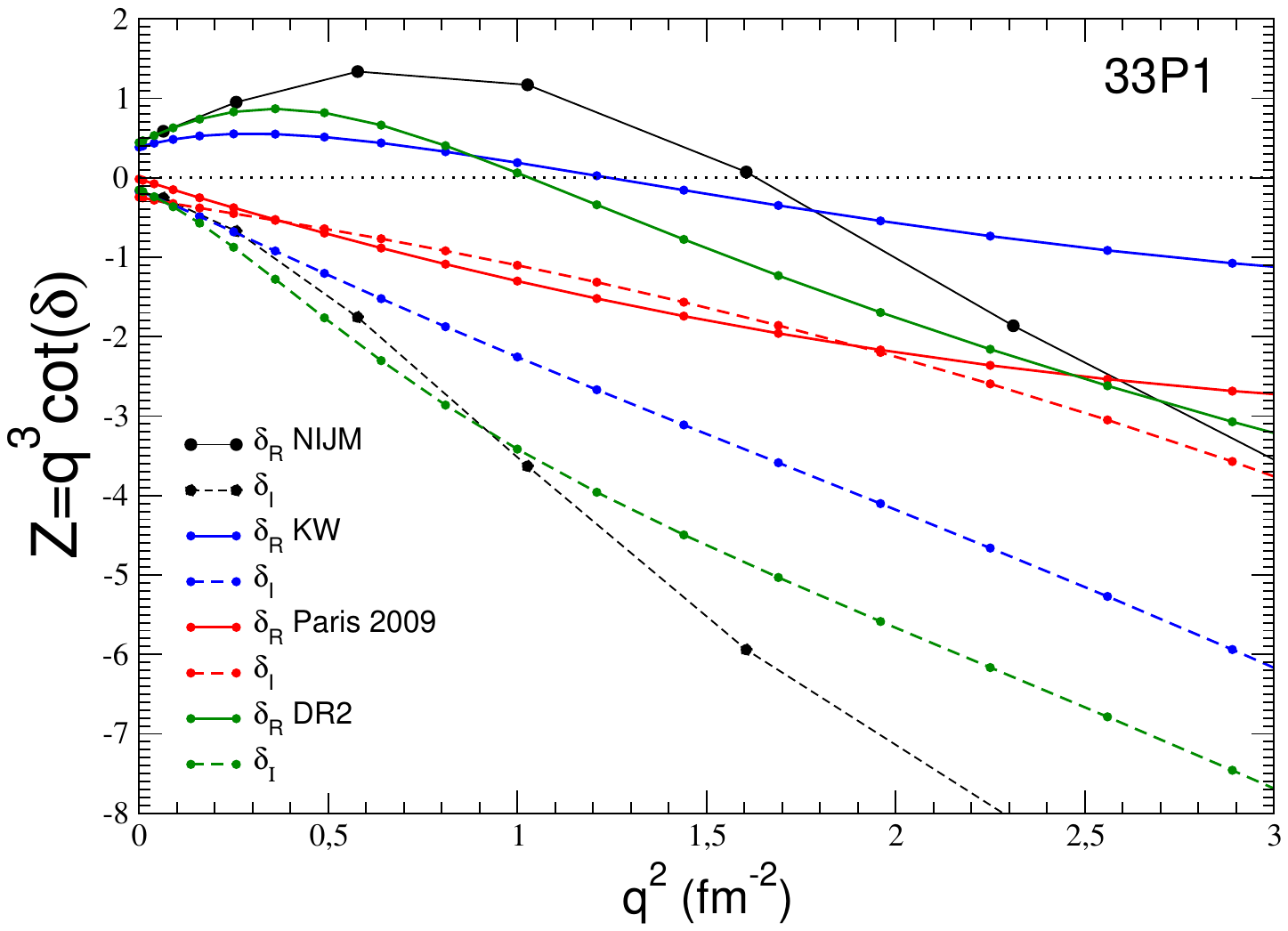} 
\includegraphics[width=8.5cm]{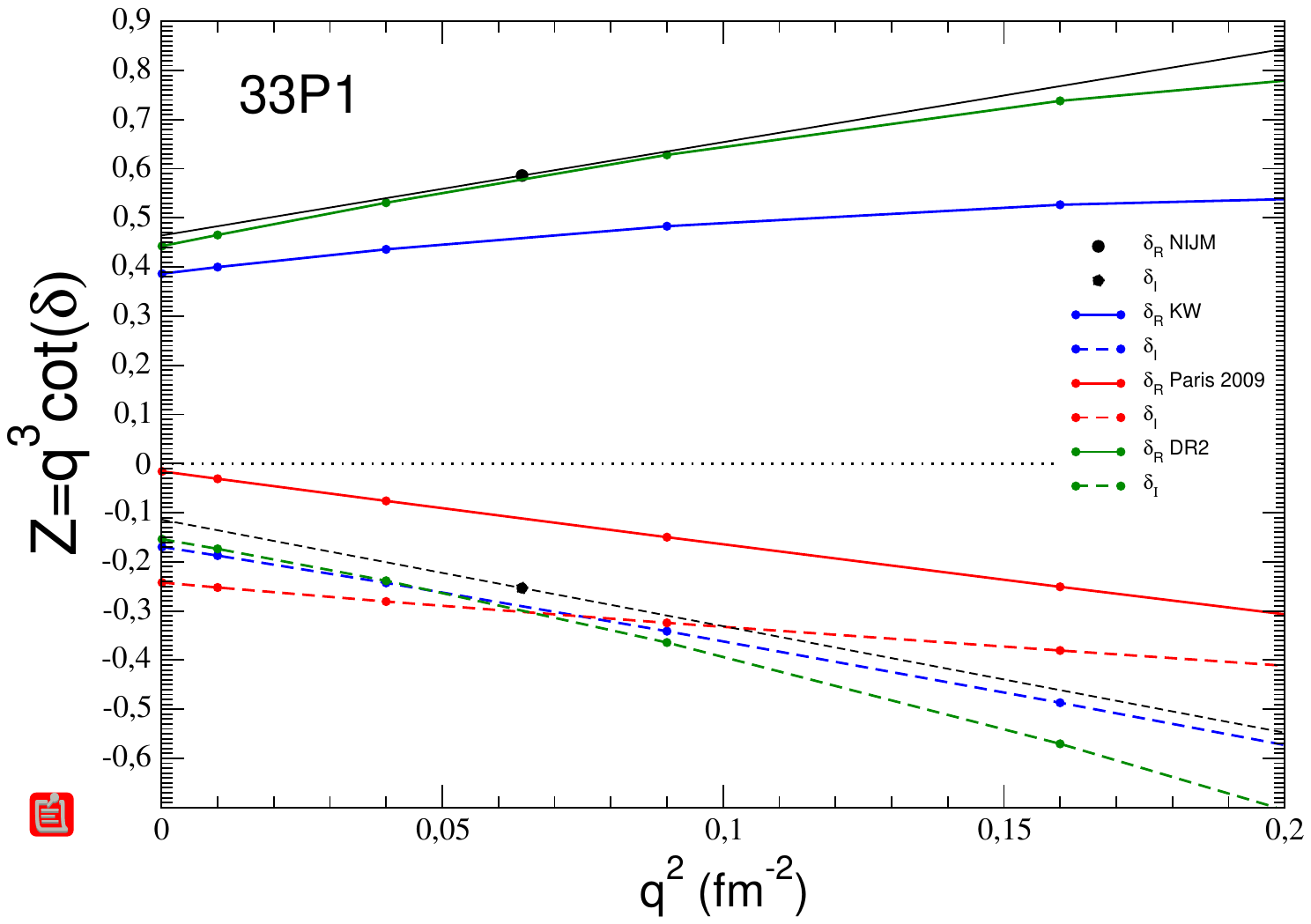}
\caption{Effective range function (\ref{Z}) for the  \={N}N   3P1 state as function of the center of mass momentum squared (in fm$^{-2}$),
with the same convention as in  figure \ref{Zq2_1S0}.}\label{Zq2_3P1}
\end{center}
\end{figure*}

\bigskip
{\bf The effective range functions for the P-waves are displayed in Figures \ref{Zq2_1P1}  to \ref{Zq2_3P1}.}
In this representation the low-energy part is magnified with respect to the phase shifts and one can see
that, as it was the case for S-waves, 
sizeable differences among the models themselves and with respect to the NPWA emerge.

\bigskip
{\bf  Figure \ref{Zq2_1P1} shows $Z(q)$  for the $^1$P$_1$ state}. 
For T=0,  the results of the real part  (left upper panel) 
have a  similar qualitative behaviour: monotonously increasing from the origin until a maximum value and decreasing
with a zero crossing towards negative region.
However,  although  the Paris, DR and KW models are close to each other up to $q^2$=0.5  
fm$^{-2}$ ($p_{Lab}\approx$ 300 MeV/c),
they differ significantly  at high energy, specially with respect  the  NPW results.

For T=1 (lower panels) the dispersion is even larger and starts at low energy.
Notice that the DR2 model displays a fast increasing near the origin 
which suggest a near-threshold resonant state.
It manifests also in the lower right panel with an ERE breaking below $q^2=0.05$ fm$^{-2}$.

Despite these differences in the model predictions, it is worth noticing  the remarkable stability of  
the real part of the scattering volumes for both isospin states. 
They remain all between a 15\% difference band, as one can see in Table \ref{Table_LEP_P}.
A possible reason of this stability will be discussed at the end of the section.
The essential differences between models are in fact given by their absorptive parts.  
These could, in principle, be settled by additional measurements of the full fine structure in antiprotonic atoms.  
Unfortunately this  measurement still waits for its turn \cite{Gotta_Review_PPNP_2004}.

\bigskip
{\bf  The results for $^3$P$_0$  wave are shown  in Figure \ref{Zq2_3P0}}.
Again, for T=0 the real phase shifts a general qualitative agreement is observed among DR, KW and Paris models
in a more extended energy region, with a departure from NPW already at $q=0$.
The validity of the ERE expansion (upper right panel) extends up to $q^2=0.2$.
Except for the J\"{u}lich model,  the corresponding 
scattering volumes have a large real part, attributed in \cite{CDPS_NPA535_91} to the existence of a near-threshold bound or resonant state, and are in a very close agreement ($<$2\%).

For T=1, the NPW resonance-like  structure displayed in Fig \ref{delta_3P0}  at $p_{Lab}\approx$ 600 MeV/c   leaves
no trace in the effective range function $Z$. However a similar structure -- absent at the level of  phase shifts --  is seen in $Z$  at  $q^2=0.5$ fm$^{-2}$, 
breaking any possible agreement with the other optical models. 
The corresponding scattering lengths (real part) remain close within 15\% and the imaginary part is very small.

\bigskip
{\bf The $^3$P$_1$ effective range functions  $Z(q^2)$ are displayed in Figure  \ref{Zq2_3P1}} for both isospin channels.
The T=0 state (upper panel) is the most stable partial wave, for the real as well as for the imaginary part.
This is probably due to the $^{13}$P$_1$ potential,  repulsive  in all the models, which also explains the small imaginary part

The quantitative disagreements start only above  $q^2\approx$1 fm$^{-2}$ ($p_{Lab}\approx $400 MeV/c)
and increase with the energy.
The ERE (right panel) works perfectly in all the domain  and the LEPs (both $a_1$ and $r_1$) are in close agreement, with
an almost vanishing imaginary part.

For T=1, the  main difference comes  Paris 2009, which displays
a different qualitative behaviour in all the energy domain, including the LEPs. 
As mentioned, this particular feature is due to a quasi-bound state generated in this model at E=-3.6-i12.4 MeV. 
The other optical models are in quite a good agreement.

\begin{table}[htbp]
\begin{tabular}{ l    ll                                    ll                    } \\
           & $a_0$          &$r_0$              &$a_0$         &$r_0$    \\    \hline     
T=0        &\multicolumn{2}{l}{\hbox{$^{11}$S$_0$}} & \multicolumn{2}{l}{\hbox{$^{13}$SD$_1$}}    \\  \hline
Nijm*      &-0.17 -1.01i    & -6.9-2.9 i        & --         & --               \cr
J\"{u}lich &-0.21 -1.23i    & --                & 1.42-0.88i & --          \\
Paris 09   & $\;$1.27 -1.18i& -0.53+0.14i       & 1.20-0.80i & --         	 \cr	
KW         &-0.03 -1.35i    & -4.7-7.9i         & 1.23-0.77i & --            \cr 
DR2        & $\;$0.10 -1.07i& -11-6.2i          & 1.28-0.78i & --         \\   \hline   
T=1               &\multicolumn{2}{l}{\hbox{$^{31}$S$_0$}} & \multicolumn{2}{l}{\hbox{$^{33}$SD$_1$}}  \\  \hline
Nijm*      & 1.02 -0.60i   &  0.7-1.2i          &  --        & --             \cr
J\"{u}lich & 1.05 -0.58i   &   --               & 0.44-0.96i & --         \\
Paris 09   &  0.76 -0.56i  &  0.9-3.9i          & 0.61-0.44i & --         \\
KW         &  1.07 -0.62i  &  0.7-1.9i          & 0.78-0.80i & --       \\
DR2        &  1.20 -0.57i  &  0.6-1.6i          & 0.89-0.71i & --         \\
\end{tabular}
\caption{S-wave \={N}N low energy parameters (in fm) for the considered optical models:  J\"{u}lich results are taken from Tab 3 of Ref.  \cite{Haidenbauer_JHEP_2017}, 
KW and DR2 from \cite{CRW_ZPA343_1992},  Paris 2009 have been recomputed and are in agreement with \cite{Benoit_Private}.
The values of Nijmegen are obtained by extrapolating the phase shifts from Figures \ref{delta_1S0} and \ref{delta_3SD1}.}\label{Table_LEP_S}
\end{table}

\begin{table*}[htbp]
\begin{tabular}{ l                      ll     ll          ll        ll  } \\
           & $a_1$       & $r_1$   & $a_1$          & $r_1$         & $a_1$   &  $r_1$     &$a_1$  & $r_1   $\\    \hline     
T=0        &\multicolumn{2}{l}{\hbox{$^{11}$P$_1$}}&\multicolumn{2}{l}{\hbox{$^{13}$P$_0$}}&\multicolumn{2}{l}{\hbox{$^{13}$P$_1$}} &\multicolumn{2}{l}{\hbox{$^3$PF$_2$}}  \\  \hline
Nijm*      & -3.34-1.22i & 9.3-1.2i & -3.06-7.23i & -1.7-1.5i  & 4.36-0.00i & -3.5-0.0i  &   --         & --  \cr
J\"{u}lich & -2.87-0.36i & --       & -2.83-7.82i & --         & 4.61-0.05i & --         &  -0.74-1.13i & --  \\
Paris 09   & -3.62-0.34i & 3.8-0.8i & -8.78-4.99i & 0.23-1.1i  & 5.12-0.02i & -3.4-0.02  &  -0.49-0.87i & --  \cr	
KW         & -3.36-0.62i & 3.7-1.6i & -8.83-4.45i & 0.25-0.97i & 4.73-0.08i & -3.5-0.1i  &  -0.46-1.09i & --  \cr 
DR2        & -3.28-0.78i & 4.2-2.3i & -8.53-3.50i & 0.63-1.0i  & 5.14-0.09i & -3.4-0.1i  &  -0.59-0.85i & --  \\   \hline   
T=1        &\multicolumn{2}{l}{\hbox{$^{31}$P$_1$}}&\multicolumn{2}{l}{\hbox{$^{33}$P$_0$}}&\multicolumn{2}{l}{\hbox{$^{33}$P$_1$}}   &\multicolumn{2}{l}{\hbox{$^3$PF$_2$}}  \\  \hline
Nijm*      &  0.66-0.18i & 3.3-20i  & 2.33-0.92i  & -10-0.7i   &-2.02-0.70i &  4.7-2.8i  &  --          &  -- \cr
J\"{u}lich &  0.80-0.34i & --       & 2.18-0.19i  &  --        &-2.04-0.55i &  --        & -0.48-0.34i  &  --   \\
Paris 09   &  1.00-0.77i &-3.7-9.8i & 2.74-0.00i  & -5.2-0.01i & 0.28-4.11i & -3.0-2.0i  & -0.13-0.21i  &  -- \\
KW         &  0.71-0.47i &-8.3-21i  & 2.43-0.11i  & -5.8-0.43i &-2.17-0.95i &  2.7-3.5i  & -0.30-0.45i  &  -- \\
DR2        &  1.02-0.43i &-11-10i   & 2.67-0.15i  & -5.4-0.53i &-2.02-0.70i &  4.6-3.9i  & -0.04-0.53i  &  --  \\
\end{tabular}
\caption{P waves \={N}N low energy parameters (in fm$^3$) for the considered optical models:   J\"{u}lich results are taken from Tab 3 of Ref. \cite{Haidenbauer_JHEP_2017}, 
KW and DR2 from \cite{CRW_ZPA343_1992}, Paris 2009 have been recomputed and are in agreement with \cite{Benoit_Private}.
The values of Nijmegen are obtained by extrapolating the phase shifts from Figures \ref{delta_1S0} and \ref{delta_3SD1}.} \label{Table_LEP_P}
\end{table*}

\bigskip
The low energy parameters ($a_L$ and $r_L$) of the examined partial waves are collected in Table  \ref{Table_LEP_S}  for S-waves and  Table \ref{Table_LEP_P} for P-waves. 
In view of these results some general remarks can be drawn. They are in order:
\begin{enumerate}
\item Despite the huge differences in the phase shifts  described in this and the previous sections, there is
a remarkable stability in the ''qualitative'' zero-energy  predictions, mainly the scattering lengths and volumes.
We mean by that their ''repulsive'' or ''attractive character", more precisely the sign of their real part, the almost vanishing 
imaginary part of $^{13}$P$_1$  and $^{33}$P$_0$ states, or the relatively small values of the $^3$PF$_2$.
This is specially true if one take into account that none of these models
has been adjusted in order to reproduce the zero energy protonium results and that the potential themselves are very different.

For the S-waves, only the $^{11}$S$_0$ result of Paris and DR2 potentials have different sign.

For the P-waves, the only exception in this general qualitative agreement is the $^{33}$P$_1$ state, again due to Paris potential
which has a different sign.
As we have already noticed, the reason for this difference is in both cases related to a near threshold quasi-bound state in the corresponding PW.

Furthermore, in most of the sates, this agreement is not only qualitative but there is a reasonable quantitative agreement (between 10-20\% with respect their averaged) in the numerical values,
except in some particular states and models that we will detail below.

\item  A possible explanation for this astonishing stability in the real parts could be the  one-pion exchange dominance, as it
was suggested by Ericson and Weise for the NN case (see Sect  3.8 of  Ref. \cite{Ericson_Weise_1988}).
Indeed,  for a tensor uncoupled state, the integral form for the scattering ''length'' can be written as
\begin{equation}\label{aL} 
a_L=   \lim_{q\to0}\;\;   {1\over q^{2L+2 }} \;   \int_0^{\infty} dr \;\hat{j}_{L}(qr) \; v(r) \; u_{L}(r) 
 \end{equation}
where $v={m\over\hbar^2}V$ is the corresponding potential, $\hat{j}_{L}$ the reduce regular spherical Bessel and $u_{L} $  the reduced radial solution that behaves asymptotically as
\[ u_{L}(r) = \hat{j}_{L}(qr)  +\tan\delta_L \;\hat{n}_{L}(qr)  \]
According to these authors, a good approximation of $a_L$ (for L$>$0) is provided  by the Born approximation of the one-pion potential tail $v_{\pi}$,
that is:
\begin{equation}\label{aLB} 
 a^B_L(\pi)= \lim_{q\to0}\;\;   {1\over q^{2L+2 }} \;   \int_0^{\infty} dr \; \mid \hat{j}_{L}(r) \mid^2 \; v_{\pi}(r) 
\end{equation}
By  inserting  the one-pion potential
\begin{equation}\label{Vpi} 
V_{\pi}(x)= c_{\pi} \left[  \sigma\cdot\sigma + S_{12} \chi_T(x) \right]  \;Y(x)\; \tau\cdot\tau  
\end{equation}
with $x= {m_{\pi}  r\over \hbar}$,
\[ c_{\pi}={m_{\pi}\over3} {g^2\over4\pi}\left({m_{\pi}\over 2M}\right)^2,\] 
\[ Y(x)= {e^{-x}\over x}, \]
and
\[  \chi_T(x)= 1+ {3\over x} + {3\over x^2}  \quad \]
into eq. (\ref{aLB}) one gets:
\begin{small}
\begin{eqnarray*}\label{aLB_pi} 
&&a^B_L(\pi) = {c_{\pi} \over (2L+1)!!^2} \; \left({ M\over\hbar^2}\right)  \; \left({\hbar\over m_{\pi}}\right)^{2L+3}  (\tau\cdot\tau) \cr
&&\bigl\{ (\sigma\cdot\sigma)  (2L\!+\!1)! + S_{12} \bigl[ (2L\!+\!1)!  + 3 [(2L) ! + (2L\!\!-\!\!1)!] \bigr]  \bigr\}  
\end{eqnarray*}   
\end{small}  

The first remark about the later expression  is the "$\tau\cdot\tau$ rule", i.e the fact that the ratio of the scattering lengths of two isospin components
of the same  PW is given by the value of  $\tau\cdot\tau$ operator:  $\tau\cdot\tau$=-3 for T=0 and $\tau\cdot\tau$=1 for T=1.
Indeed the real part of the scattering volumes displayed in Table \ref{Table_LEP_P} (uncoupled states)  roughly fulfil this requirement. 
There are two exceptions: the results of Nijmegen-J\"{u}lich $^3$P$_0$ (in relative sizes) and the Paris $^3$P$_1$ (in relative sizes and signs).

By using the numerical values $m_{\pi}$=138.039 MeV, M=938.9183 MeV  (averaged  pion and N masses) and $g^2/4\pi$=14.4, 
one obtains for the \={N}N states the results displayed in Table \ref{Table_LEP_VpiB}.
The $^3$P$_2$ results is decoupled from the $^3$F$_2$ tensor partner.
With the recommended NPW $\pi NN$ coupling constant $g^2/4\pi$=13.9, a reduction factor  0.965 must be used.

\begin{table}[htbp]
\begin{center}
\begin{tabular}{ l    rr                                    rr                    } \\
                        &$\sigma\cdot\sigma$&$\tau\cdot\tau$&$S_{12}$& Re[$a_L$] \\    \hline     
%
$^{11}$P$_1$ &  -3                             & -3                   &  0            & -3.09         \cr
$^{31}$P$_1$ & -3                              &  1                   &  0            &  1.03         \\  \hline  
$^{13}$P$_0$ & 1                               & -3                   & -4            & -9.27       	 \cr	
$^{33}$P$_0$ & 1                               &  1                   & -4            &  3.09           \\  \hline  
$^{13}$P$_1$ & 1                               & -3                   & 2             &  6.18          \\   
$^{33}$P$_1$ & 1                               &  1                   &  2            & -2.06        \\ \hline  
$^{13}$P$_2$ & 1                               & -3                   & -2/5         &  0            \\
$^{33}$P$_2$ & 1                               &  1                   & -2/5         &  0             \\
\end{tabular}
\caption{ \={N}N  scattering volumess (fm$^3$) as predicted by the pion dominance from \cite{Ericson_Weise_1988}}\label{Table_LEP_VpiB}
\end{center}
\end{table}
For $^1$P$_1$ state, the Born pion values are close (15\%) to the  full results from Table \ref{Table_LEP_P}, except for KW where the difference is twice larger.
 
For $^{33}$P$_0$, the differences are of the same order. Only the $^{13}$P$_0$ J\"{u}lich  result shows a large discrepancy in the $\tau\cdot\tau$  rule.

For $^{33}$P$_1$, the agreement is even better, except for the instructive Paris result which differs substantially.
Indeed the "one-pion exchange dominance"  is based on the assumption that the scattering solution $u_L$ is close to the free wave $\hat j_L$ in the
dominant part of the integral (\ref{aL}). 
In case of the existence of a bound state, as in the Paris model, $u_L$ has a node and change its sign with respect to the free solution.

For $^{33}$PF$_2$, the Born results from eq. (\ref{aLB})  cannot directly be applied since they
were stablished for tensor uncoupled states. However for the single $^3$P$_2$ state they predict a vanishing Re(a) and
the small value of the full results  from Table \ref{Table_LEP_P} could be a trace of this compensation.   

To close this remark, we would like to mention that while the "one-pion exchange dominance"  is justified in the NN case, where it was stablished,   
it has an uncertain applicability in the \={N}N physics. Apart
from disregarding the annihilation physics, this approach will fail in presence of one or  several bound or resonant states, as
it is the case in most of $V_{\bar NN}$ models.
We have seen an illustrative example in the $^{33}$P$_1$ case with Paris results having different sign. However
the same breakdown of the "one-pion exchange dominance" can happen if there are two bound states, although keeping the same sign. 
This can be the case of the $^{13}$P$_1$ state with KW or the $^{13}$P$_0$ with J\"{u}lich where the  $\tau\cdot\tau$-rule is badly violated.

\item The imaginary part of  S-waves is also remarkably stable within quite narrow limits  Im[a($^{11}$S$_0$)]=1.18$\pm$ 0.17 fm,
Im[a($^{31}$S$_0$)]=0.60$\pm$0.03 fm and\par
Im[a($^{13}$SD$_1$)]=0.82$\pm0.05$ fm. 
Only  the $^{33}$SD$_1$  state presents some dispersion  essentially due to Paris result, with Im[a($^{33}$SD$_1$)]=0.73$\pm0.30$ fm. 

The imaginary part of P-waves is much less stable, although some common features are shared  like the small values for the $^{13}$P$_1$, due to its  repulsive character.

\item Of particular interest is  the $^{13}$P$_0$ state, with a very large real part  $\sim 9$ fm$^3$ shared by
DR, KW and Paris models (J\"{u}lich results are 3 times smaller), and confirmed by protonium data.
This large and negative value was attributed in \cite{CDPS_NPA535_91}  to the existence of a near-threshold state.
However it finds also a "natural" explanation in terms of the ''pion dominance'' in Table \ref{Table_LEP_VpiB}, which predicts Re(a)=-9.3 fm$^3$. 
This, at first glance, puzzling situation can be reconciled if one takes into account the result of Ref. \cite{NN3P0_chiral_PRC1997}, according to which, in the chiral  limit ($m_{\pi}$=0),
the NN $^3$P$_0$ state (so T=1) has a zero energy virtual state, with a diverging scattering volume.
The existence of such a NN bound state, as well as some related consequences in nuclear matter, is prevented by the short range NN repulsion, which is however absent in the \={N}N case
and leave open such a possibility.

\item The most stable states are those with a repulsive potential.
These are the $^{13}$P$_1$  and $^{33}$P$_0$.
They have in common a small imaginary part both in $a_1$ and $r_1$ since they are little sensitive to annihilation dynamics.\par
For $^{13}$P$_1$, all models agree with a real part of  $4.9\pm0.3$  fm$^3$ and an imaginary part smaller than 0.1 fm$^3$
For $^{33}$P$_0$, they all  agree  with a real scattering volume  $2.5 \pm 0.25$ fm$^3$ and  a small imaginary part $\sim0.1$ fm$^3$. 
The Paris potential is a particular case:  $V_{^{33}P_1}$ is attractive with a quasi-bound state previously discussed.
However $V_{^{13}P_1}$ is even more
attractive than the former but has a positive Re[$a_{^{13}P_1}$] as for the repulsive models.
This suggest the existence of a second bound state for the $^{13}$P$_1$ state which plays the role of an effective repulsion.

\item Concerning the effective range values $r_L$  there is no any trace of stability in the model predictions, which translate
the fact that beyond the zero energy region the examined \={N}N optical models display larger differences.
\end{enumerate}

\begin{table*}[htbp]
\begin{tabular}{| l l |     c                                      |               c         |     c                  |      c              | c  |}             \hline
state     &       &  Exp                          &  Paris 2009       & J\"{u}lich    & KW            & DR2           \\ \hline
$^1$S$_0$ &\={N}N &                               & $\;$1.02 - i 0.87 & 0.42 - i 0.91 & 0.52 - i 0.99 & 0.65 - i 0.82 \\
          &\={p}p &$\;\;$0.493(92) - i 0.732(146) & $\;$0.92 - i 0.67 & 0.50 - i 0.71 & 0.57 - i 0.77 & 0.68 - i 0.64 \\
$^3$SD$_1$&\={N}N &                               & $\;$0.91 - i 0.62 & 0.93 - i 0.92 & 1.01 - i 0.79 & 1.09 - i 0.75 \\
          &\={p}p &    0.933(45) - i 0.604(51)    & $\;$0.82 - i 0.50 & 0.90 - i 0.74 & 0.92 - i 0.63 & 0.98 - i 0.59 \\
S-averaged&\={N}N &                               & $\;$0.94 - i 0.68 & 0.80 - i 0.92 & 0.89 - i 0.84 & 0.98 - i 0.77 \\
          &\={p}p &   0.823(57) - i 0.636(75)     & $\;$0.85 - i 0.54 & 0.80 - i 0.74 & 0.83 - i 0.67 & 0.90 - i 0.60 \\ \hline
$^3$P$_0$ &\={N}N &                               &-3.02 - i 2.50     &-0.32 - i 4.01 &-3.20 - i 2.28 &-2.93 - i 1.83 \\ 
          &\={p}p &$\!\!$ -5.68(123) - i 2.45 (49)&-2.74 - i 2.46     &-0.32 - i 3.85 &-2.81 - i 1.99 &-2.53 - i 1.62 \\ \hline
\end{tabular}
\caption{Isospin averaged ($a_{\bar NN}$)  and  \={p}p scattering lengths 
are compared with those obtained from hydrogen atom level shifts and widths, in fm  for  S  and fm$^3$ for P states. 
The \={p}p values including Coulomb and $\Delta m$ corrections are taken from \cite{CRW_ZPA343_1992} for DR2 and KW, from \cite{PARIS_PRC79_2009}
for Paris and from \cite{Haidenbauer_JHEP_2017} for J\"{u}lich model.
The statistical averaged value for  S-wave  is defined as ($^1$S$_0$+3~$^3$S$_1$)/4 and is given with averaged errors.}\label{Hydro}
\end{table*}

\subsection{Hydrogen atoms}

The measurement of level shifts and widths in Hydrogen atoms is an alternative way to access the \={p}p scattering lengths and volumes. 
A formula derived by Trueman  \cite{Trueman_NP26_1961} finds a connection between 
the protonium complex level shifts and the  Coulomb corrected \={p}p scattering lengths. 
In the case of antiprotonic hydrogen, due to large Bohr radius ($B\approx$ 57 fm), this relation is essentially linear \cite{CRW_ZPA343_1992}. 

The \={p}p scattering lengths are obtained by coupling both T components by Coulomb and $\Delta m$ corrections.
One obtains however a reasonable approximation, denoted $a_{\bar N N}$ 
to distinguish it from the exact value $a_{\bar p p}$, by neglecting this coupling and isospin-averaging the results of Tables \ref{Table_LEP_S} and \ref{Table_LEP_P}, i.e:
\begin{equation}\label{aNNB}
2\; a_{\bar N N}=   a_{T=0} + a_{T=1} 
\end{equation}
Table \ref{Hydro} shows the comparison between the computed values -- $a_{\bar NN}$ and $a_{\bar pp}$ -- 
and those  extracted from the atomic measurements \cite{Gotta_Balmer} via the Trueman relation.
Notice that the inclusion of Coulomb and $\Delta m$ can represent up to a $\approx 30\%$ difference between $a_{\bar NN}$ and $a_{\bar pp}$ values.

For S-waves the discrepancies existing in Table \ref{Table_LEP_S} within the different models, mainly concerning the $^{11}$S$_0$ state,
are smeared out in the T and S-averaged value, which is found to be in a nice agreement among them and with the experimental value.
A remarkably good  agreement is also observed in the, non trivial, $^3$SD$_1$ state.
The major problem to improve the situation for S-waves is the $^{11}$S$_0$ near-threshold quasi-bound state, present in Paris and probably DR2 models but absent in KW and  
J\"{u}lich ones. 
It results into a negative value of the corresponding scattering length and that generates a factor 2 in the real parts.

For P-waves, little is known experimentally. 
The measurement of the isolated $^3$P$_0$ \={p}p level shift \cite{Gotta_Review_PPNP_2004} 
seems to confirm the large value of the $^{13}$P$_0$ scattering volume displayed in Table \ref{Table_LEP_P},
predicted by Paris, KW and DR2 models.
In fact the large value of Re[$^{13}$P$_0$]$\approx$-9 fm$^3$ that they predict, and that is averaged with a positive Re[$^{33}$P$_0$]$\approx$2.5 fm$^3$, 
is still underestimated for reproducing the experimental result. One would rather need Re[$^{13}$P$_0$]$\approx$ -13 fm$^3$.
This is clearly in tension with the J\"{u}lich prediction which is one order of magnitude smaller than the  other models and  the experimental value.

Since the large values of Re[$a_{13P0}$] are predicted by the "pion dominance" described in the previous section,
one could find an explanation of this disagreement in the particular form of the one-pion potential (eq. 2.1 of \cite{Haidenbauer_JHEP_2017}),
which includes a non-local relativistic correction and results into a  smaller effective $\pi NN$ coupling constant.

As polarisation experiments are missing, the atoms offer a unique possibility to study the spin structure of interactions. 
Again, there are sizeable differences between the models. 
Unfortunately these happen  also in the absorptive parts which are vital for the PUMA experiment. 
These should be resolved on the side of theory and more important on the side of experiments.  
The priority, we believe, should be given to the full resolution of the atomic fine structure in Hydrogen and Deuterium.  
In particular the $2P$ state in Hydrogen displays a  clear $^3$P$_0$ state, 
 indicated in Table \ref{Hydro} and three other states lumped together and difficult to resolve. 
 See  Ref \cite{Gotta_Review_PPNP_2004} for a dedicated review.
 An improvement of this  resolution would be extremely helpful to eliminate the model  differences in the partial waves. 

\section{Conclusion}\label{Conclusion}

We have compared the  strong \={N}N phase shifts
obtained in the Nijmegen Partial Wave Analysis  \cite{ZT_NNB_Nijm_PW_2012}, 
used to construct  the chiral EFT J\"{u}lich  optical potential at N3LO \cite{Haidenbauer_JHEP_2017}, 
with some of the currently used \={N}N optical models in configuration space: Dover-Richard (DR2) \cite{DR1_PRC21_1980,DR2_RS_PLB110_1982}, 
Kohno-Weise \cite{KW_NPA454_1985}  and  Paris (updated version from 2009) \cite{PARIS_PRC79_2009}.
For all these models we have computed the strong phase shifts and extracted the low energy parameters (scattering lengths and effective ranges).
The corresponding potentials are included in the Appendix.
This comparison is limited to the S and P waves.

In spite of  providing very close elastic, annihilation and charge-exchange integrated cross sections (Figure \ref{sigmas}),
these models are not phase-equivalent:  
 large and systematic differences have been observed in almost all the partial waves, among them and with respect to the NPWA.

In the low energy region one observes some stability in the scattering lengths and volumes
(Tables \ref{Table_LEP_S} and \ref{Table_LEP_P}),
in particular the "repulsive" or "attractive" character, i.e.  the sign of Re[$a_L$],
which is  respected by all  models in almost all partial waves.
For P-waves this stability could be related to the "one-pion exchange dominance", that is scattering volumes 
roughly determined by the pion Born term.
It is  however also manifested in S-waves, like  the surprising stability of the low energy parameters of the tensor coupled $^{13}$SD$_1$ state. 
Exceptions are  the $^{11}$S$_0$ and  $^{33}$P$_1$ partial waves, due to the presence of a neartheshold quasi-bound state in DR2 and Paris, 
and the $^{13}$P$_0$ result of J\"{u}lich model which underestimates the protonium experimental measurements.
Despite these isolated differences, the isospin- and spin-averaged values for S-wave
are in close agreement among the models themselves as well as with the measured quantities 
(Table \ref{Hydro}).
The later concern mainly the \={p}p measurements  and so are unable to disentangle a selected isospin component.

The differences worsen when increasing the energy, as it is already manifested
with the dispersion in the effective range values and more explicitly 
in the phase-shifts (Figures  \ref{delta_1S0}, \ref{delta_3SD1}, \ref{delta_1P1}, \ref{delta_3P0}, \ref{delta_3P1} and \ref{delta_3PF2}) and zoomed in the corresponding effective range functions.
These increasing differences cannot be explained by the relativistic 
kinematics implemented in  the Nijmegen Partial Waves Analysis, relating $p_{Lab}$ to the center of mass momentum, or in the J\"{u}lich relativistic dynamical equation. 

Our main conclusion in this work is that if the Nijmegen Partial waves analysis must be considered
as a reference, as it was the case for the J\"{u}lich model \cite{Haidenbauer_JHEP_2017}, 
none of the examined optical potentials is compatible with these results and require quite a severe adjustment.
This could be easily achieved if one limits to  $p_{Lab}\leq$400 MeV/c, the main obstacle
lies in the position of the near-threshold quasi-bound states.

On the other hand, we have pointed out some anomalous behaviours of the Nijmegen Partial Waves Analysis,
also reported into  the J\"{u}lich potential.
They manifest as a resonant-like structures of the phase shifts in the $^{31}$S$_0$ and $^{33}$P$_0$ states
which takes place at the same -- relatively high -- energy and which are difficult to interpret  as having a dynamical origin, 
in particular in terms of resonant states.
Furthermore they coincide with an almost zero of the S-matrix modulus (or inelasticity parameter) 
that can introduce a bias in the analysis or a spurious change from one solution to another. 
The existence of such structures in the phase shifts constitutes one of the major differences with respect to the examined optical models.
It would be of the greatest interest to clarify this point or to better understand the underlying
dynamics of these states. It would also be interesting to decrease the lowest energy value ($p_{Lab}$=100 MeV/c) and eventually incorporate the protonium zero-energy data.
This will not provide a magic solution of the observed  discrepancies
but will clearly facilitate the agreement of the models, in particular above $p_{Lab}=$500 MeV/c.

To have at our disposal a model-independent extraction of the strong \={N}N phase shifts, the non trivial part of the interaction, is of paramount importance to the field.
In this respect it would be also suitable to have at our disposal
an independent Partial Wave analysis, as it has been always the case in the simpler NN case. 

All  the examined models roughly reproduce the experimental \={p}p elastic, inelastic and charge-exchange total cross sections, including some differential cross sections.
Unfortunately, these observables  are computed at relatively high energy, result from a coherent and incoherent sum of many partial waves and hide the existing differences among them that have been evidenced here. 

It would be of the highest interest to the community to develop, in parallel with more ambitious projects, an experimental program  
to measure the most complete set of \={N}N observables at energies $p_{Lab}<$200 MeV/c allowing to determine
the main partial waves (S,P,D) that control the low energy structure calculations. 
In this energy domain we are not only faced to a  bad "quality of data", but to a total lack of experimental results.

As far as we will not have at our disposal a reliable determination of the \={N}N strong phase shifts
for the lowest partial waves, would they be limited to  a restricted energy domain of  few tens of MeV, 
any prediction concerning  more complex systems, like those of interest in PUMA project,  could be strongly model dependent.

\vspace{1.5cm}
\section*{Acknowledgement}

We are grateful for the support of the "Espace de Structure et de r\'eactions Nucl\'eaires Th\'eorique" (ESNT, https://esnt.cea.fr) 
organizing a workshop and welcoming the visit of S. Wycech at CEA Saclay
This work was started during our visit to National Center for Nuclear Research in Warsaw.
We thank the staff members of the theory group for their warm hospitality.
We acknowledge the support of the CNRS/IN2P3 French-Polish COPIN agreement.
We are  thankful to Beno\^it Loiseau, Johann Haidenbauer and Ruprecht Machleidt for enlightening discussions and for providing us with their respective potentials.

\vspace{1.cm}
\appendix


\section{Potentials in configuration space}\label{App_Pots}

\bigskip
Although not being observable we believe it could be instructive to compare
the potentials of the different models in a given partial wave.
The J\"{u}lich model being in momentum space and non local is not  included.
The Paris potential is E-dependent and, except for the tensor-coupled states, 
we selected some positive and negative arbitrary values of $E$. 
For the NPWA,  the inner part corresponds to the square well defining the boundary conditions at $r$=1.2 fm. Beyond this value it is
continued with the one- plus two-pion (N2LO) exchange potentials.

As one can see in the following figures, the \={N}N potentials exhibit quite  dramatic differences, making even difficult
to asses wether  the \={N}N interaction in a given PW is attractive or repulsive. 
This is in sharp contrast with the NN case. 

\subsection*{$^1$S$_0$}  

\bigskip
This partial wave is globally attractive in both isospins for all models,
and much  stronger than for the NN case, specially in T=1.
However in the NPWA, there is no any need of short-range attraction in T=0.
Paris potential presents two peculiar differences with respect to the other potentials: 
the strong short range repulsion, claimed to be imposed by phenomenology,  and the repulsive peak at 1 fm,
which cannot be justified in terms of pion- or omega-exchanges since they are shared by all models.

\begin{figure}[htbp]
\begin{center}
\includegraphics[width=4.1cm]{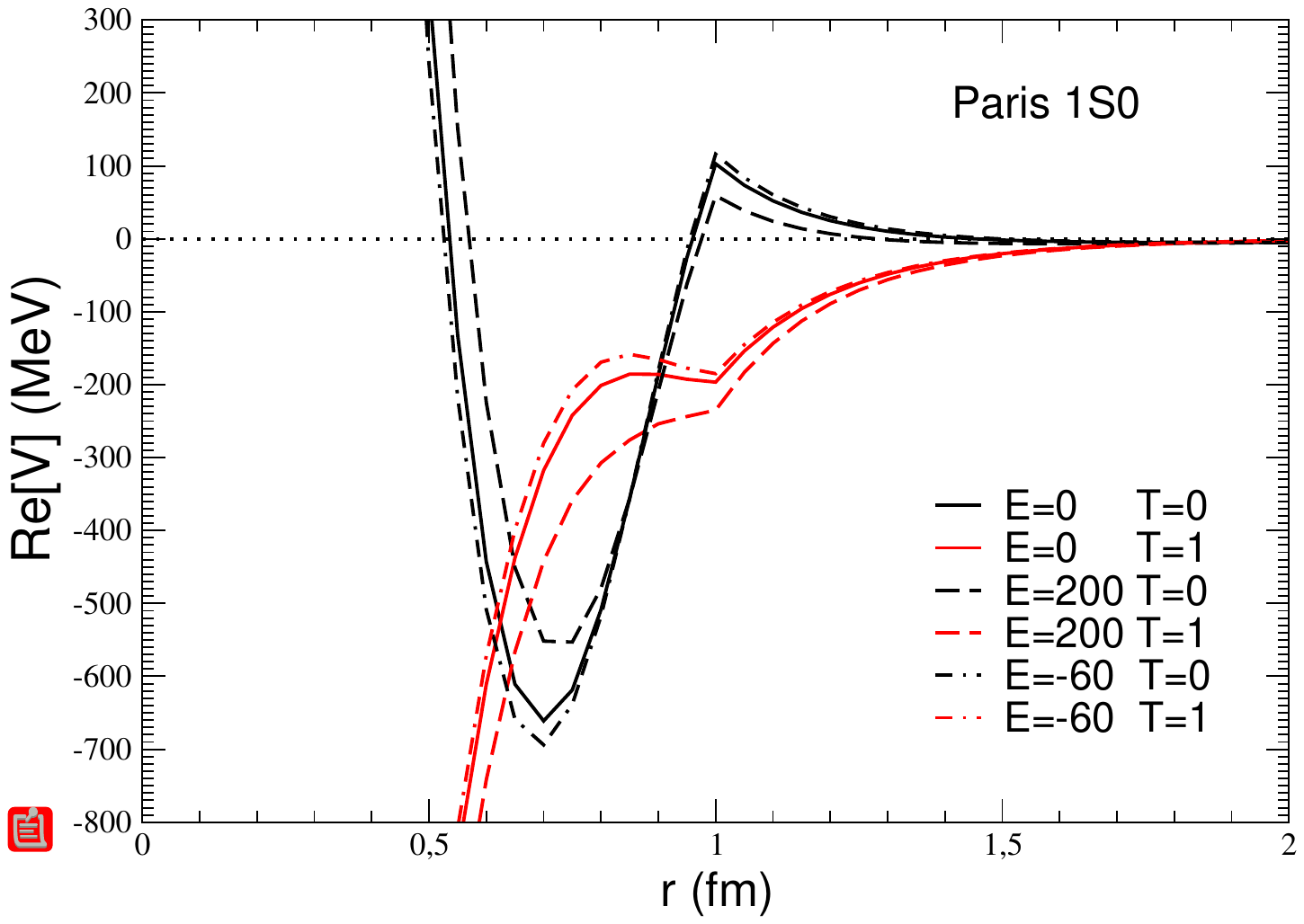}
\includegraphics[width=4.1cm]{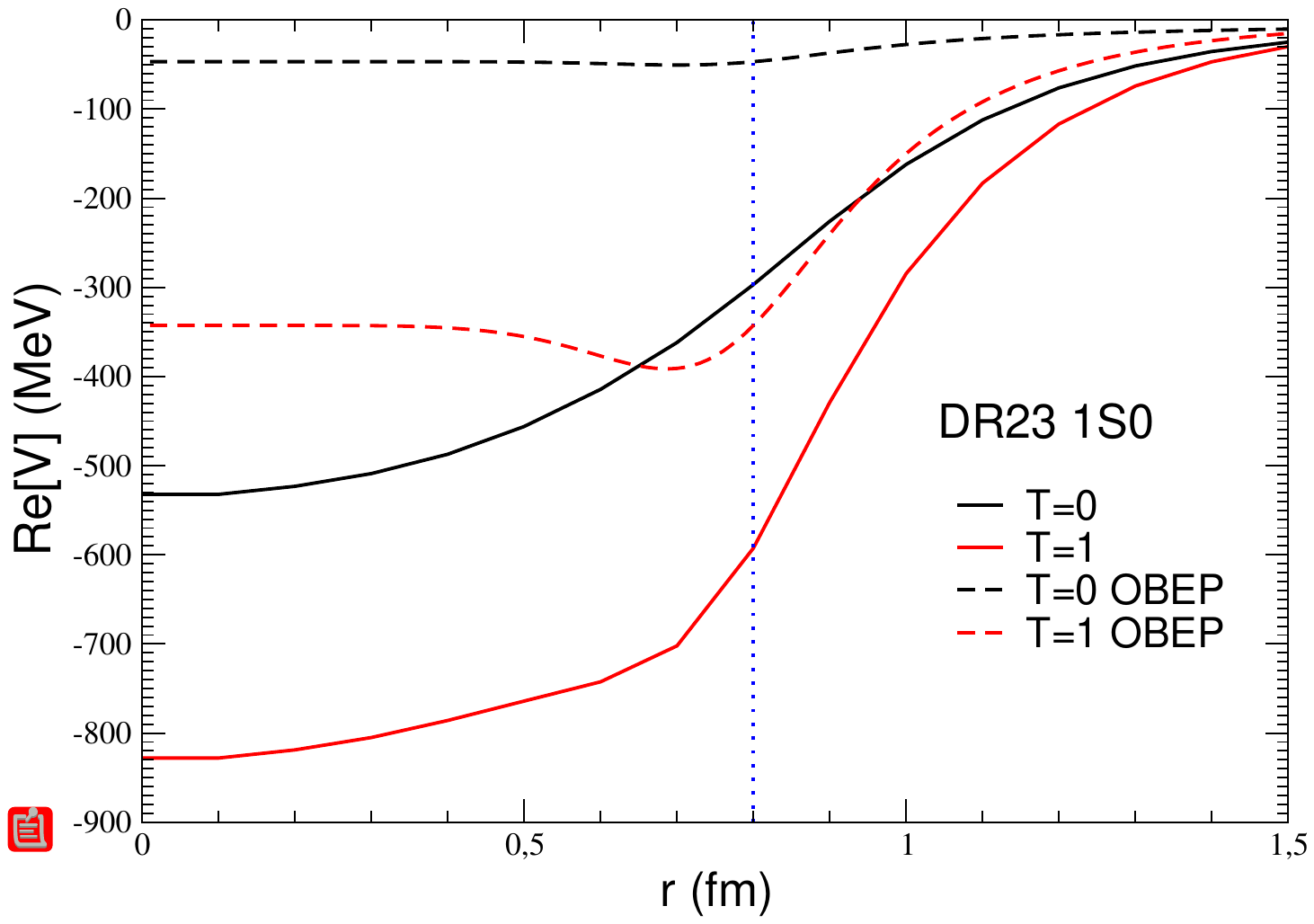}

\includegraphics[width=4.cm]{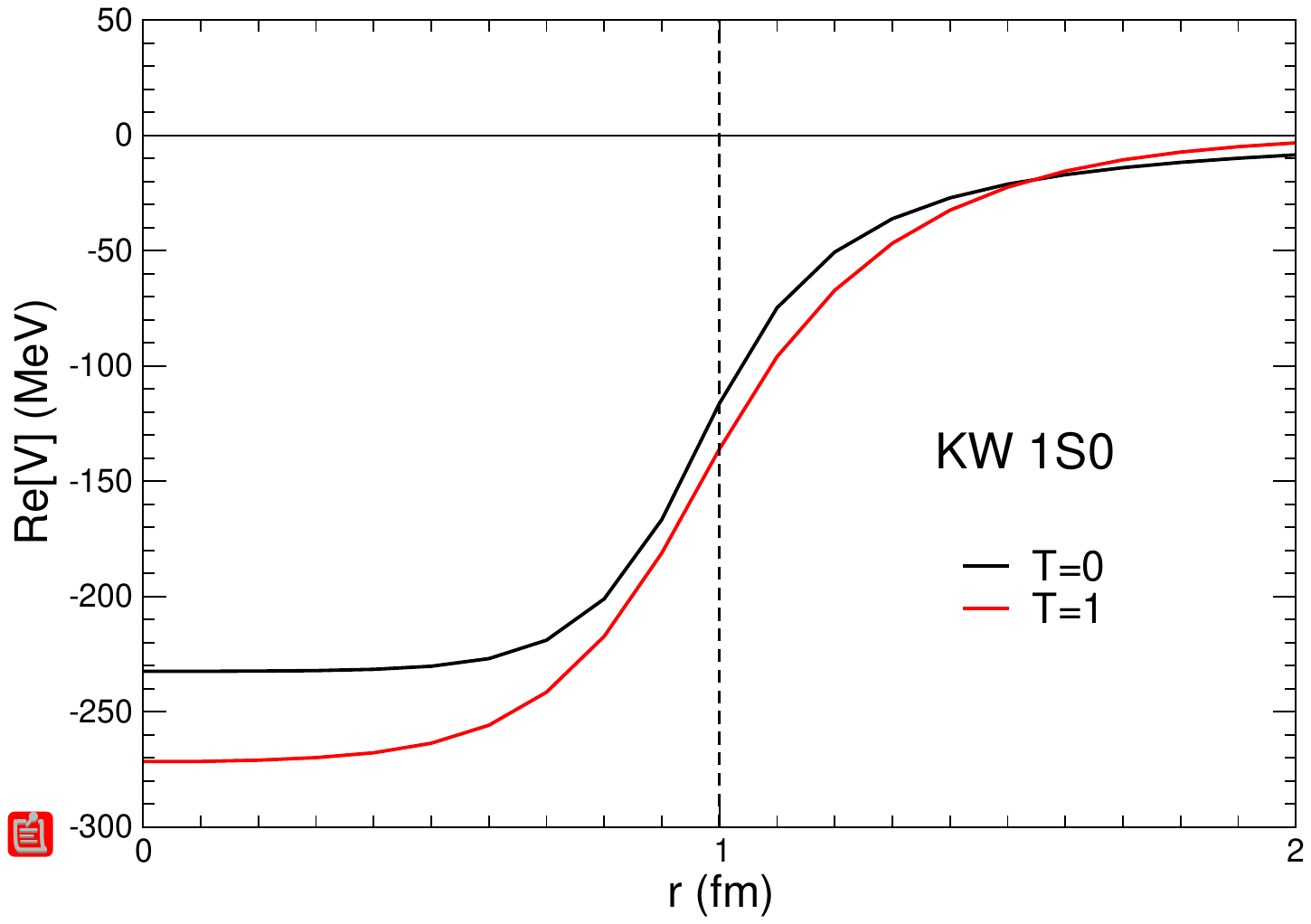}
\includegraphics[width=4.cm]{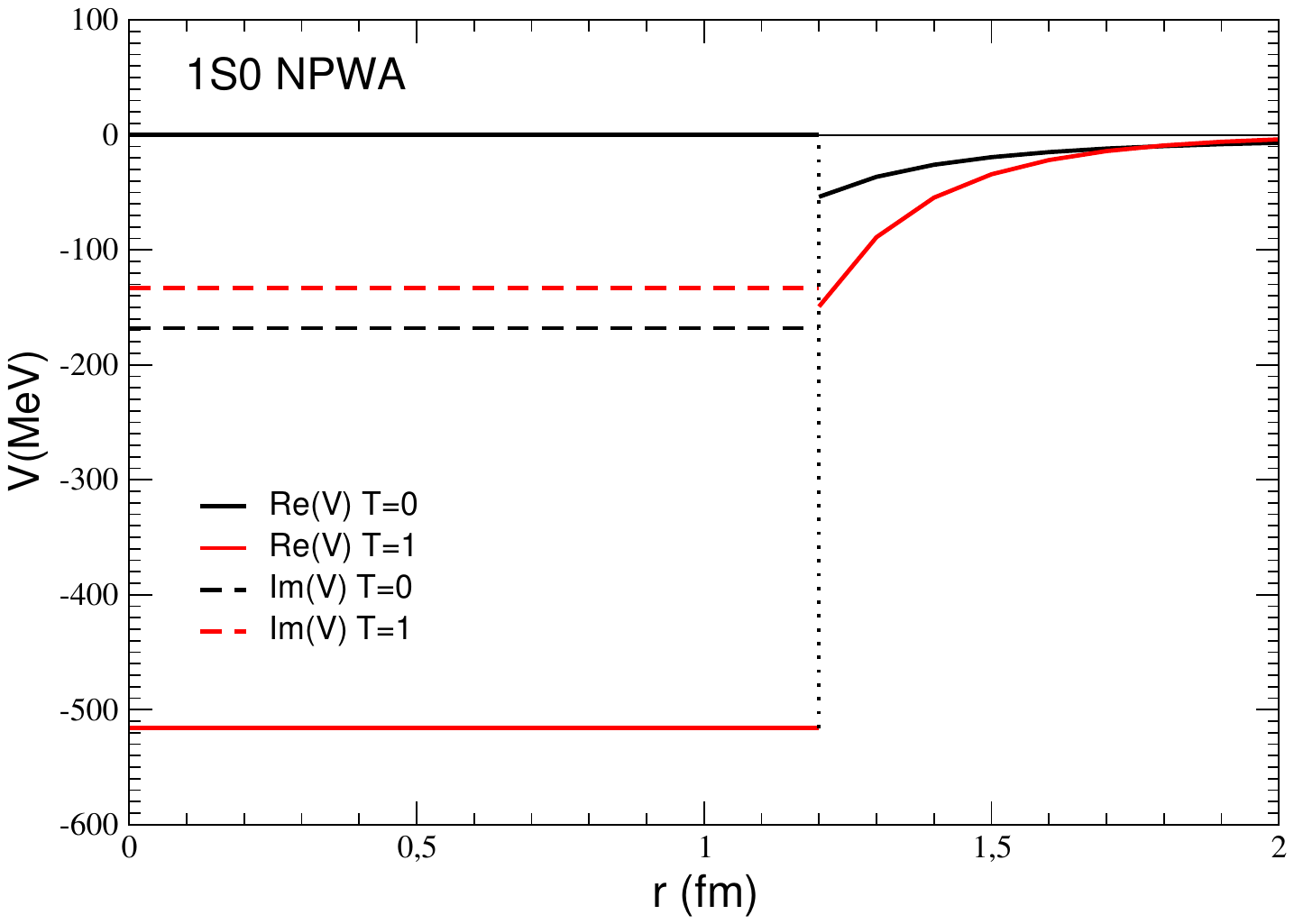}
\end{center}
\caption{Real parts of $^1$S$_0$ potentials for both isospins (T).}\label{U_1S0}
\end{figure}

\subsection*{$^3$SD$_1$}

\bigskip
The S-wave tensor-coupled state  presents also some striking differences: the $^{13}$S$_1$ potentials are strongly attractive wells, going from 500
MeV to several GeV depth, while the NPWA is limited to 130 MeV.
$V_{^{33}S_1}$ is also deeply attractive in all models but  turns to be slightly repulsive ($\approx$ 50 MeV)in the NPWA.

\begin{figure}[htbp]
\begin{center}
\includegraphics[width=4.cm]{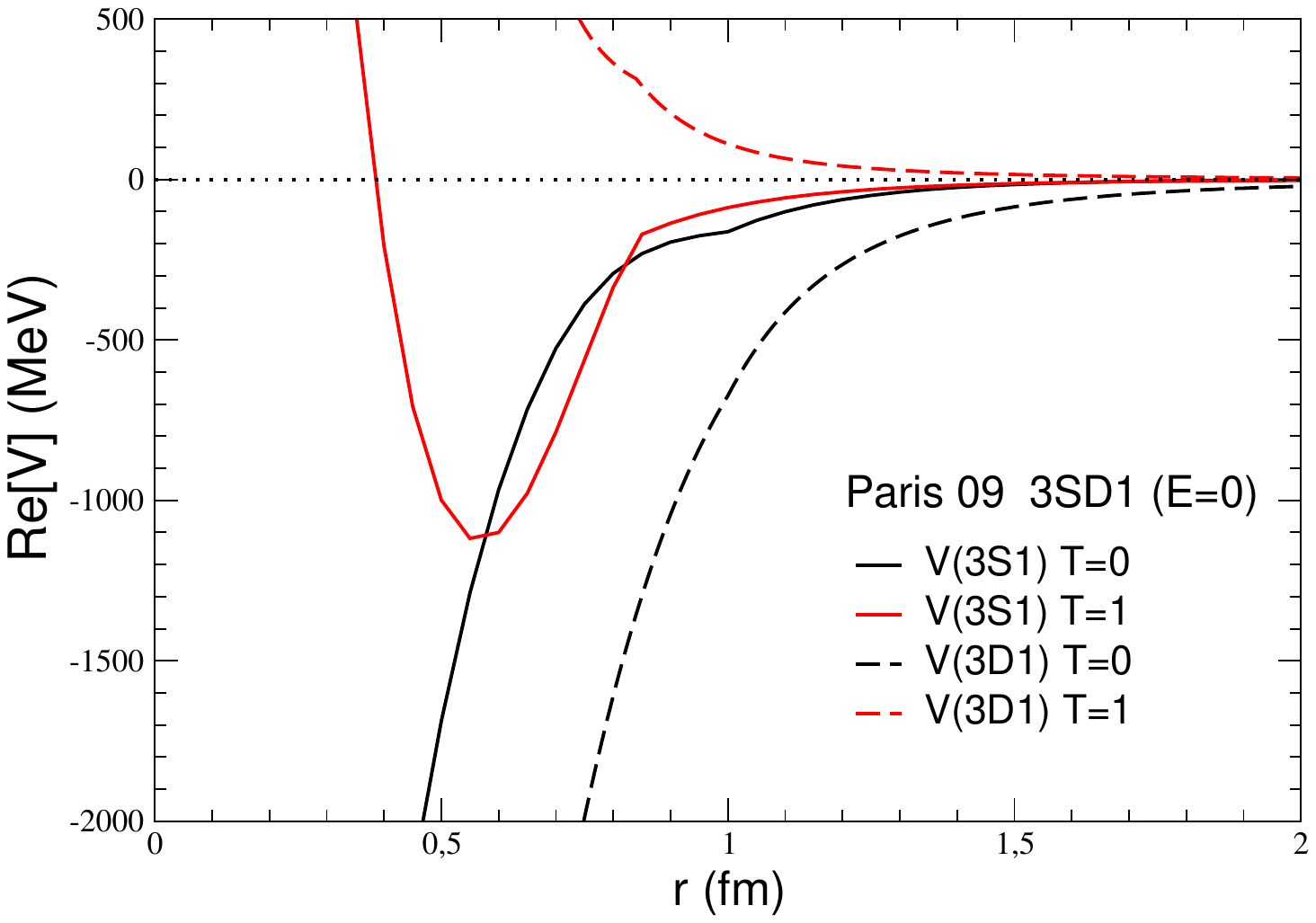}
\includegraphics[width=4.cm]{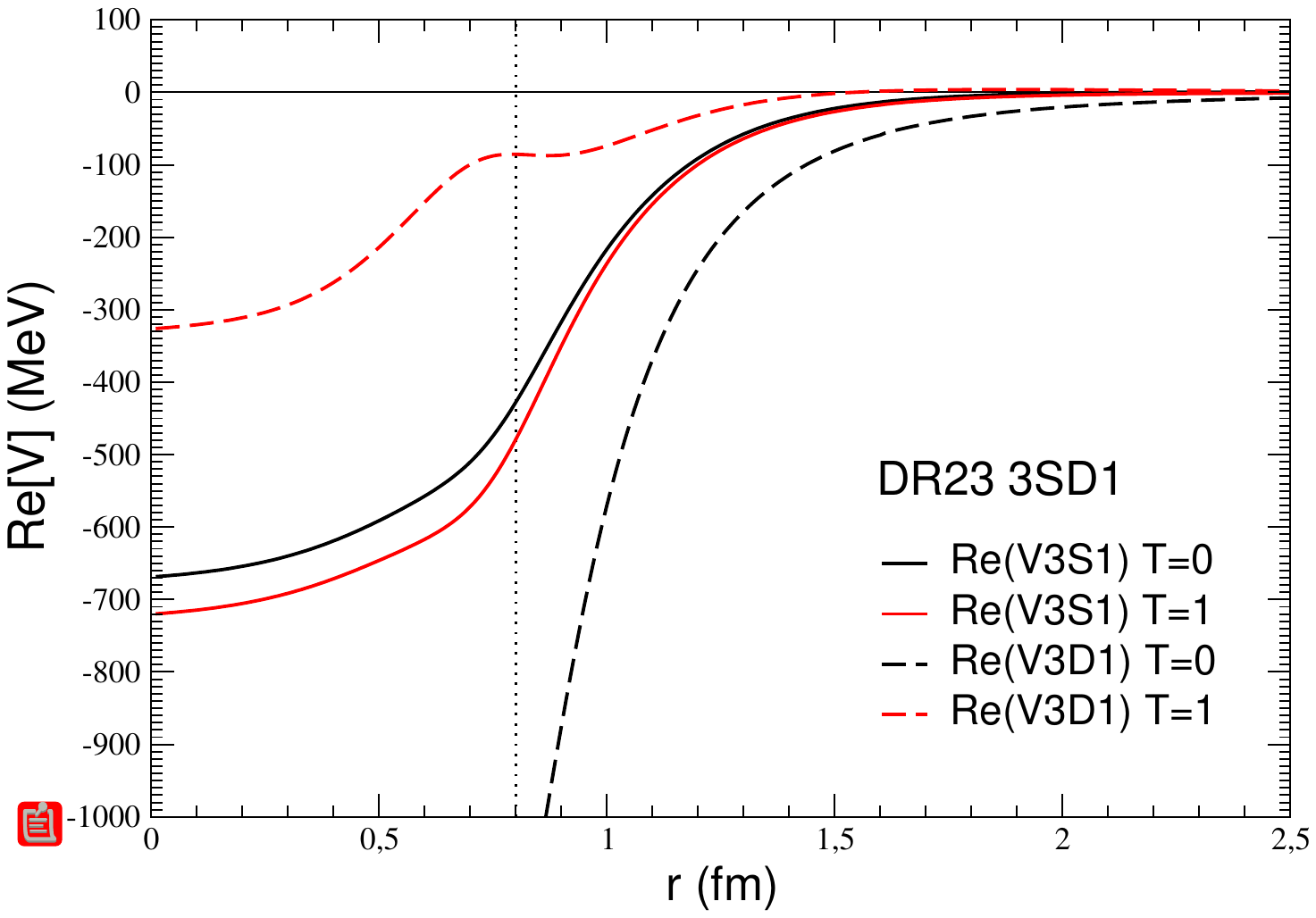}

\includegraphics[width=4.cm]{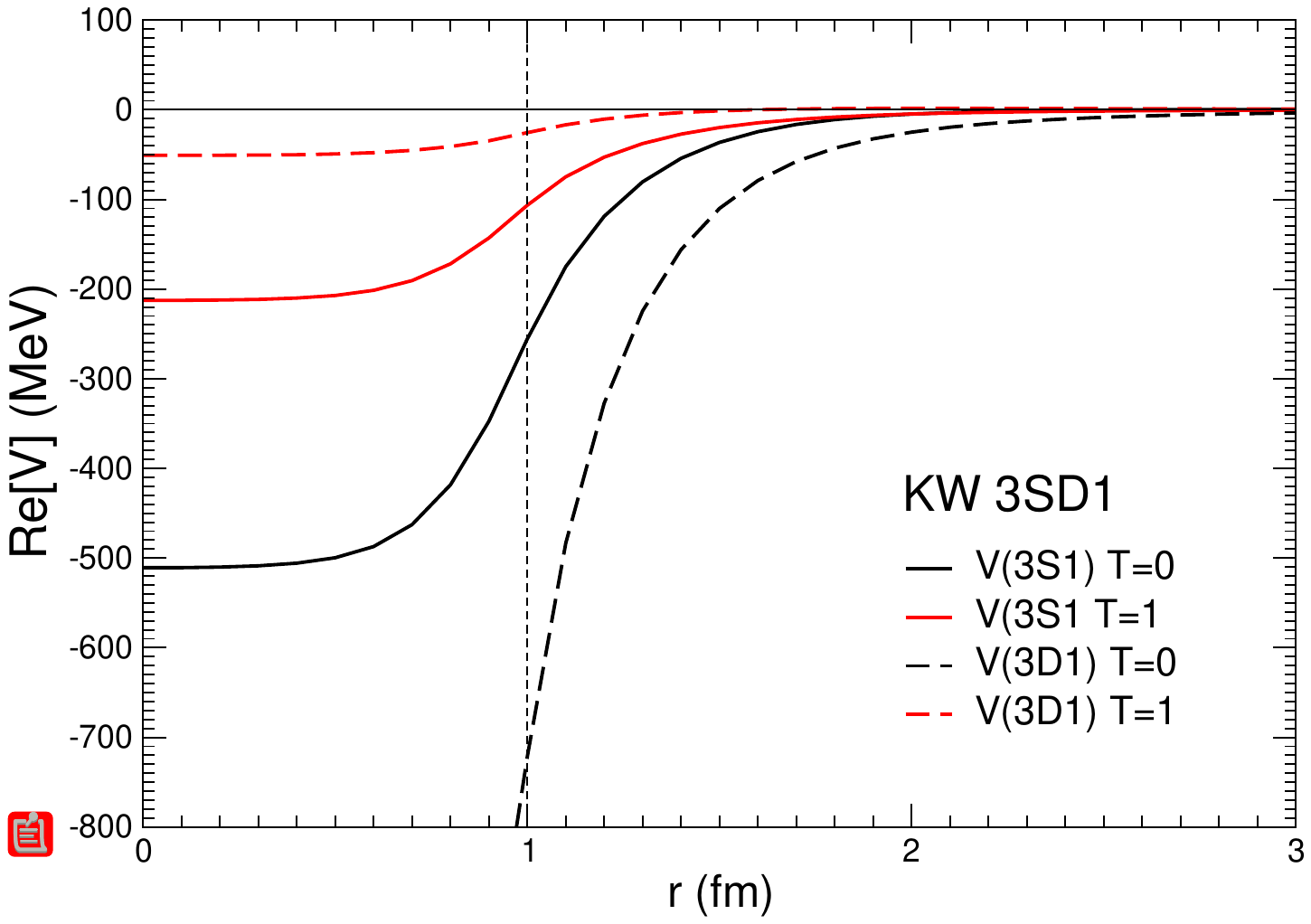}
\includegraphics[width=4.cm]{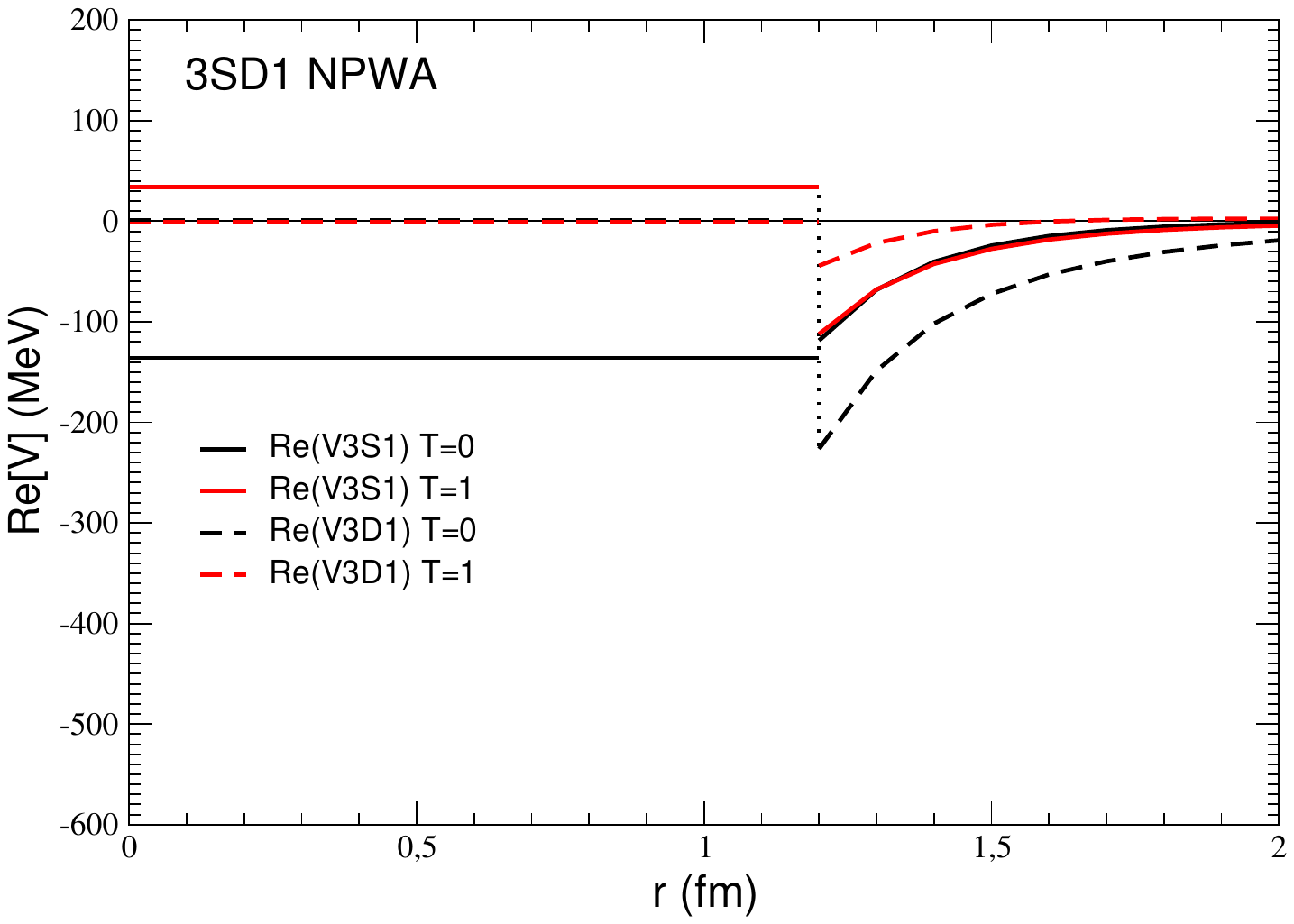}
\end{center}
\caption{Real parts of $^3$S$_1$ and $^3$D$_1$ potentials for both isospins (T).}\label{U_3SD1}
\end{figure}

The  $^3$S$_1\to ^3$D$_1$ transition potentials have in common that they are all very strong but they  display also  sizeable differences. 
Notice that in the DR and KW models the couplings dont vanish in the limit $r\to0$, what spoils the usual $r^{L+1}$ behaviour of the (reduced) radial wave functions. 

\begin{figure}[htbp]
\includegraphics[width=4.cm]{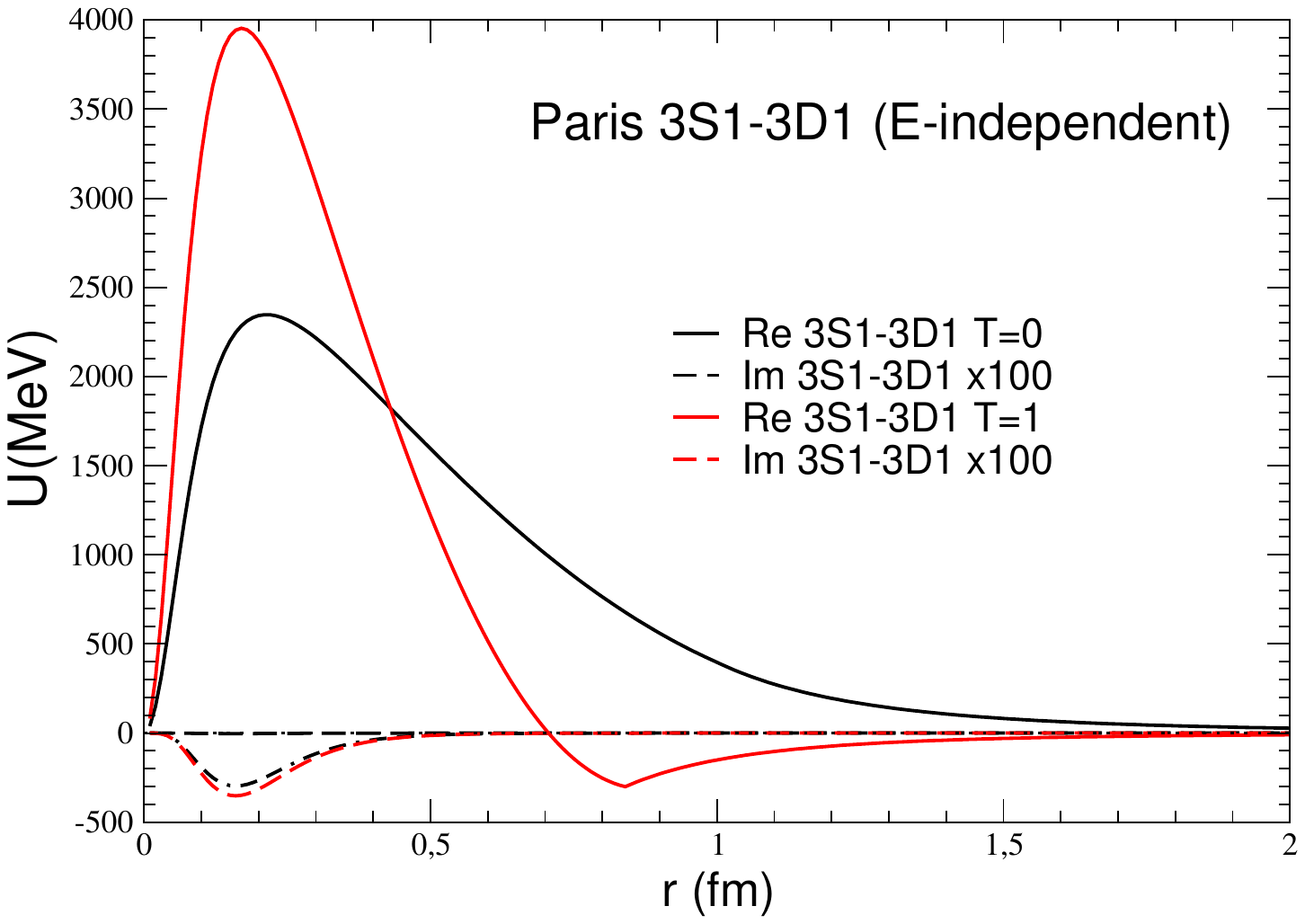}
\includegraphics[width=4.cm]{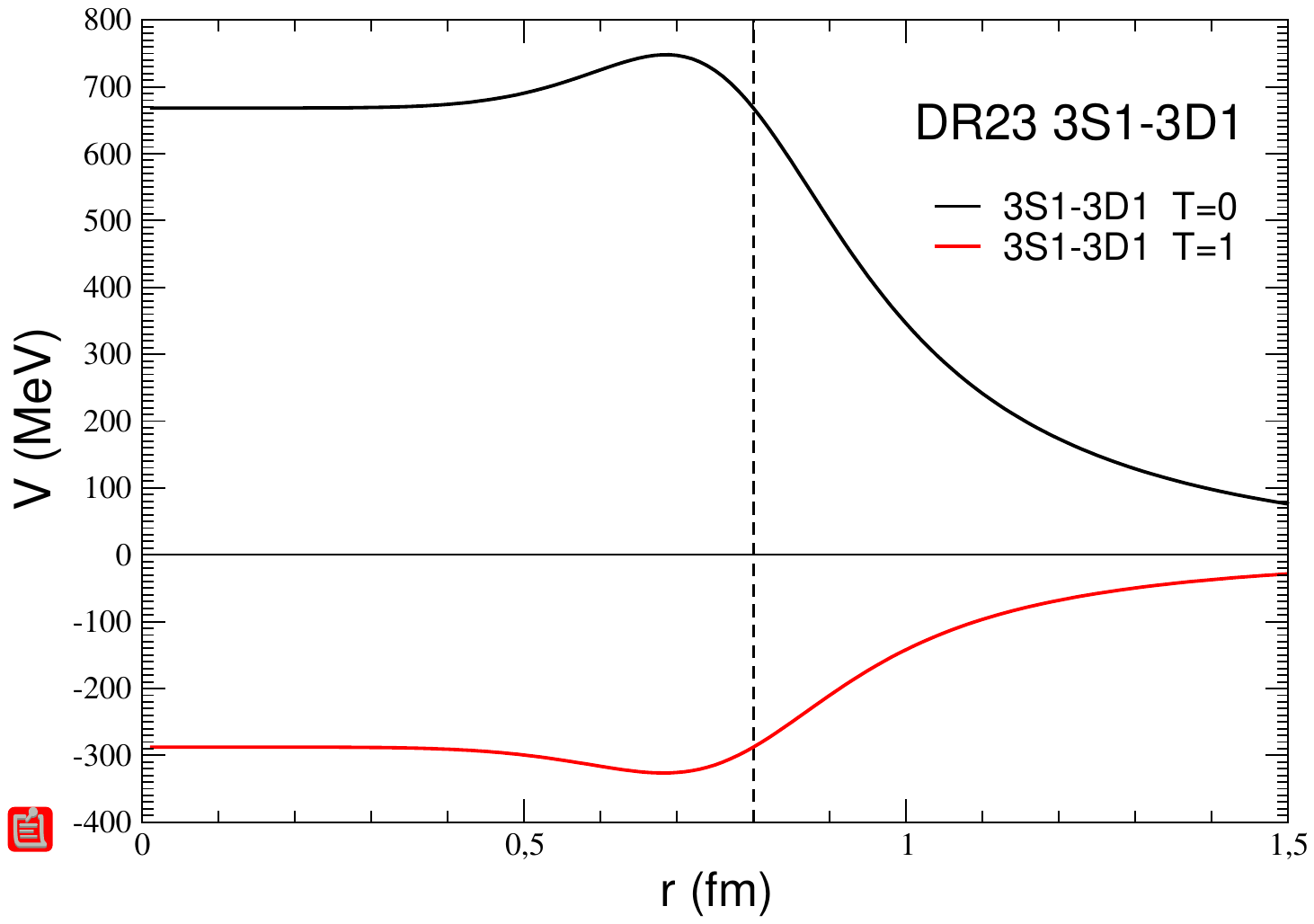}
\includegraphics[width=4.cm]{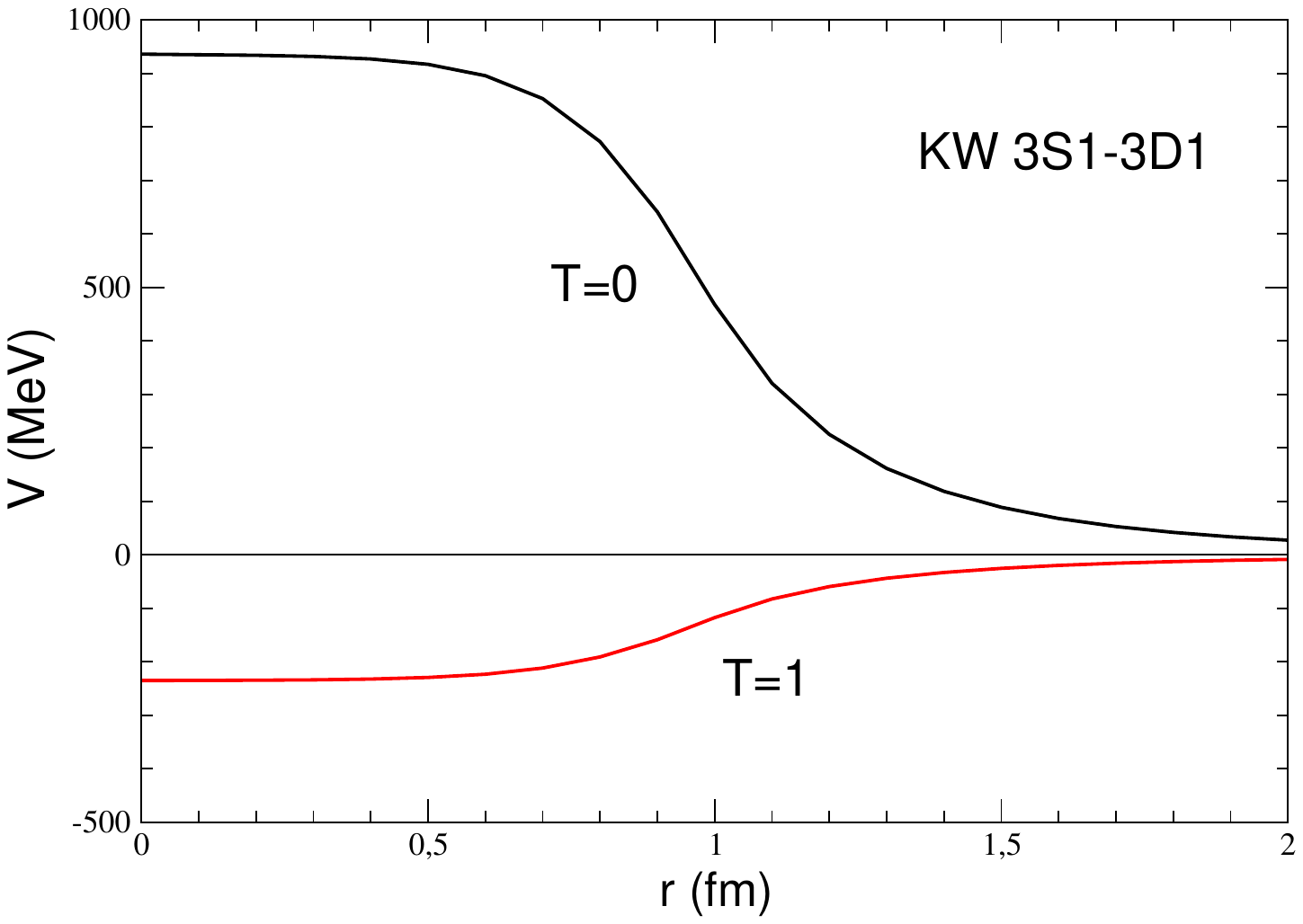}
\caption{$^3$S$_1 \to ^3$D$_1$ transition  potentials  for both isospins (T). They are real in DR and KW models.}\label{V_3S1_3D1}
\end{figure}

\vspace{-.cm}
\subsection*{$^1$P$_1$}

\bigskip
Apart form the centrifugal barrier, this potential is very close  to the $^1$S$_0$ one in all models.
Their difference is due to the, attractive, Quadratic Spin-Orbit term ($Q_{12}$), 
present in Paris and DR2 models but  absent in KW.
In the short range part of NPWA, the vanishing $^{11}$S$_0$ potential  displayed in Figure \ref{U_1S0},
vanishes also in the $^{11}$P$_1$ state, indicating that there is no any $Q_{12}$ contribution.
However the strong (500 MeV) attraction present in the $^{31}$S$_0$ state has
now totally disappeared indicating rather an unexpected repulsion.

\vspace{-0.cm}
\begin{figure}[!ht]
\begin{center}
\includegraphics[width=4.cm]{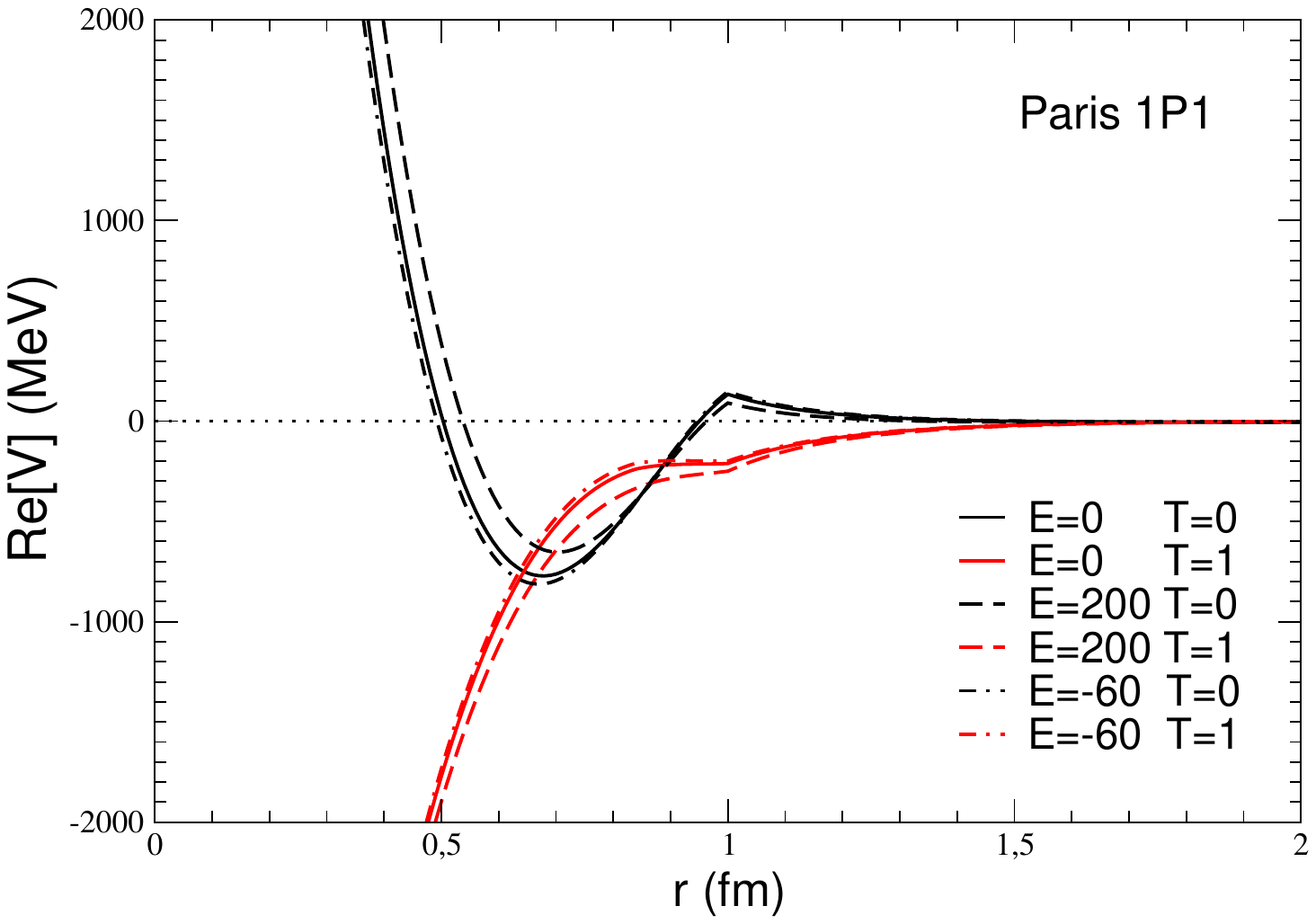}\hspace{0.cm}
\includegraphics[width=4.cm]{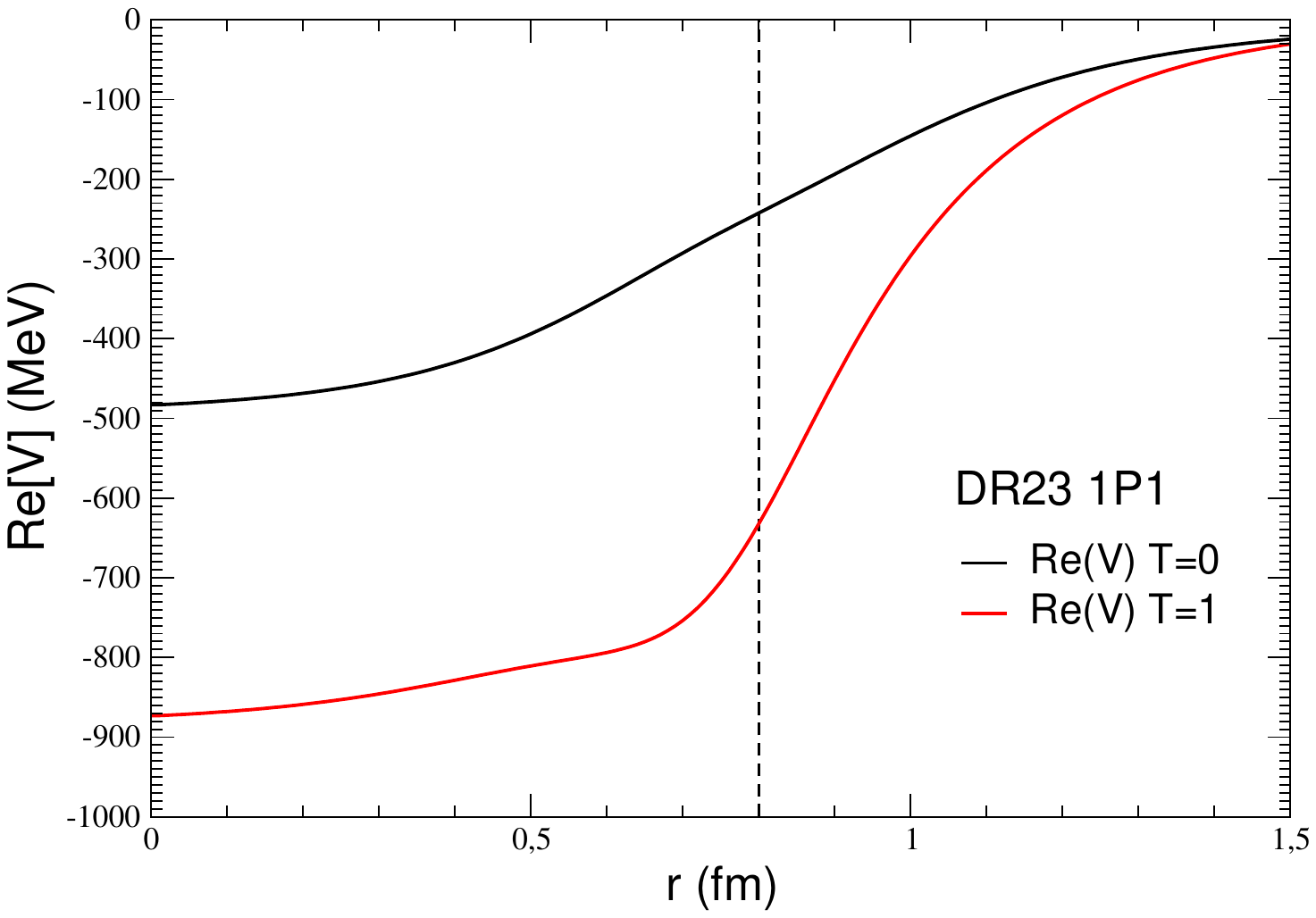}

\includegraphics[width=4.cm]{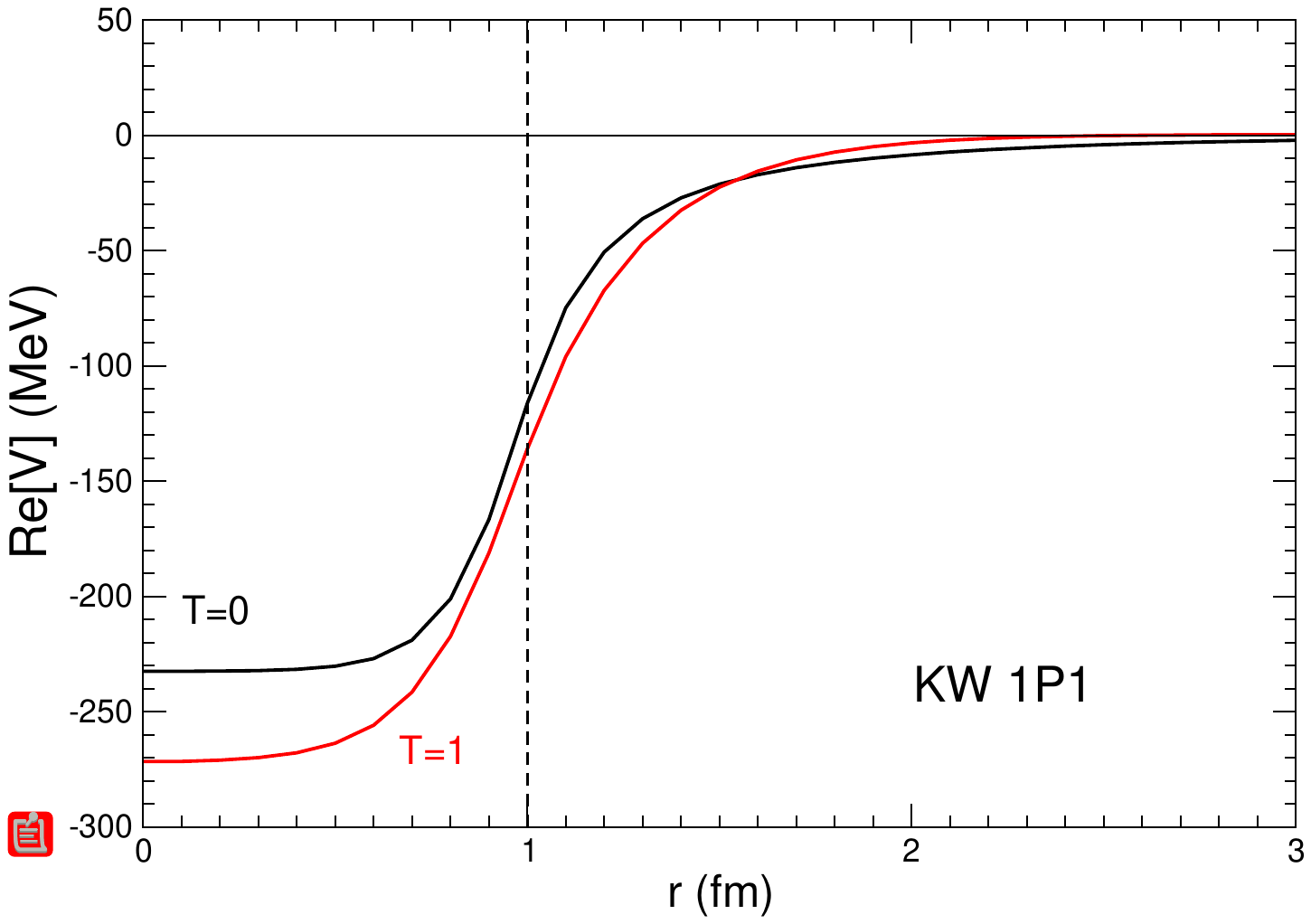}
\includegraphics[width=4.cm]{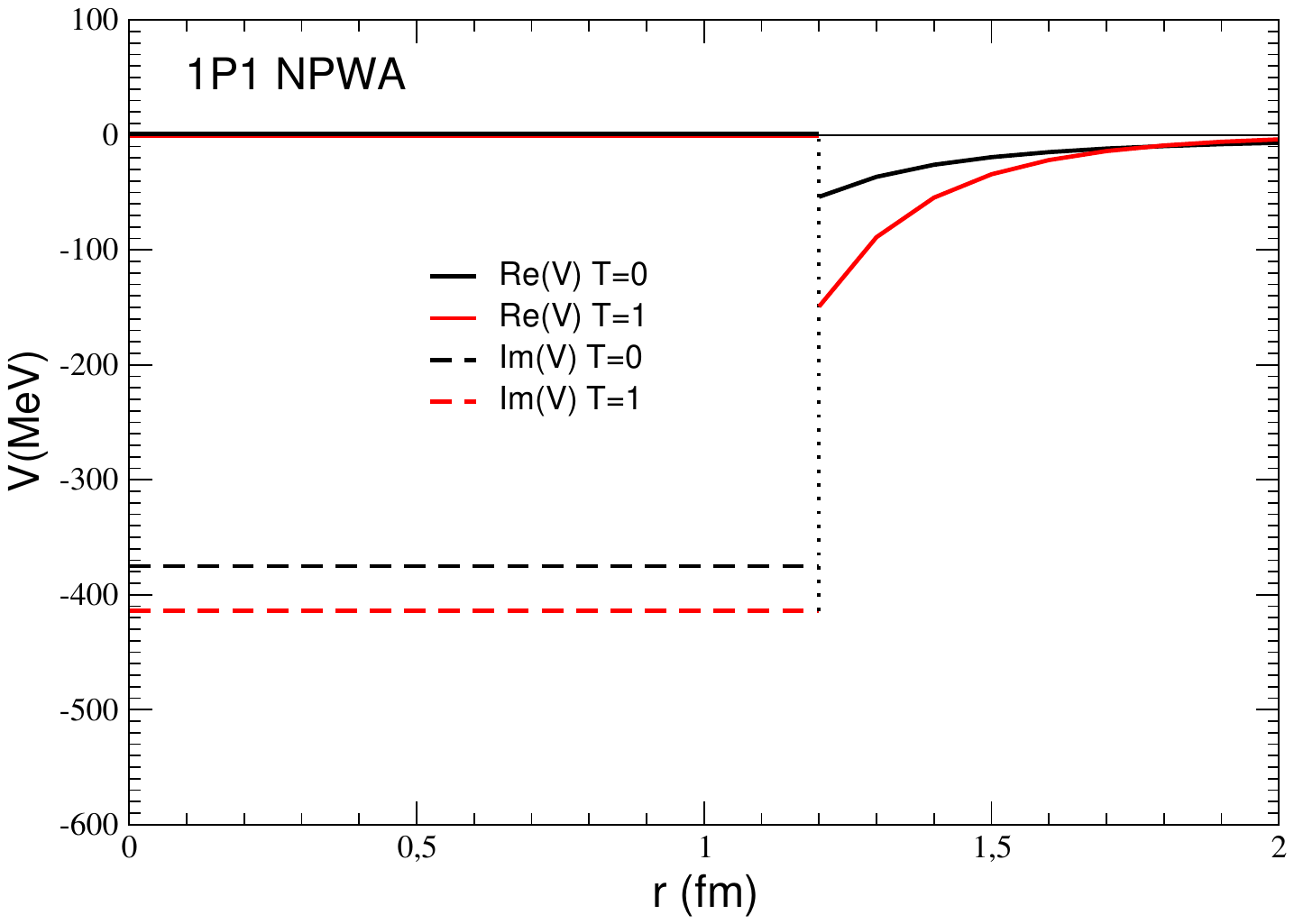}
\end{center}
\caption{Real parts of $^1$P$_1$ potentials for both isospins (T).}\label{U_1P1}
\end{figure}

\vspace{.cm}
\subsection*{$^3$P$_0$}

\bigskip

All models agree with a huge attraction in T=0 state, $\sim$ 1 GeV  at $r$=1 fm. 
The NWPA does not require such a large attraction and the fit is done with  a potential depth of $\approx$ 100 MeV in the internal region,
although matched with a pion potential of 350 MeV.

For T=1, and in view of the repulsive scattering lengths, there is also a general agreement in the repulsive character of the interaction,
although the direct inspection of the potentials requires some caution.

In KW model, the $^{33}$P$_0$ potential is repulsive everywhere, while DR2 has an attractive pocket below $r$=0.7 fm which is  fully compensated by the centrifugal term.
Paris potential (at E=0) has also a deep attractive pocket (-260 MeV) between $r$=0.5 fm and $r$=0.7 fm. It is almost totally compensated
by the centrifugal barrier, but there remains a shallow attractive pocket (-35 MeV)  between 0.56 and 0.63 fm. 
Under these dynamical conditions there is no room for developing a resonance, especially taking into account  the repulsive E-dependent amplitude at positive energies.
The examined models are, thus, globally repulsive.

However, the NPWA  requires an overall attractive short range contribution of $\approx$ 150 MeV, though matched at r=1.2 fm with a  repulsive $V_{\pi}$.

\begin{figure}[htbp]
\begin{center}
\includegraphics[width=4.cm]{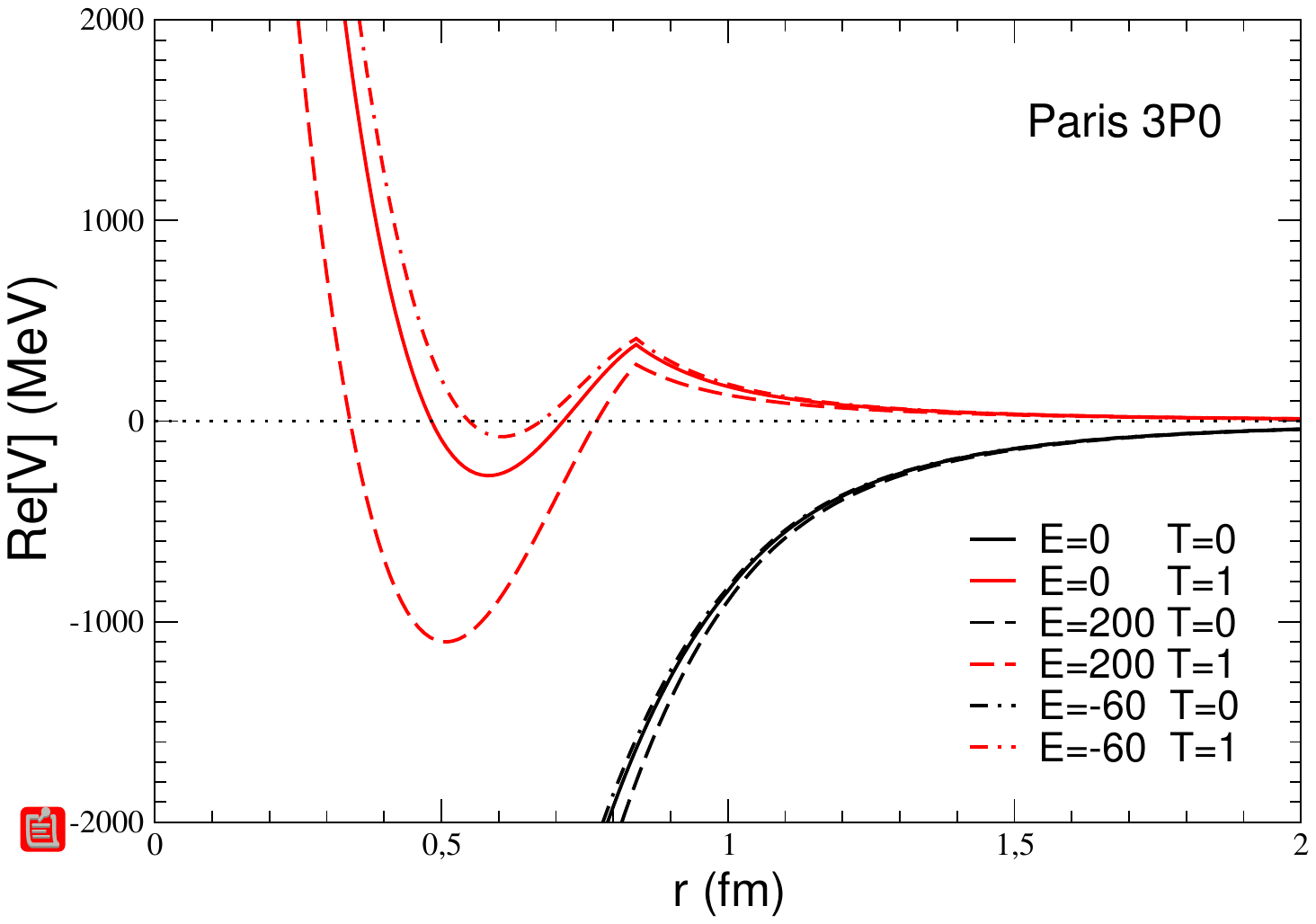}
\includegraphics[width=4.cm]{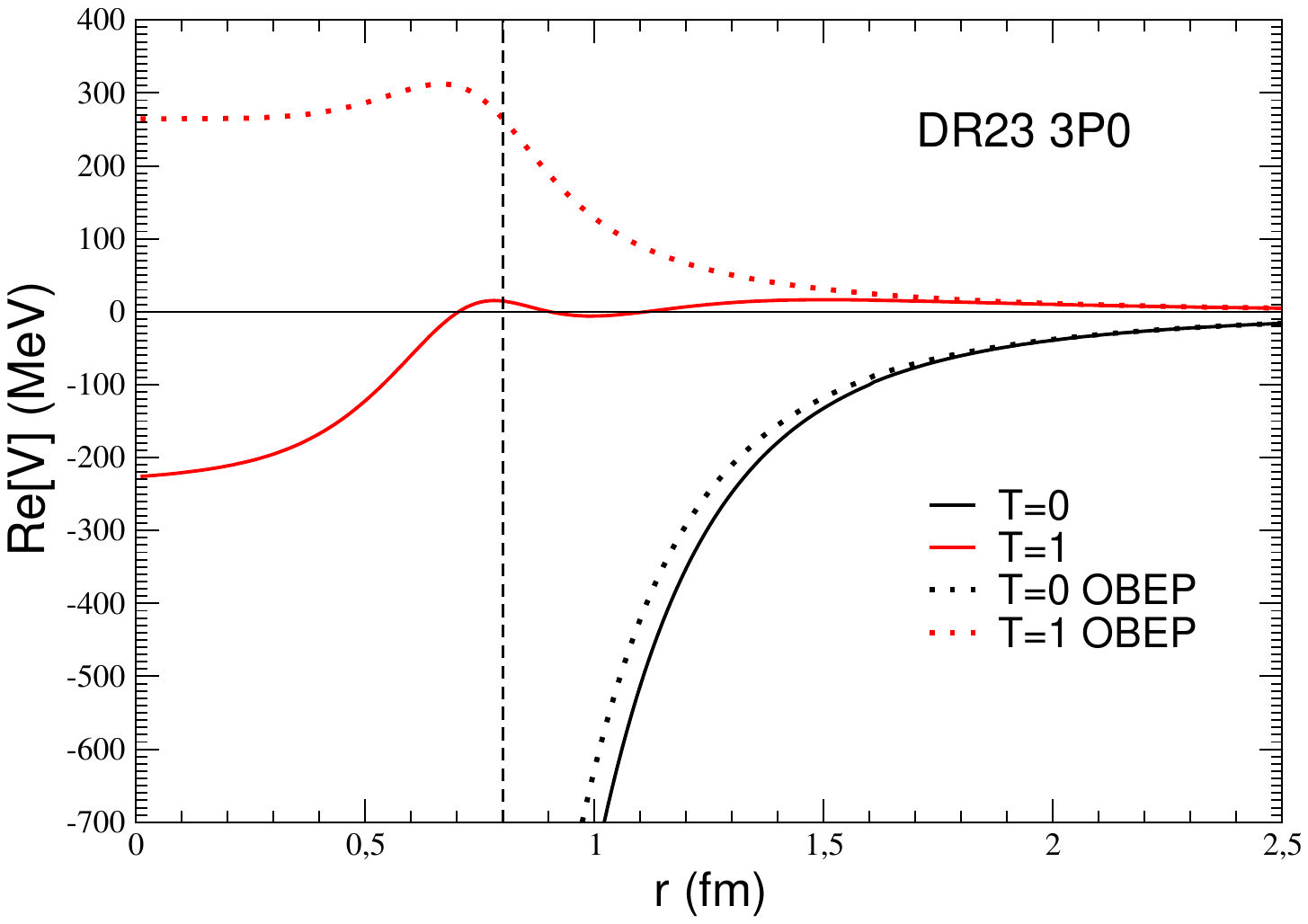}

\includegraphics[width=4.cm]{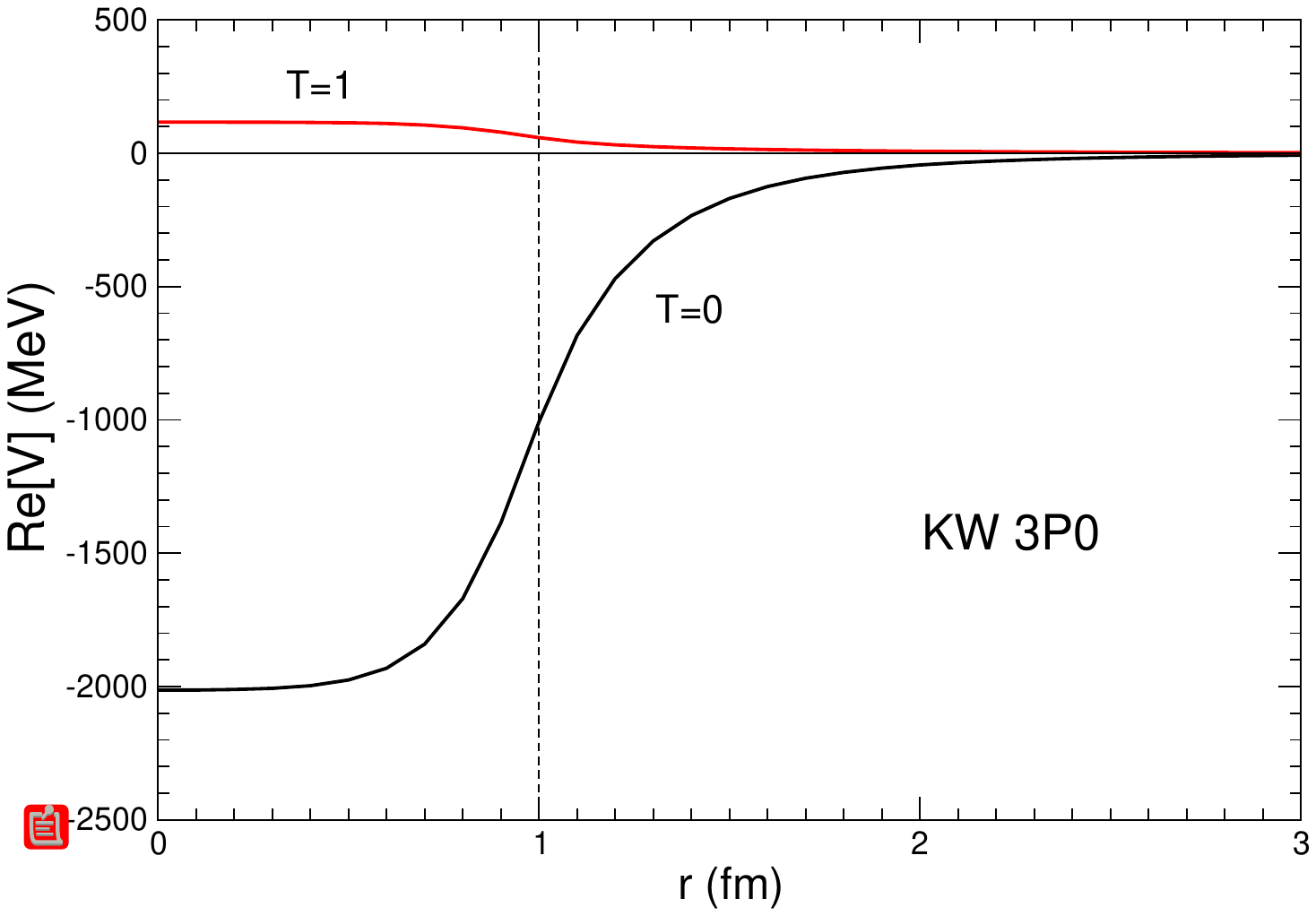}
\includegraphics[width=4.cm]{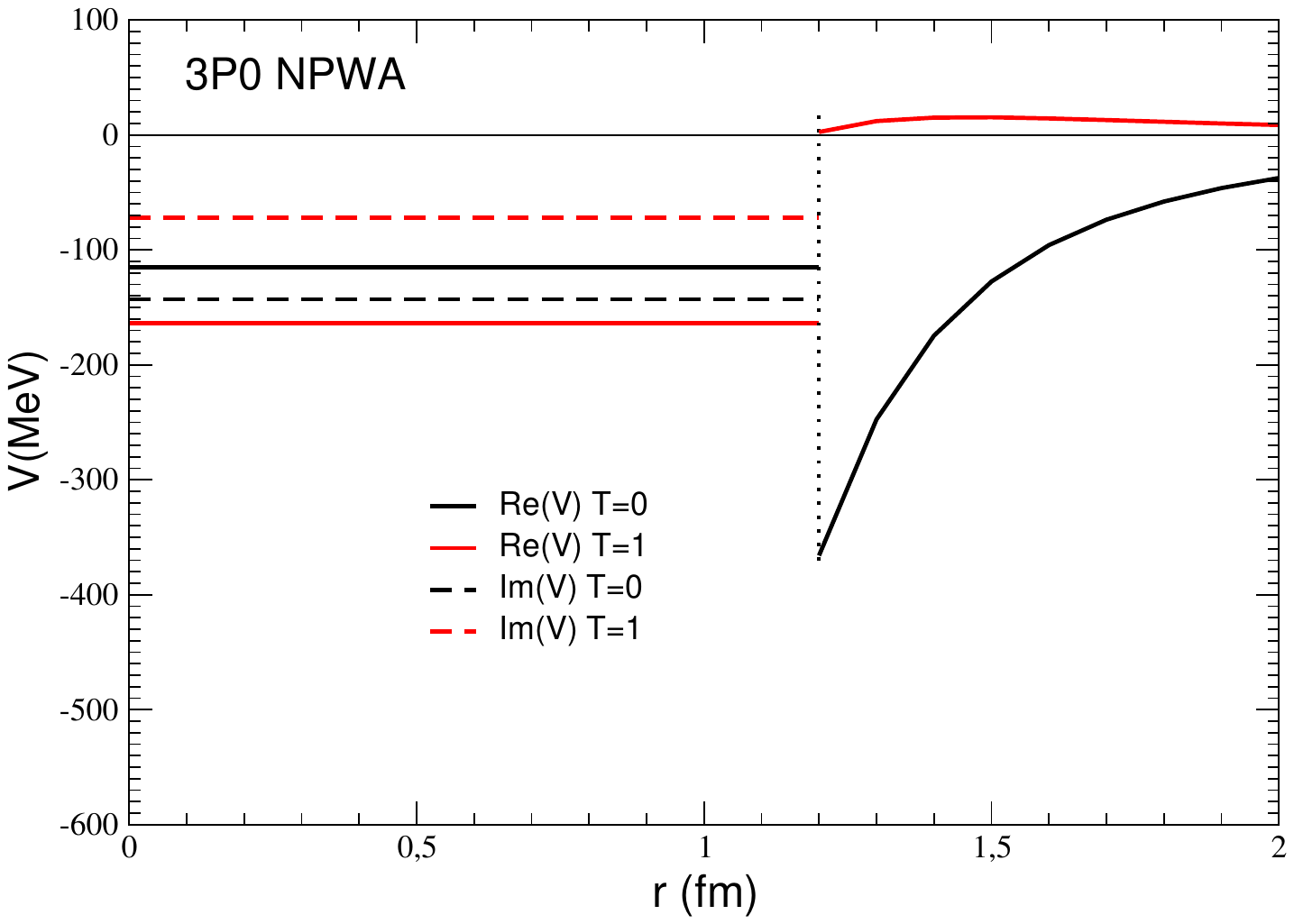}
\end{center}
\caption{Real parts of $^3$P$_0$ potentials for both isospins (T).}\label{U_3P0}
\end{figure}

\vspace{-.5cm}
\subsection*{$^3$P$_1$}

\bigskip

This PW has repulsive scattering length in both isospins states. 
For T=0, KW, DR are indeed repulsive (once the centrifugal barrier is included) 
but NPW  has an attractive pocket and Paris remains strongly attractive (2 GeV at $r$=0.5 fm). 
\vspace{-.5cm}
\begin{figure}[htbp]
\begin{center}
\includegraphics[width=4.cm]{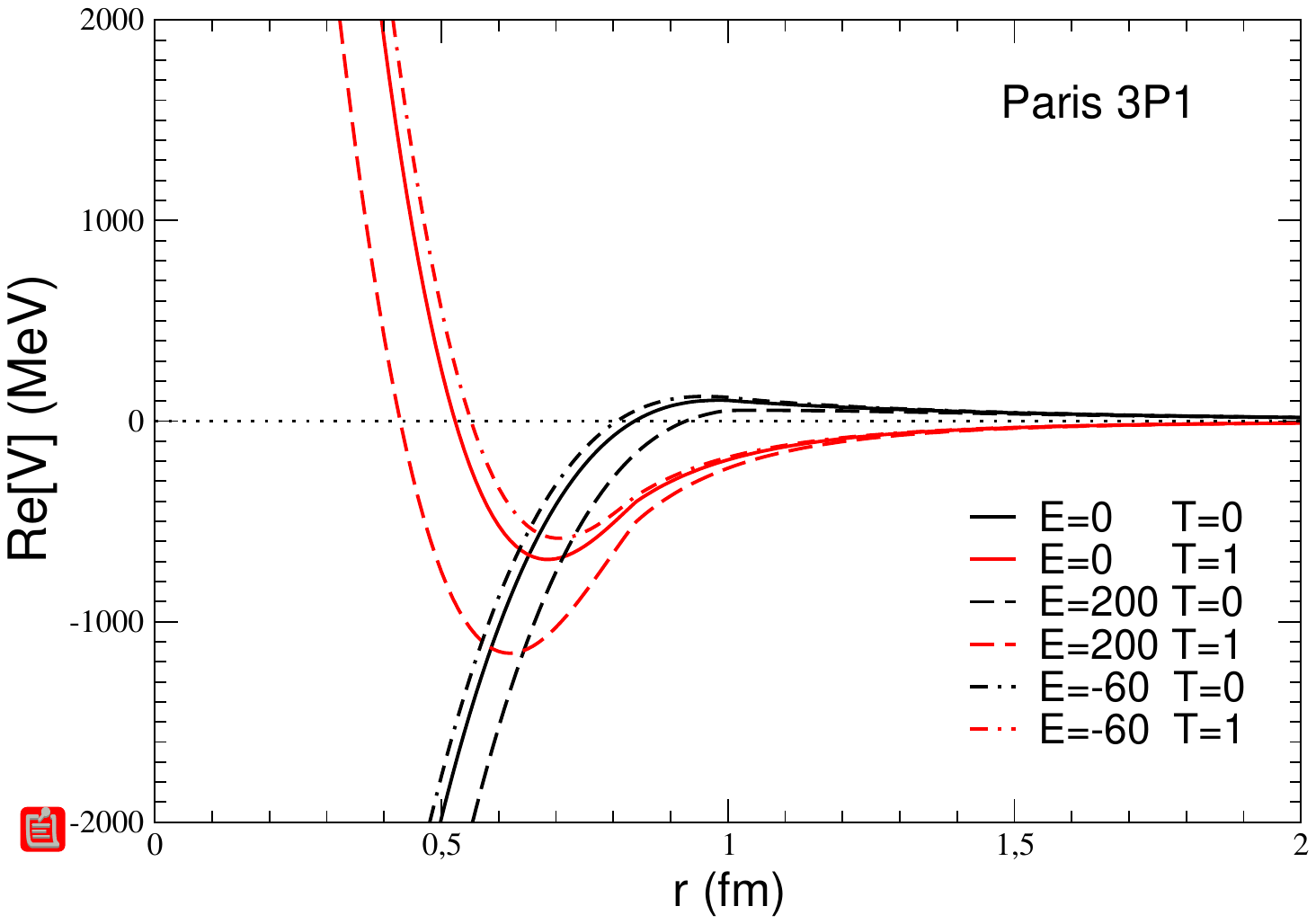}
\includegraphics[width=4.cm]{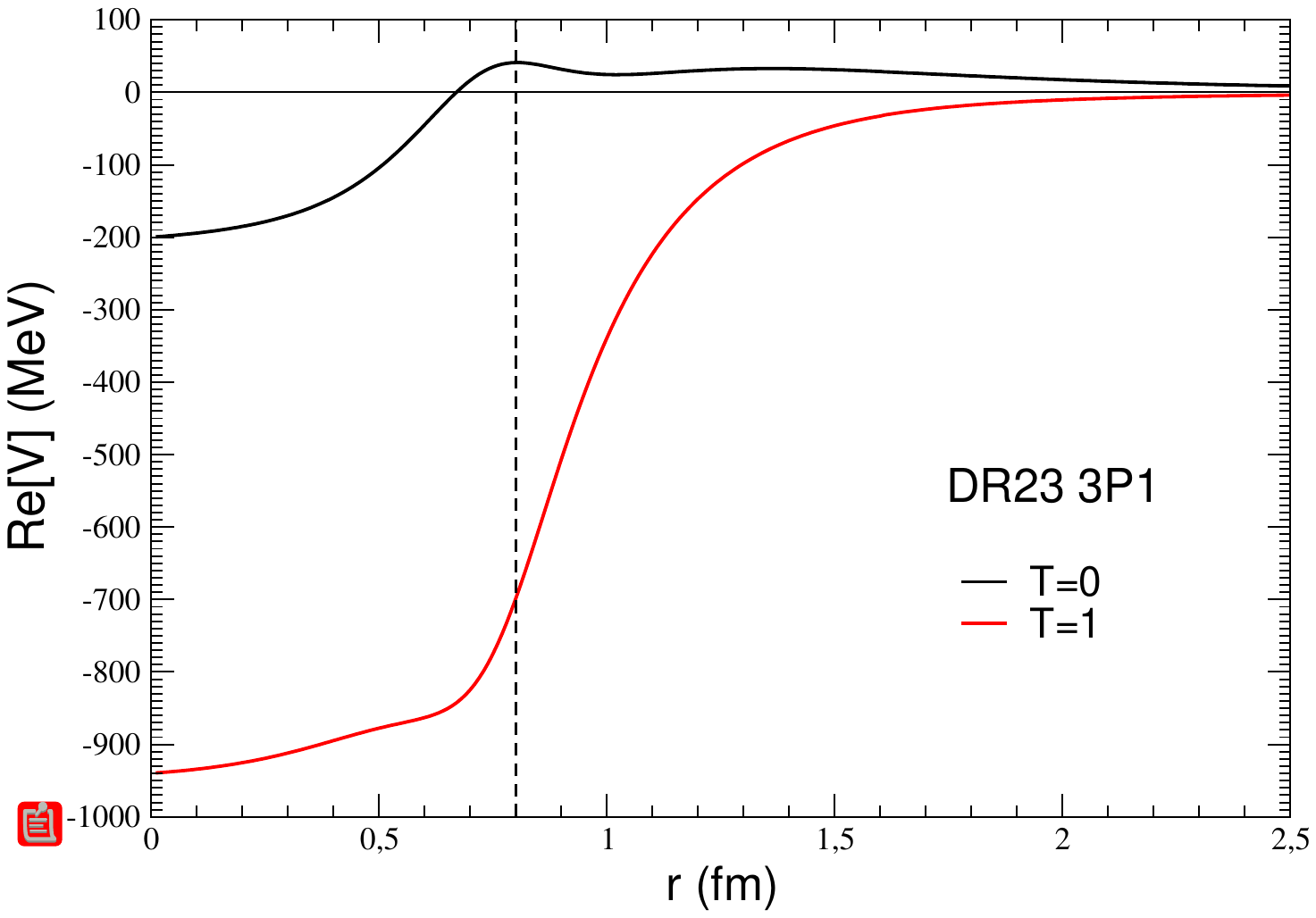}

\includegraphics[width=4.cm]{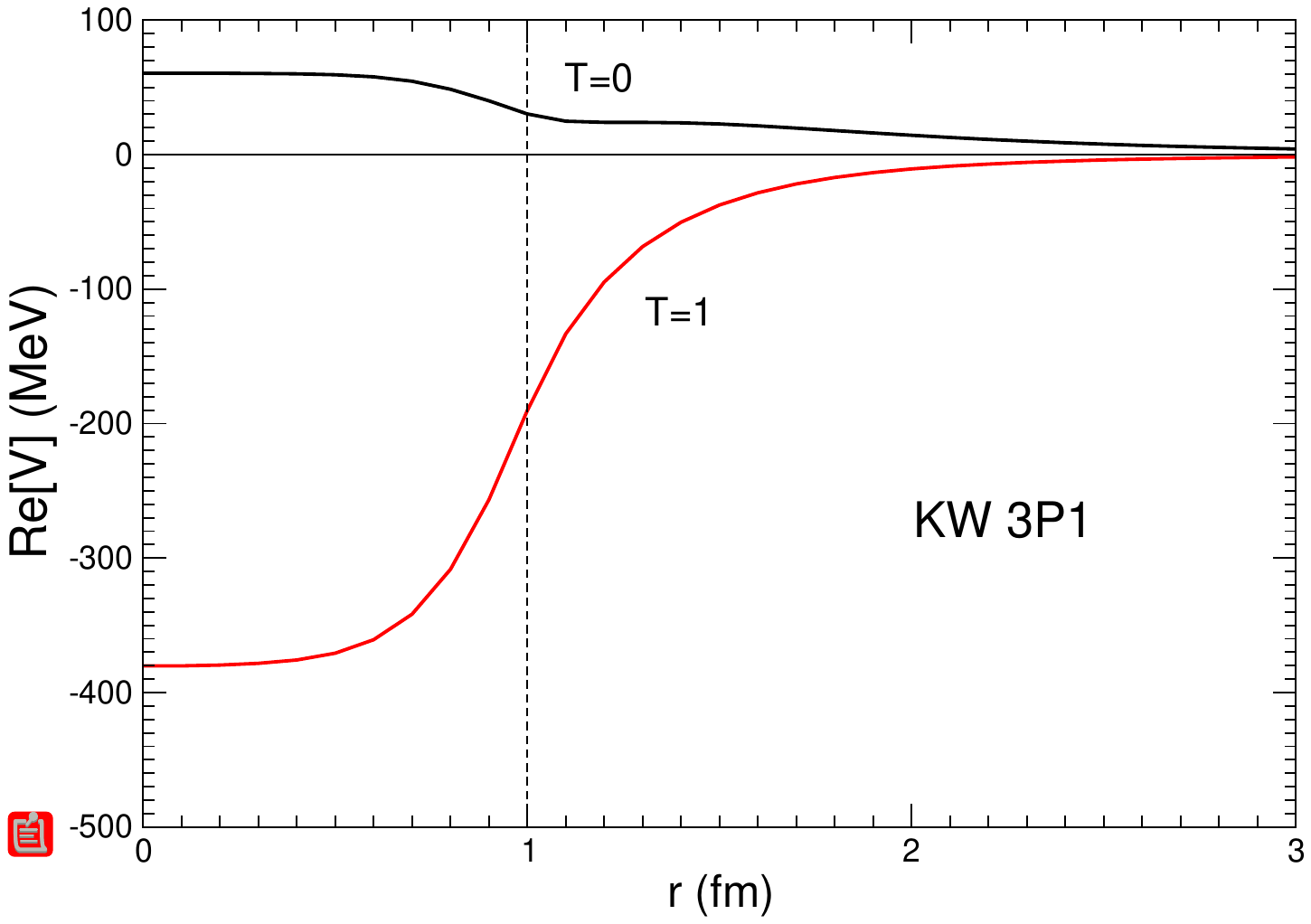}
\includegraphics[width=4.cm]{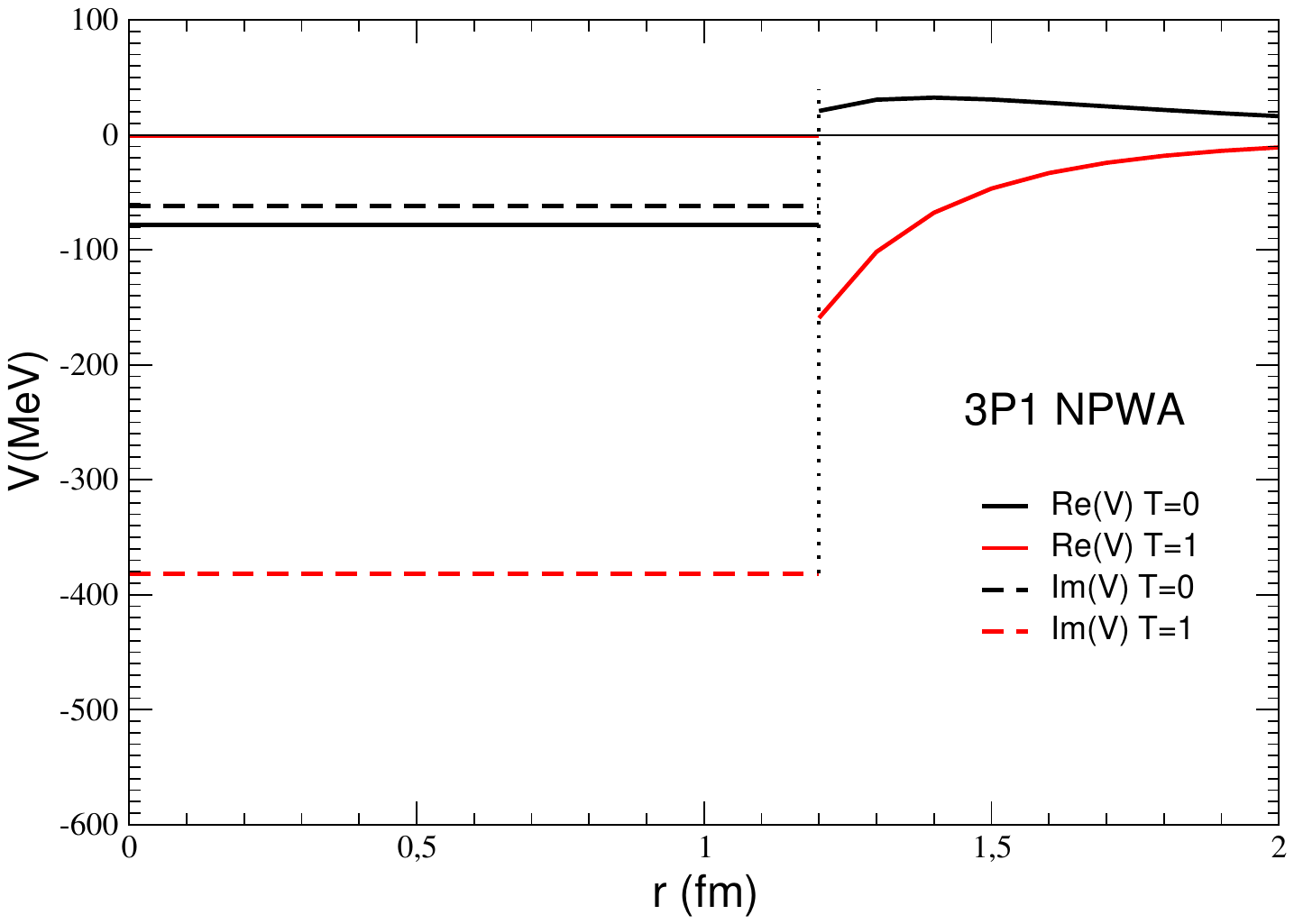}
\end{center}
\caption{Real parts of $^3$P$_1$ potentials for both isospins (T).}\label{U_3P1}
\end{figure}

\subsection*{$^3$PF$_2$}

The P-waves tensor coupled state
is  the one exhibiting the largest differences among models.
The $^3$P$_2$ component is attractive in bot isospin states for all models but large  differences in the strength are observed among them.
The $^3$F$_2$ is attractive and huge in all models but vanishes in the NPWA.
For T=1, the potential is repulsive in Paris 09 but attractive in all the other models:
 the unique case where NPWA requires an  attraction stronger than in  all  other potentials
\begin{figure}[htbp]
\begin{center}
\includegraphics[width=4.cm]{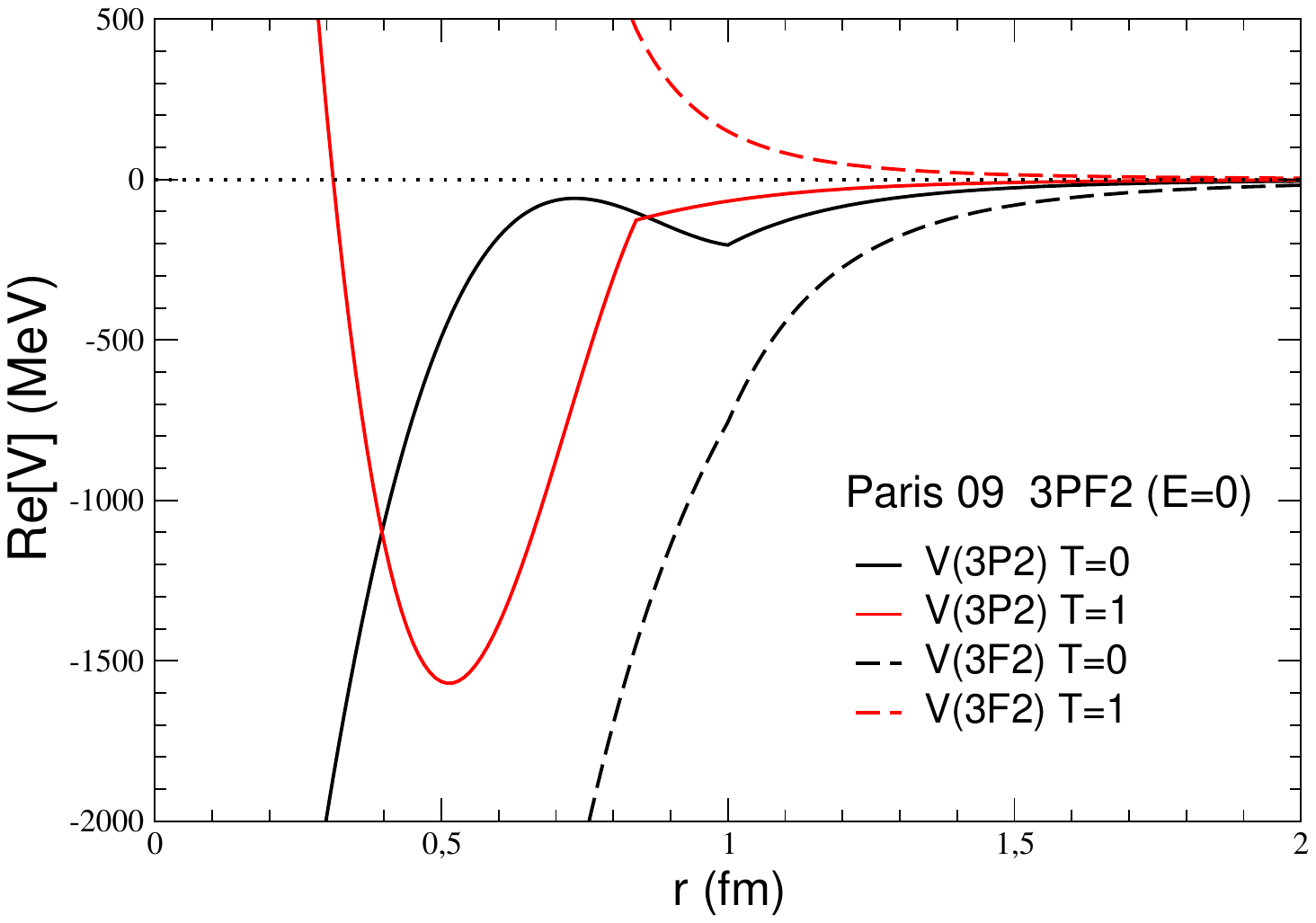}
\includegraphics[width=4.cm]{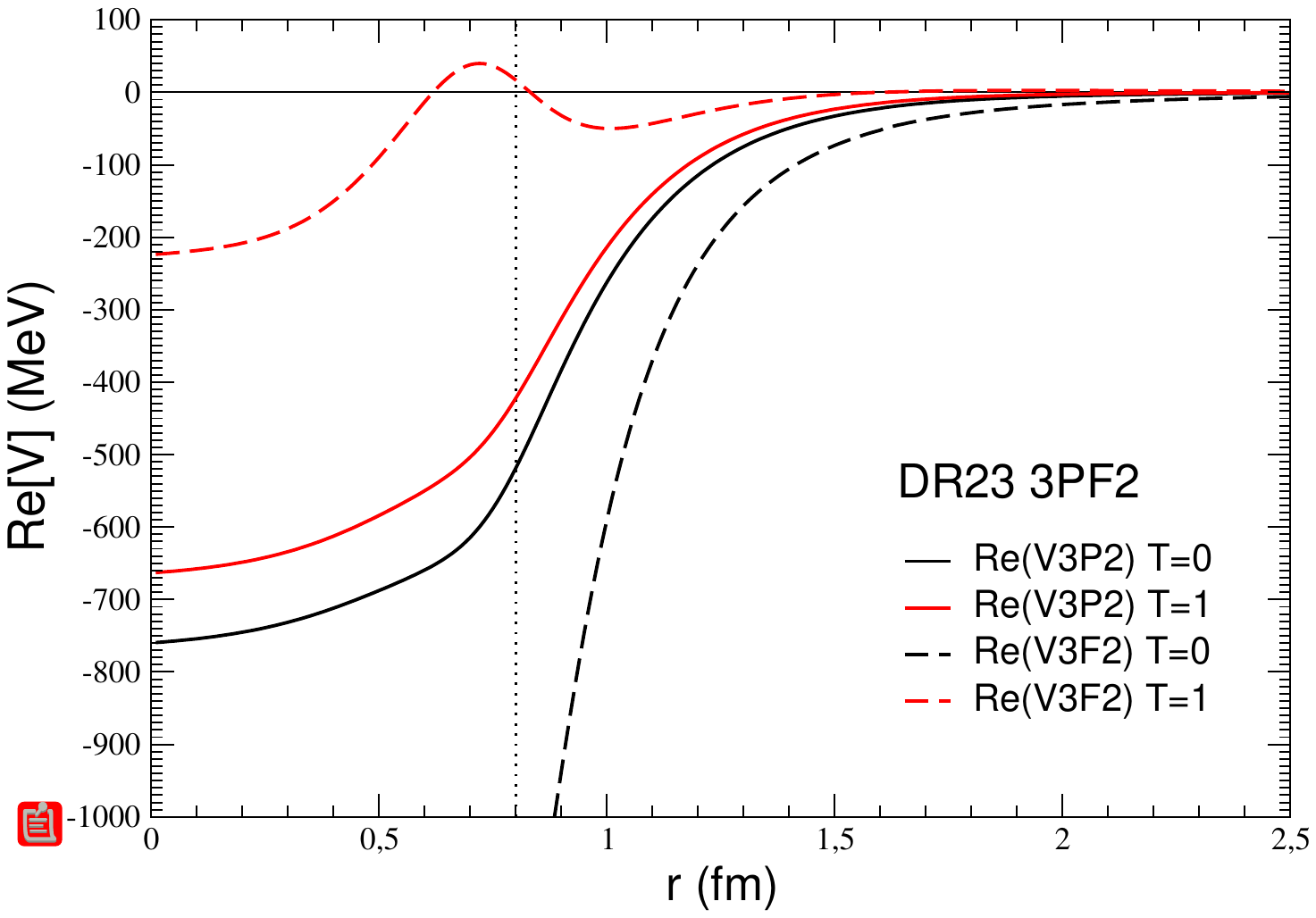}

\includegraphics[width=4.cm]{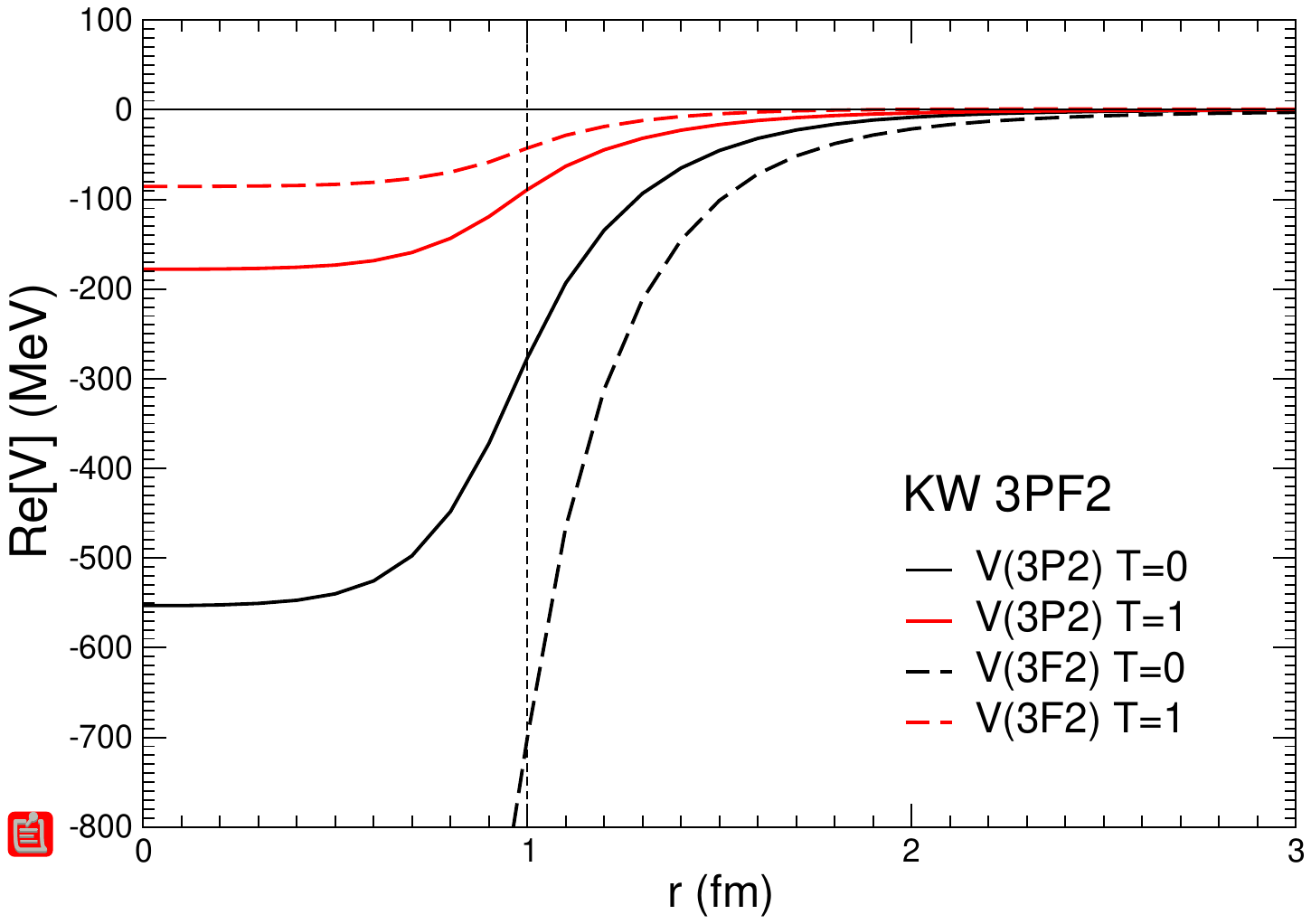}
\includegraphics[width=4.cm]{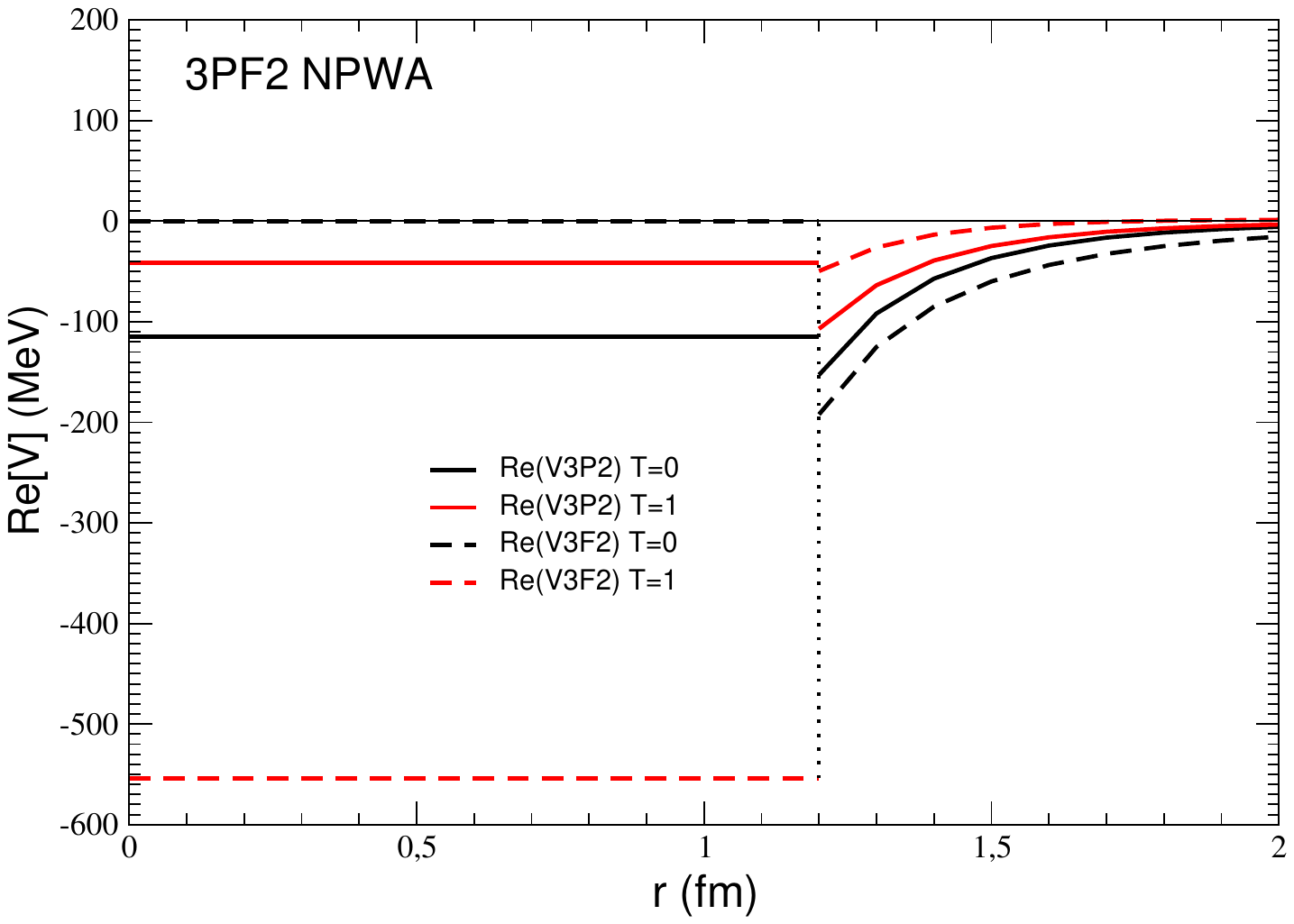}
\end{center}
\caption{Real parts of $^3$P$_2$ and $^3$F$_2$ potentials for both isospins (T).}\label{U_3PF2}
\end{figure}

\vspace{0cm}

Concerning the $^3$P$_2\to ^3$F$_2$  transition potentials, the same remarks as for $^3$S$D_1$ partial wave are in place.
\begin{figure}[htbp]
\includegraphics[width=4.cm]{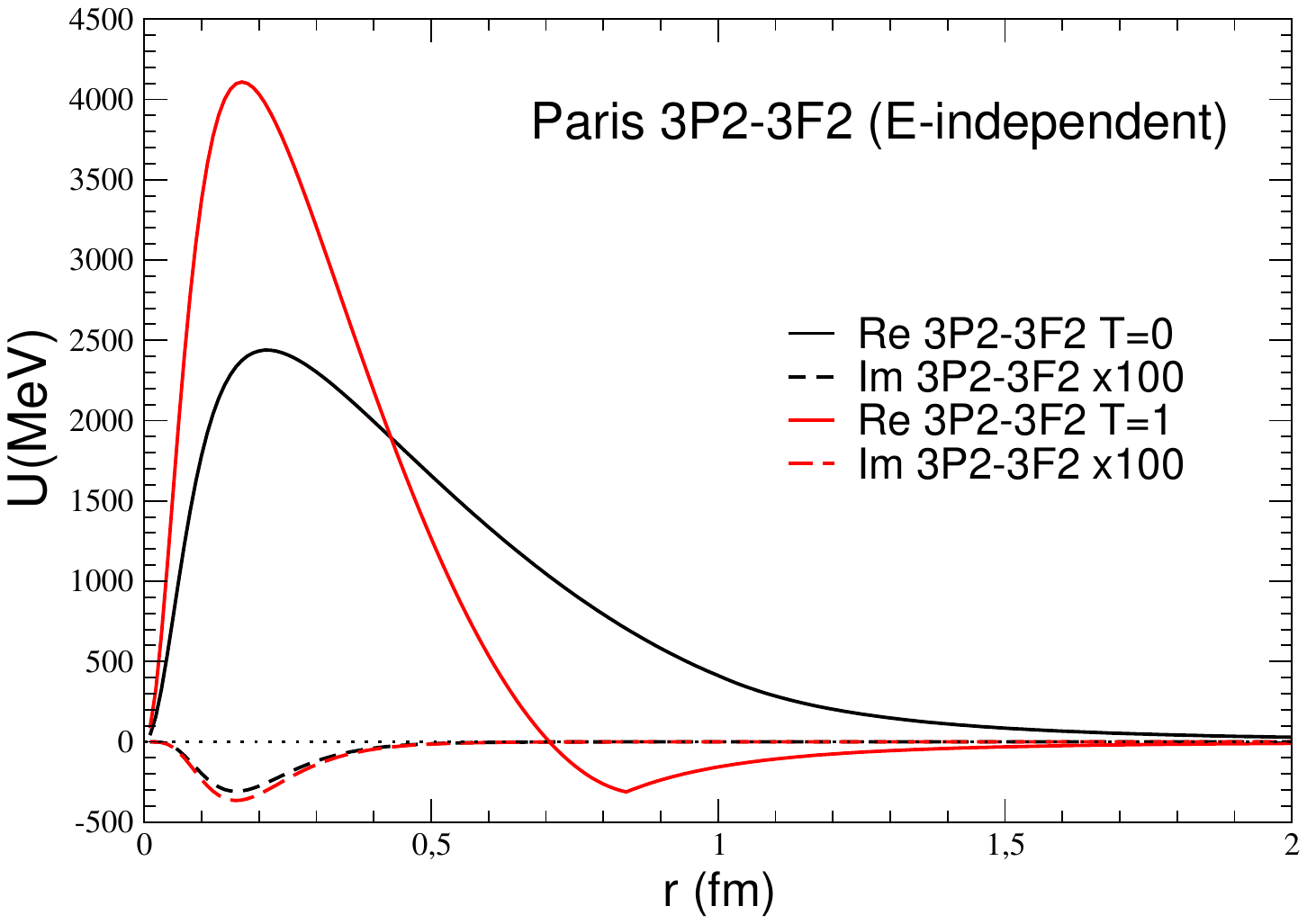}
\includegraphics[width=4.cm]{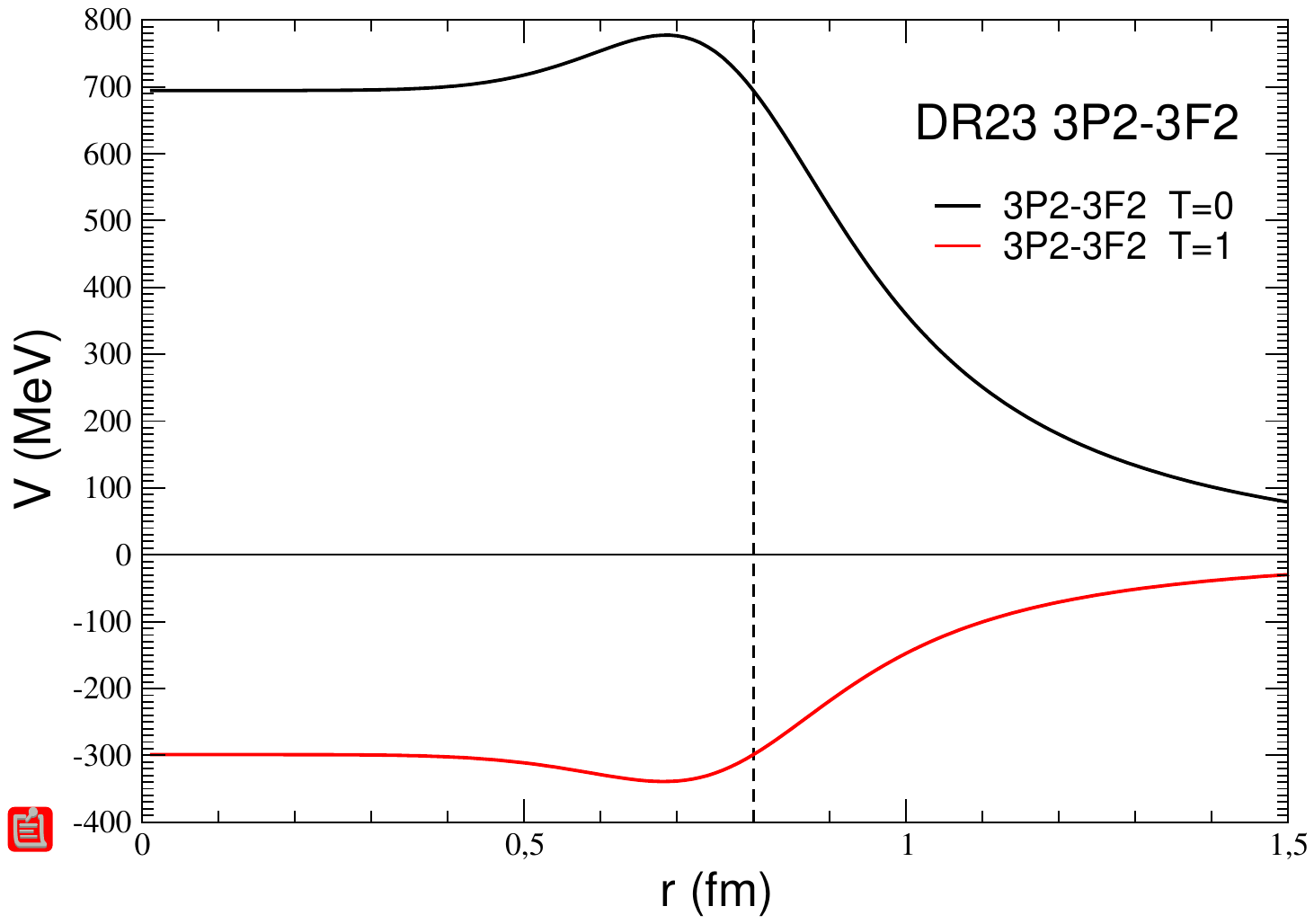}
\includegraphics[width=5.4cm]{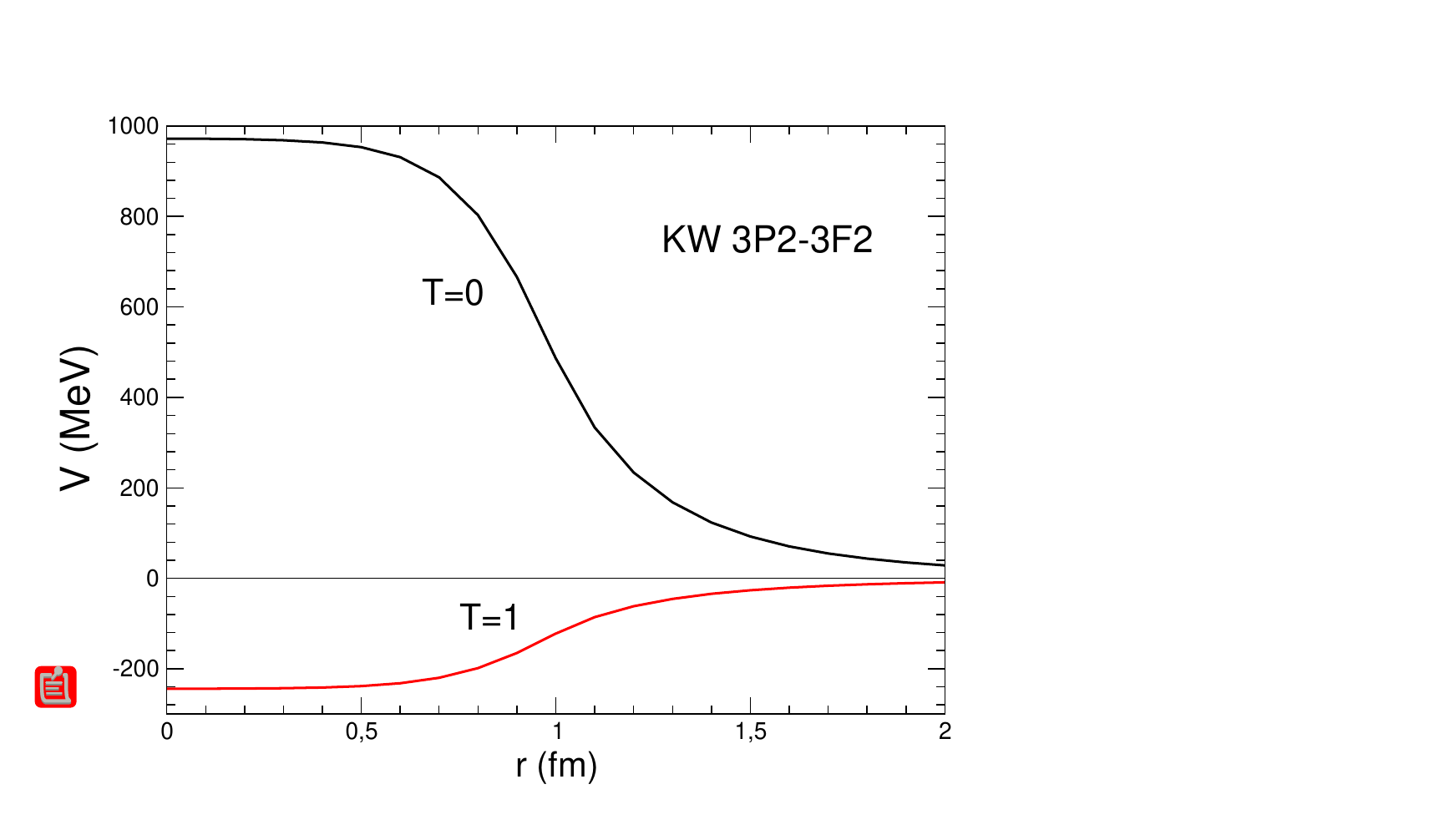}
\caption{$^3$P$_2 \to ^3$F$_2$ transition  potentials  for both isospins (T). In DR and KW models, they are real.}\label{V_3P2_3F2}
\end{figure}

\end{document}